\documentclass[fleqn,usenatbib]{mnras}

\usepackage{newtxtext,newtxmath}

\usepackage[T1]{fontenc}

\DeclareRobustCommand{\VAN}[3]{#2}
\let\VANthebibliography\thebibliography
\def\thebibliography{\DeclareRobustCommand{\VAN}[3]{##3}\VANthebibliography}

\usepackage{graphicx}	
\usepackage{amsmath}	

\defcitealias{maia+19}{Paper I}
\defcitealias{santos+20}{Paper II}
\defcitealias{dias+21}{Paper III}
\defcitealias{dias+22}{Paper IV}
\defcitealias{bica+22}{Paper V}
\defcitealias{oliveira+23}{Paper VII}
\defcitealias{saroon+23}{Paper VIII}

\title[The SMC Southern Bridge in 8D]{The VISCACHA survey -- IX. The SMC Southern Bridge in 8D}

\author[Parisi et al.]
{M. C. Parisi$^{1,2}$\thanks{E-mail: cparisi@unc.edu.ar},
R. A. P. Oliveira$^{3,4}$,
M. Angelo$^{5}$,
B. Dias$^{6}$,
F. Maia$^{7}$,
S. Saroon$^{6}$,
C. Feinstein$^{8,9}$,
\newauthor
J. F. C. Santos Jr.$^{10}$,
E. Bica$^{11}$,
B. Pereira Lima Ferreira $^{10}$,
J. G. Fern\'andez$-$Trincado$^{12}$,
\newauthor
P. Westera$^{13}$
D. Minniti$^{6,14,15}$,
E. R. Garro$^{16}$,
O. J. Katime Santrich$^{17}$,
B. De Bortoli$^{8,9}$,
S. Souza$^{3}$,
\newauthor
L. Kerber$^{17}$,
A. P\'erez$-$Villegas$^{18}$
\\
$^{1}$Observatorio Astron\'omico, Universidad Nacional de C\'ordoba, Laprida 854, X5000BGR, C\'ordoba, Argentina.\\
$^{2}$Instituto de Astronom{\'\i}a Te\'orica y Experimental (UNC-CONICET), Laprida 854, X5000BGR, C\'ordoba, Argentina.\\
$^{3}$Universidade de S\~ao Paulo, IAG, Rua do Mat\~ao 1226, Cidade Universit\'aria, S\~ao Paulo 05508-900, Brazil\\
$^{4}$Astronomical Observatory, University of Warsaw, Al. Ujazdowskie 4, 00-478 Warszawa, Poland\\
$^{5}$Centro Federal de Educa\c c\~ao Tecnol\'ogica de Minas Gerais, Av. Monsenhor Luiz de Gonzaga, 103, 37250-000 Nepomuceno, MG, Brazil\\
$^{6}$Instituto de Astrof\'isica, Facultad de Ciencias Exactas, Universidad Andres Bello, Av. Fern\'andez Concha 700, Santiago, Chile\\
$^{7}$Instituto de F\'isica, Universidade Federal do Rio de Janeiro, 21941-972, Rio de Janeiro, RJ, Brazil\\
$^{8}$Facultad de Ciencias Astron\'omicas y Geof\'isicas, Universidad Nacional de La Plata, Paseo del Bosque s/n, 1900 La Plata, Argentina\\
$^{9}$Instituto de Astrofísica de La Plata (CCT La Plata, UNLP-CONICET), Paseo del Bosque s/n, 1900 La Plata, Argentina\\
$^{10}$Departamento de F\'isica, ICEx - UFMG, Av. Antônio Carlos 6627, 31270-901 Belo Horizonte, Brazil\\
$^{11}$Universidade Federal do Rio Grande do Sul, Departamento de astronomia, CP15051, Porto Alegre91501-970,Brasil\\
$^{12}$Instituto de Astronom\'ia, Universidad Cat\'olica del Norte, Av. Angamos 0610, Antofagasta, Chile\\
$^{13}$Universidade Federal do ABC, Centro de Ci\^{e}ncias Naturais e Humanas, Avenida dos Estados, 5001, 09210-580, Santo Andr\'{e}, SP,
Brazil\\
$^{14}$Vatican Observatory, V00120 Vatican City State, Italy\\
$^{15}$Departamento de F\'sica, Universidade Federal de Santa Catarina, Trindade 88040-900, Florianópolis, Brazil\\
$^{16}$ESO - European Southern Observatory, Alonso de Cordova 3107, Vitacura, Santiago, Chile\\
$^{17}$Universidade Estadual de Santa Cruz (UESC), Departamento de Ciências Exatas, Rodovia Jorge Amado km 16, 45662-900 Ilh\'eus, Brazil.\\
$^{18}$Instituto de Astronomía, Universidad Nacional Autónoma de México, Apartado Postal 106, C. P. 22800, Ensenada, B. C., Mexico\\
}
\date{Accepted XXX. Received YYY; in original form ZZZ}

\pubyear{2023}

\begin{document}
\label{firstpage}
\pagerange{\pageref{firstpage}--\pageref{lastpage}}
\maketitle

\begin{abstract}
The structure of the Small Magellanic Cloud (SMC) outside of its main body is characterised by tidal branches resulting from its interactions mainly with the Large Magellanic Cloud (LMC). Characterising the stellar populations in these tidal components helps to understand the dynamical history of this galaxy and of the Magellanic system in general. We  provide full phase-space vector information for  Southern Bridge clusters. 
We performed a photometric and spectroscopic analysis of twelve SMC clusters, doubling the number of SMC clusters with full phase-space vector information known to date. We reclassify the sample considering 3D distances and 3D velocities. We found that some of the clusters classified as   Southern Bridge objects according to the projected 2D classification actually belong to the Main Body and Counter-Bridge in the background. The comparison of the kinematics of the genuine foreground Bridge clusters with those previously analysed in the same way reveals that  Southern Bridge clusters are moving towards the LMC and  share the kinematics of the Northern Bridge. Adding to our sample clusters from the literature with CaT metallicity determinations we compare the age-metallicity relation of the Southern Bridge with the one of the Northern Bridge. We reinforce the idea that both regions do not seem to have experienced the same chemical enrichment history and that there is a clear absence of clusters in the Northern Bridge older than 3Gyr and more metal-poor than -1.1, which would not seem to be due to a selection effect. 
\end{abstract}




\section{Introduction}
\label{sec:introduction}

The Large and Small Magellanic Clouds (LMC and SMC, respectively) constitute one of the most interesting pairs of interacting dwarf galaxies. Due to their proximity to the Milky Way (MW) (LMC: 49.59$\pm$0.54 kpc, \citealt{Pietrzynski19}; SMC: 62.44 $\pm$ 0.81 kpc \citealt{Graczyk+20}) and the fact that their individual stars can be resolved, these galaxies have made it possible to study in detail not only the chemical and dynamical evolution but also how interaction tidal processes can shape the morphology of this kind of galaxies 
\citep[e.g.][] {zivick+18,deleo+20,patel+20,f_trincado+20,nidever+23}. \\

Numerous evidences that the Magellanic Clouds (MCs) are interacting with each other and with the MW can be found imprinted in different characteristics of their stellar populations \citep[e.g.][]{dobbie+14a,subramanian+17,deleo+20}. For example, since the MCs are gas-rich galaxies, close encounters between them would have produced bursts of star and cluster formation \citep[e.g.][]{glatt+10,bitsakis+18,rubele+18} and subsequent chemical enrichment \citep[e.g.][]{pt98}. Also, complex patterns of velocities can be observed in the stellar populations due to galaxy-galaxy interaction \citep[e.g.][]{niederhofer+18,Niederhofer+21,zivick+18,james+21} and stellar overdensities are suggested to be the products of stretching and tidal stripping processes  \citep[e.g.][]{pieres+17,ElYoussoufi+2021,cullinane+23}.

Besides, several tidal trails can be found around the MCs \citep[e.g.][]{mackey+18,belokurov+19,nidever+19,gaia+21b,ElYoussoufi+2021}. In the particular case of the SMC, the galaxy we will focus on in this paper,  the Magellanic Bridge \citep[e.g.][]{omkumar+21,dias+21}, the Counter-Bridge \citep[e.g.][]{ripepi+17,muraveva+18,omkumar+21,dias+21,Niederhofer+21}, and the West Halo \citep{dias+16,zivick+18,niederhofer+18,tatton+21} are examples of tidal trails, all of them located outside of the Main Body \citep{dias+16,parisi+22} and interpreted as the consequences of the LMC-SMC interaction \citep[e.g][]{besla+12,diaz+12}.  Therefore, studying the stellar populations spatially separated into these tidal components can help to understand the dynamical history of the SMC in the context of its interactions with its partner the LMC, and the MW. Any theoretical model that attempts to explain the formation, evolution, and history  of the MCs must be able to reproduce the properties of their stellar populations.  \\

Star clusters are excellent tracers of all the processes that their host galaxies have undergone, and the star cluster system of the SMC is not an exception. \citet{dias+14} first introduced the framework of analysing the properties of the star clusters separating them into samples according to the coincidence of their projected positions with the aforementioned tidal components (Main Body, Wing/Bridge, Counter-Bridge and West Halo, see Fig. \ref{fig:distrib}). This approach was subsequently applied to some works yielding interesting results \citep[e.g.][]{dias+16,debortoli+22,parisi+22}.

In that sense, our group, as part of the international collaboration VISCACHA (VIsible Soar photometry of star Clusters in tApii and Coxi HuguA) Survey\footnote{\url{http://www.astro.iag.usp.br/~viscacha/}} and its spectroscopic follow-up, has been systematically observing star clusters in the SMC outskirts in order to analyse their structures, kinematics, dynamics and chemical evolution. The obtained photometric (SAMI/SOAR) and spectroscopic (GMOS/Gemini-S) data allowed the determination of accurate distances, radial velocities (RV), ages and metallicities of a number of clusters from the catalogue of \citet{bica+20}. Using this information, along with proper motions (PM) from Gaia eDR3\footnote{\url{https://archives.esac.esa.int/gaia}} \citep{gaia+21a}, we analysed the Magellanic Bridge, Counter Bridge \citep[][hereafter Paper III]{dias+21} and  the West Halo \citep[][hereafter Paper IV]{dias+22} in the 8D parameter space (6D phase-space vector, age, and metallicity), discovering that the Bridge and Counter-Bridge are roughly aligned along the LMC-SMC direction and they are both moving away from the SMC in opposite directions along the LMC-SMC axis.\\

The Magellanic Bridge is a large structure that extends towards the LMC, inicially thought only to contain gas and young stellar populations \citep{putman+03,harris07}. Subsequent studies revealed that there is a second branch (which we call the Southern Bridge) not aligned with the LMC and having an old stellar content traced by RR Lyrae \citep{belokurov+17,Jacyszyn+17}. More recently, our work about the Bridge \citepalias{dias+21} provided an even more interesting scenario, showing the existence of a third branch  (see also some initial partial evidence by \citealt{nidever+13,omkumar+21}). This third branch (which we refer to as Northern Bridge) would be traced by star clusters that move from the SMC towards the LMC. We marked in Fig. \ref{fig:distrib} the projected location on the sky of the Southern Bridge and Northern Bridge, as well as the other previously mentioned tidal components. The $N$-body simulation of \citet{diaz+12} suggested that the Counter-Bridge was formed together with the Magellanic Bridge ($\sim$ 250 Myr ago) as its tidal counterpart. In \citetalias{dias+21} we found the first star cluster belonging to the Counter-Bridge. Additionally, in  VISCACHA's \citetalias{oliveira+23} \citep{oliveira+23} two groups of star clusters were identified: one younger (formed in-situ) and one older (tidally dragged from the SMC) than the Magellanic Bridge. 

Continuing this line of research, we now focus on the Southern Bridge. We also include two clusters from the West Halo and one cluster from the Counter-Bridge for which we have data. In Sections \ref{sec:obs} and \ref{sec:parameters} we describe the observations, data reduction and the procedures we employed for the cluster parameter determinations. We discuss in Section \ref{sec:discussions} the structure and kinematics of the studied cluster sample and we summarise our work and conclusions in Section \ref{sec:conclusions}. 

\section{Observation}
\label{sec:obs}

\begin{figure}
    \centering
    \includegraphics[width=\columnwidth]{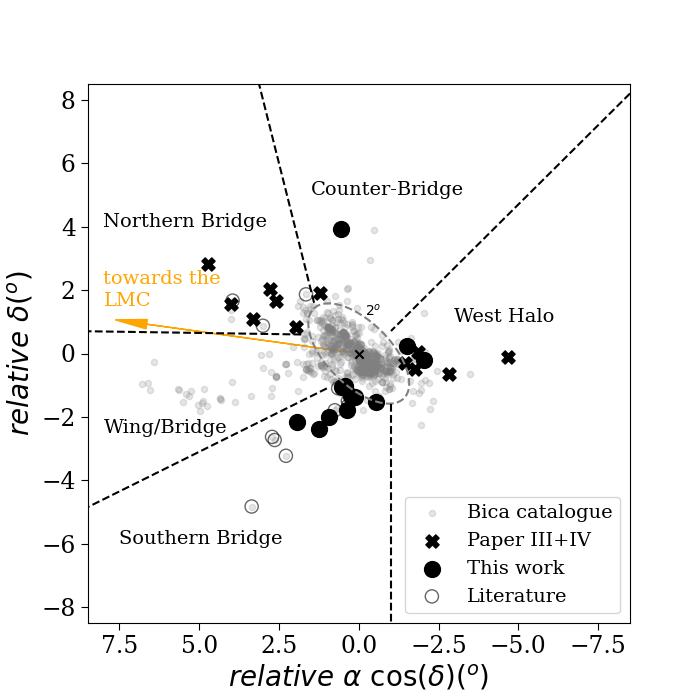}
    \caption{Distribution of SMC clusters included in the present work  as well as those studied previously by the VISCACHA collaboration \citepalias{dias+21,dias+22} superimposed to the objects from \citet{bica+20} catalogue. Also clusters from the literature having CaT metallicities are included. The ellipse is used to delimit the projected location of the SMC main body. Thick dashed lines split the regions outside the main body into different 2D SMC components defined by \citet{dias+14,dias+16}. 
    }
    \label{fig:distrib}
\end{figure}

The SMC clusters sample analysed in this paper includes nine clusters from the Southern Bridge, two from the West Halo and one from the Counter-Bridge, according to the classification by \citet{dias+16}. In this way we complement our previous works (\citetalias{dias+21} and  \citetalias{dias+22}), wherein clusters of the Wing/Bridge, Counter-Bridge and West Halo were observed and analysed in a homogeneous way. The distribution of the clusters kinematicaly analysed by VISCACHA can be seen in Fig. \ref{fig:distrib}. Clusters from \citet{parisi+09,parisi+15} and \citet{geisler+23} with available CaT metallicities are also included in this figure.

The photometric images were obtained with the SOAR adaptive optics module (SAM) of the SOAR 4-meter telescope in the context of the VISCACHA project,  under the programs SO2016B-015, SO2017B-014, SO2019B-019, CN2020B-001 and SO2021B-017. SAM is a ground-layer adaptive optics (GLAO) instrument using a Rayleigh laser guide star (LGS) at $\sim$7 km from the telescope \citep{tokovinin+16}, and was employed with its internal CCD detector (SAMI) binned into a 2x2 configuration, yielding a resolution of 0.091 arcsec.pixel$^{-1}$ over its $3.1 \times 3.1$\,arcmin$^2$ field of view.
Long and short exposures were acquired for each target in the $V$ and $I$ filters of the Johnson-Cousins photometric system; \citet{Stetson:2000}  standard star fields were also observed during each night to allow photometric calibration.

For the particular case of cluster HW\,79, the obtained $I$ exposure was not deep enough due to technical problems. Therefore we used the magnitudes in the $g$ and $i$ bands from the Survey of the MAgellanic Stellar History (SMASH\footnote{\url{https://datalab.noirlab.edu/smash/smash.php}}, \citealt{nidever+17}), which were obtained from PSF photometry performed on their pre$-$images using DAOPHOT \citep{s87}. Quality control of the SMASH sources was ensured by filtering out those with DAOPHOT parameters $\rm{abs(SHARP)} > 1.0$ and $\chi^2 > 3.0$, and SExtractor stellar probability $< 0.8$, as instructed in \citet{martinezdelgado19}. An additional cut leaving only stars with three or more detections in the $g$ band ($ndetg \geq 3$) was made to avoid stars located in the gaps between the DECam CCDs.

On the other hand, the GMOS Trinational Project between Chile, Brazil and Argentina is in charge of carrying out the spectroscopic follow-up of the VISCACHA Survey. Therefore, cluster pre$-$images in the $g$ and $r$ filters and spectra for a number of red giant stars in the field of the clusters have been obtained in the CaT region. The instrument GMOS on Gemini-S was used to obtain a total of $\sim$ 370 spectra in the 12 observed clusters, under the programs GS-2019B-Q-303, GS-2020B-Q-321 and GS-2021B-Q-134. We use the next instrumental configuration: the R831 grating, the CaT blocking filter, slits having a $1.0^{\prime\prime}$ width and central wavelength at $\sim$8540 \AA. This configuration allowed to obtain spectra with a resolution of $\approx$ 2000 that cover a spectral range of $\sim$ 7345 - 9705 \AA. Since the CaT technique is suitable to red giant stars \citep{dch98,cole+04,dias+20b}, we selected the spectroscopic targets from the cluster colour-magnitude diagrams (CMDs) red giant branches (RGBs). All the relevant information about the observations are presented in Table \ref{tab:loggmos}.

\begin{table*}
\caption{Log of observations.}
\label{tab:loggmos}
\centering
\begin{tabular}{llcccc}
\hline
cluster       & date             & grism/filter & exp.time (s) & airmass & FWHM (") \\
\hline
\multicolumn{6}{c}{SAMI/SOAR photometry}\\
\hline
B\,4; OGLE\,317                 &   2021-11-08              &   V, I    &   3x400, 3x600    &   1.38 &   0.63, 0.50 \\
B\,98ne, RZ\,120                &   2021-11-08              &   V, I    &   3x400, 3x600    &   1.39 &   0.65, 0.54 \\
BS\,75                          &   2016-11-02              &   V, I    &   6x200, 7x300    &   1.40 &   0.85, 0.65 \\
BS\,80                          &   2021-11-08              &   V, I    &   3x400, 3x600    &   1.40 &   0.61, 0.49 \\
ESO\,51SC9                     &   2017-10-22              &   V, I    &   6x200, 6x300    &   1.33 &   0.64, 0.49 \\
HW\,20; RZ\,39                  &   2016-09-27              &   V, I    &   6x200, 5x300    &   1.40 &   0.66, 0.44 \\
HW\,36; RZ\,109                 &   2021-11-07              &   V, I    &   3x400, 3x600    &   1.41 &   0.70, 0.58 \\
HW\,66; ESO\,29$-$36; OGLE\,307 &   2017-10-22              &   V, I    &   5x200, 5x300    &   1.44 &   0.77, 0.64 \\
HW\,79$^* $                     &    ---                    &   ---     &   ---             &   ---  &   ---        \\
L\,14; RZ\,9                    &   2021-11-07              &   V, I    &   3x400, 3x600    &   1.36 &   0.64, 0.57 \\ 
RZ\,158                         &   2021-11-08              &   V, I    &   3x400, 3x600    &   1.45 &   0.81, 0.63 \\
RZ\,107, B\,94                  &   2016-09-27              &   V, I    &   6x200, 6x300    &   1.40 &   0.63, 0.49 \\
\hline
\multicolumn{6}{c}{GMOS/Gemini-S Pre-images}\\
\hline
B\,4            &   2021-07-29   &   g, r    &   4x50, 4x166    &   1.40 &   0.78, 0.72 \\
B\,98ne         &   2019-09-27   &   g, r    &   4x50, 4x166    &   1.40 &   1.58, 1.30 \\
BS\,75          &   2019-08-06   &   g, r    &   4x50, 4x166    &   1.60 &   1.20, 1.08 \\
BS\,80          &   2019-09-04   &   g, r    &   4x50, 4x166    &   1.53 &   1.33, 1.22 \\
ESO\,51SC9     &   2021-07-29   &   g, r    &   4x50, 4x166    &   1.30 &   0.75, 0.72 \\
HW\,20          &   2019-10-06   &   g, r    &   4x50, 4x166    &   1.51 &   1.18, 1.16 \\
HW\,36          &   2019-10-06   &   g, r    &   4x50, 4x166    &   1.56 &   1.28, 0.94 \\
HW\,66          &   2019-09-27   &   g, r    &   4x50, 4x166    &   1.62 &   1.34, 0.96 \\
HW\,79          &   2019-09-27   &   g, r    &   4x50, 4x166    &   1.71 &   1.17, 1.10 \\
L\,14           &   2021-07-29   &   g, r    &   4x50, 4x166    &   1.48 &   0.77, 0.62 \\
RZ\,107         &   2019-09-26   &   g, r    &   4x50, 4x166    &   1.59 &   0.88, 0.85 \\
RZ\,158         &   2019-10-06   &   g, r    &   4x50, 4x166    &   1.64 &   1.31, 1.23 \\
\hline
\multicolumn{6}{c}{GMOS/Gemini-S Multi-object spectroscopy}\\
\hline
B\,4            &   2021-14-10   &   R831+CaT    &   4x902    &   1.39 &   0.92 \\
B\,98ne         &   2021-01-09   &   R831+CaT    &   5x903    &   1.54 &   0.98 \\
BS\,75          &   2019-12-07   &   R831+CaT    &   4x878    &   1.39 &   1.12 \\
BS\,80          &   2021-01-05   &   R831+CaT    &   4x903    &   2.19 &   0.82 \\
ESO\,51SC9     &   2021-10-15   &   R831+CaT    &   4x902    &   1.33 &   0.96 \\
HW\,20          &   2021-01-04   &   R831+CaT    &   4x903    &   1.69 &   0.76 \\
HW\,36          &   2021-01-04   &   R831+CaT    &   4x903     &   1.79 &  0.74 \\
HW\,66          &   2021-01-11   &   R831+CaT    &  4x903    &   1.64 &    0.85 \\
HW\,79          &   2020-12-24   &   R831+CaT    &   4x903    &   1.68 &   0.80 \\
L\,14           &   2021-10-15   &   R831+CaT    &   4x902   &   1.45 &    1.04 \\
RZ\,107         &   2020-12-21   &   R831+CaT    &   4x903    &   1.45 &   0.92 \\
RZ\,158         &   2020-12-25   &   R831+CaT    &   4x903    &   1.82 &   0.84 \\
\hline
\end{tabular}\\
Notes: The FWHM for the GMOS/Gemini-S pre-images and SAM/SOAR images were measured on the reduced and combined images. The FWHM for the spectroscopic observation is a reference in the $V$ band. $^*$ Due to technical problems it was not possible to obtain a deep $I$ image for HW\,79. For this cluster photometry from the SMASH Survey was used (see text for details).
\end{table*}

\section{Data reduction and cluster parameter determinations}
\label{sec:parameters}

\subsection{Data reduction of photometric and spectroscopic data}
\label{subsec:reduction}

Photometric data processing was performed using standard IRAF routines for mosaic CCD data reduction: MSCRED for bias subtraction and  flat-field correction and CRUTIL for cosmic rays removal. Astrometric calibration was done by MSCCMATCH, using the Gaia eDR3 catalogue as astrometric reference; typical RMS residuals of these calibrations are inferior to 0.1 arcsec. Exposures were then co-added into the final mosaics by using the World Coordinate System to register them to a global frame. Photometry was carried out by applying our own developed IDL code, which is based on STARFINDER \citep{diolaiti+00}, to perform PSF photometry in all co-added exposures. Photometric calibration was done by employing a set of \citep{stetson00} standard star fields, observed nightly, and the MCPS catalogue (Magellanic Clouds Photometric Survey; \citealt{zaritsky+02}), when available. More details about these procedures are explained in detail in the VISCACHA Paper I \citep{maia+19}. 

The spectroscopic GMOS/Gemini-S data were reduced with the use of dedicated IRAF tasks available in  version 1.14 of the Gemini IRAF package\footnote{\url{https://www.gemini.edu/observing/phase-iii/reducing-data/gemini-iraf-data-reduction-software}}. The detailed scripts (developed by M. Angelo) are publicly available\footnote{\url{http://drforum.gemini.edu/topic/gmos-mos-guidelines-part-1/}} and the general steps are briefly described here. The GMOS frames were pre-reduced with the usual procedures of bias and overscan subtraction and flat-field division (GSREDUCE task). Bad pixels across the detector's arrays (together with the unilluminated gaps between the CCDs) were corrected by means of local linear interpolations (GMOSAIC and FIXPIX tasks) with the use of dedicated bad pixel masks. The pixel versus $\lambda$ calibrations (GSWAVELENGTH task) were derived from CuAr arc-lamp spectra and the catalogued emission line lists available in the Gemini IRAF package. The RMS of the fitting residuals were typically smaller than $\sim$0.05\,\AA\,(i.e., typically smaller than $\sim$10\% of a pixel size, for an employed dispersion of 0.76\,\AA/pix). Bright emission sky lines from the list of \cite{Hanuschik2003} were used as references along with the arc-lamps in order to properly find the absolute zero-points in the wavelength solutions. 

After that, the spectra were ($i$) corrected for differences in quantum efficiency between the detector chips (GQECORR task), ($ii$) cleaned of cosmic rays with the use of the Laplacian Edge Detection algorithm \citep[LACOSMIC task;][]{vanDokkum2001}, ($iii$) corrected for distortions along the spatial direction (GSTRANSFORM task), ($iv$) background subtracted (GSSKYSUB task), ($v$) extracted to the usual flux\,$\times$\,wavelength format (GSEXTRACT task) and ($vi$) normalized to the continuum level (CONTINUUM task) by fitting low-order polynomials to the $8450-8710$\,\AA\, spectral range.

\subsection{Radial velocity measurement}
\label{subsec:rv}

Radial velocities (RVs) for the present sample were determined from cross$-$correlation  (IRAF's FXCOR task) between the science spectra and a set of theoretical templates taken from Paula Coelho's library \citep{coelho14}. The synthetic templates were first degraded in spectral resolution, in order to properly match the GMOS/Gemini-S science spectra ($R\approx2000$), then converted to the same wavelength dispersion scale and finally continuum-normalized. The selected models present log$\,(g)=1.0\,$ and 4700\,$\leq$\,$T_{\textrm{eff}}\,(K)$\,$\leq$\,5300, values that are representative of stars in the RGB phase. We also allowed for a relatively large range in metallicity ($-$1.3\,$\leq$\,[Fe/H]\,(dex)\,$\leq$\,$-$0.5), totalizing 12 templates with these selection criteria.

For each \textit{science$-$template} comparison, individual RV values are derived from fitting the peak of the cross-correlation function. The final RV of each star, together with its associated uncertainty, corresponds to the mean value obtained with the ensemble of synthetic templates. 

\subsection{Spectroscopic metallicity determination}
\label{subsec:met}

The equivalent widths (EWs) of the CaT lines were measured from the spectra in the rest frame by fitting a Gaussian plus Lorentzian functions to the line profiles \citep{cole+04}. We followed the prescriptions of \citet[][]{dias+20b} and built the CaT index as the sum of the EWs of the three CaT lines, $\Sigma$EW = EW$_{8498}$ + EW$_{8542}$ + EW$_{8662}$. Most of the spectra have a high enough S/N ratio (between 30 and 100) to be able to correctly fit the three lines. However, in a few cases, the weakest line is not adequately defined to make a satisfactory fit. In such cases the CaT index was computed from the EWs of the two most intense lines  ($\Sigma$EW = EW$_{8542}$ + EW$_{8662}$) and then corrected, according to  equation (5) of \citet{dias+20b}, to the value it would have if the three lines could have been measured. The $\Sigma$EW was also corrected for the effects of temperature and surface gravity through the reduced equivalent width ($W'$), which is a direct indicator of metallicity \citep{dc91}. We use  equation (4) and the value of the slope in the  $r$ filter, $\beta_r$=0.67$\pm$0.07, from \citet{dias+20b} to calculate the $W'$ of each observed star. The $r$ vs. $g-r$ CMDs and our own python script allowed us to derive the horizontal branch level (see \citealt{dias+20b} for details). Finally, metallicities were calculated with  equation (8) from \citet{dias+20b}. 

For the membership selection, we followed the  methodology inherited from \citet{grocholski+06}  that combines RVs, metallicities and distances of the observed stars to the center of the corresponding cluster. The empirical cluster size adopted here is its tidal radius ($r_t$), derived by us following the procedure described below. On the other hand, we have adopted the same cuts in RV ($\pm$10 km s$^{-1}$) and metallicity ($\pm$0.20 dex) as in our previous works. Stars that lie outside the distance, RV, and metallicity cuts are coloured in Fig. \ref{fig:memb} in blue, cyan, and green, respectively, and have been discarded as likely cluster members. We consider as cluster members those stars that are consistent with the three cuts and they are identified by red symbols in Fig. \ref{fig:memb}. 

The cluster's structural parameters are derived from the fitting of the \cite{king62} model to the radial density profile (RDP). The RDPs were built following the procedure described in detail in \citetalias{maia+19}. The method is based on completeness-corrected stellar counts  \citep[e.g.][]{maia+16} in annular bins of several sizes using the calibrated $V$ and $I$ images, except for the cluster HW\,79 for which only the $V$ image was used, due to the reasons explained in section \ref{sec:obs}. The RDPs and the results are shown in Fig.  \ref{fig:rad_profiles}  as well as the corresponding fitted King functions (dashed lines). Also, the derived background/foreground stellar density ($\sigma_0$) and the core and tidal radii (r$_t$ and r$_c$ respectively) are specified in the figures.

In Table \ref{tab:results} we present, for each cluster in our sample, the mean RV and metallicity (together with the corresponding errors) determined from those stars identified as cluster members according to our analysis.  

\subsection{Statistical decontamination and isochrone fitting}
\label{subsec:decontam}

For the statistical decontamination, which assigns photometric membership probability to each star, we used the method described in detail in \citet{maia+10}. This is the same procedure that we have used in all previous VISCACHA works: \citetalias{maia+19,dias+21,dias+22}; \citetalias{bica+22} \citep{bica+22}; \citetalias{oliveira+23} and \citetalias{saroon+23} \citep{saroon+23}. The CMD of a nearby control field is compared with the CMD of the cluster, both having  similar densities and reddening. The CMDs are properly binned into a uniform grid of cells, in which the local density of stars are compared. The expected number of members and the membership probability are derived from the overdensity of stars within a given cluster cell relative to the corresponding field cell and from the distances of these stars from the cluster centre. This process is iterated using several grid configurations to obtain the final membership values.

The \texttt{SIRIUS} code \citep{souza+20} was employed to fit the decontaminated CMDs using PARSEC isochrones \citep{bressan+12} through a Bayesian approach based on the Markov chain Monte Carlo sampling method. The likelihood function of each star consists of a $\chi^2$ in each CMD axis to find the best fitting isochrones, including the photometric errors. Three other quantities are also considered: the membership value (stars with higher membership have a higher weight), the number of neighbour stars on the CMD (smaller weight for regions containing more stars), and a set of priors. A prior distribution can be introduced for any well-known feature or parameter to better constrain the parameter space and overcome degeneracies. In the present case, we used both the observed red clump locus and the [Fe/H] derived from the CaT analysis as Gaussian priors, as done in \citetalias{dias+21} and \citetalias{dias+22}. The latter one is applied with a fixed standard deviation of 0.20 dex in order to not constrain too much the posterior distributions. In general, the [Fe/H] prior is consistent with the final solution, but those clusters with discrepant metallicities, even considering uncertainties, are indications that a compromise between a more metal-rich CMD shape and the CaT value is found.

In Table \ref{tab:results} we list for each studied cluster  the parameters resulting from the isochrone fits. CMDs for our sample cluster and the corresponding corner plots with the likelihood distributions are presented in Figs. \ref{fig:cmds} and \ref{fig:corner}.
Metallicity, being a common parameter between photometric and spectroscopic procedures, is an excellent point of comparison that allows quantifying the quality of photometric fits. As it is possible to see in Table \ref{tab:results} there is a systematic offset between the metallicity values determined by both methods. There are many sources of uncertainties in the model isochrones that may be contributing to the observed metallicity difference between photometry and spectroscopy, for example the mixing length \citep{joyce+18}, rotation \citep{maeder+00,nguyen+22}, overshooting and atomic diffusion \citep{cassisi+02}. The mean difference between  [Fe/H]$_{CaT}$ and  [Fe/H]$_{CMD}$ is 0.13 with a standard deviation of 0.08. This value is smaller than the mean error in the determination of the metallicity of the individual stars (0.20 dex) which corresponds to the metallicity cuts adopted in our membership analysis. Therefore the agreement between the photometric and spectroscopic determinations is good. In all subsequent analysis we decided to use the CaT metallicities.\\

\subsection{Proper motions}
\label{subsec:pm}
For each cluster in the sample, we downloaded the PMs and parallax data from the Gaia eDR3 catalogue. We centered our search on the cluster's coordinates using a radius of 5'. In this way, we cover an area larger than the one of the clusters, according to the tidal radii determined by us (see Table \ref{tab:results} and Fig. \ref{fig:rad_profiles}). We built the Vector Point Diagrams for the 12  clusters, which are shown in Fig. \ref{fig:vpd}. As in our previous papers, we selected stars with the criteria as constructed from \citet{v18}, but with a more relaxed constraint on PM errors to establish a better compromise between statistics and uncertainties. We selected stars with $\sigma_{\mu_\alpha} < 0.35~mas.yr^{-1}$, $\sigma_{\mu_\delta} < 0.35~mas.yr^{-1}$, and $\pi < 3.\sigma_\pi$, equivalent to parallaxes consistent with zero. 
In the figures, the stars available in Gaia eDR3 within the cluster radius, that fit our selection criteria, are shown with coloured symbols according to their distance to the corresponding cluster centre. Spectroscopic targets that are cluster members according to our membership procedure are marked (black circles) as well as their average PMs weighted by the individual PMs errors (red arrows).  The adopted mean PMs (red crosses) are compared with the ones of the SMC  and Magellanic Bridge regions (see \citetalias{dias+21} for details about the determination of these regions). All identified spectroscopic cluster members present the RUWE parameter (Gaia Renormalized Unit Weight Error) smaller than 1.4, which is an indicative of a good astrometric solution \citep{Lindegren+21}.

\subsection{Comparison with previous determinations}
\label{subsec:met_comp}

All clusters from our sample have been previously investigated by other authors, but not in an homogeneous way, that is one of the goals of the present work. In Table \ref{tab:literature} we have summarised the parameters that can be found in the literature so that the reader can make a detailed comparison of our determinations with those made by other works. In this paper, we have derived for the first time the [Fe/H] values for the clusters RZ\,107 and B\,98ne and the distance of the clusters RZ\,107, B\,98ne, HW\,36 and ESO\,51SC9. All our cluster sample have previous age determinations. All the works listed in Table \ref{tab:literature} are based on photometric data, therefore this constitutes the first spectroscopic study for all clusters in our sample. Consequently we have determined unprecedented  spectroscopic metallicities and RVs for these objects.

Some clusters in our sample have been previously photometrically studied using the same data but applying different methods to the one in this work, or the same method with minor differences (HW\,20 in \citetalias{maia+19}, RZ\,158 in \citetalias{bica+22}, B\,4 and L\,14 in \citetalias{saroon+23}).  In what follows we will make a small comparison of our results with those found in previous works of the VISCACHA collaboration.\\

\textbf{HW\,20}: In \citetalias{maia+19}, the r$_t$ determined for this cluster (37 $\pm$ 11 arcsec) is in excellent agreement with the value derived in the present work (35 $\pm$ 12 arcsec), which is expected considering that the same method for the radial profile fitting was used. Also in \citetalias{maia+19}, the age, metallicity, reddening and distance were derived for this cluster (see Table \ref{tab:literature}) via the Markov Chain Monte Carlo technique in a Bayesian framework on a synthetic simple stellar population CMD built using \texttt{PARSEC} isochrones \citep{bressan+12}. As can be seen in Tables \ref{tab:results} and \ref{tab:literature} there is a good agreement between the values derived by both works, considering the errors. \\

\textbf{RZ\,158:} In \citetalias{bica+22}, the same method was used as in the present work for the determination of the astrophysical parameters for this cluster, but with different prior conditions, since they did not have the CaT metallicity.   
The parameters determined by the two independent analyses are consistent with each other. There is however a difference of 0.21 dex between the value of the metallicity determined by us and the one from \citetalias{bica+22}, our value being more metal poor. This difference slightly exceeds our metallicity errors (0.19 dex), however it is important to note that in \citetalias{bica+22} the errors are substantially higher ($^{+0.43}_{-0.39}$\,dex), thus being compatible within 1-sigma. \\

\textbf{B\,4:} Although the metallicity values derived from the CMD analysis and CaT spectroscopy showed some discrepancy in the present work (0.24 dex), they are still compatible within 1.4-sigma, given their uncertainties. In \citetalias{saroon+23}, \texttt{PARSEC} isochrones were visually fitted to the statistically decontaminated CMD, finding a metallicity very close to the value from the CaT technique. 
The \citet{dias+16} derived metallicity value also favours the CaT metallicity value, even though it is substantially more metal poor. 
Our corner plot (Fig. \ref{fig:corner}) shows that the [Fe/H] prior leads to a compromise between the CaT value and an even more metal-rich solution. The decontamination was redone compared to \citetalias{saroon+23} in order to remove the large number of binaries in V$\sim$22\,mag and to better define the subgiant branch region. This strategy possibly forced the metallicity to a more metal-rich value (compared of \citetalias{saroon+23} and CaT), but resulted in a statistically better fitting on the CMD. \\

\textbf{L\,14:} The visual fit from \citetalias{saroon+23} for L\,14 is similar to the \texttt{SIRIUS} fit of the present work and the resulting parameters are compatible with each order within uncertainties. The present result comes from statistical fitting and shows a good compromise among all parameters. \\

We note that the small changes in age and metallicity for clusters B\,4 and L\,14 do not change the conclusions of \citetalias{saroon+23}. \\

\begin{table*}
    \caption{Derived parameters for the star clusters. (1) cluster name; (2,3) $ (\alpha,\delta) $ coordinates from \citet{bica+20}; (4,5) coordinates relative to the SMC centre (6) deprojected angular distance from the SMC centre $a$ following the definition by \citet{dias+14}; (7) Classification from \citet{dias+14,dias+16} (2D) and from \citetalias{dias+21,dias+22} (3D) (8) number of member stars and observed stars corresponding to the GMOS/Gemini-S spectroscopy; (9,10) $RV_{hel}$ and [Fe/H]$_{\rm CaT}$ from GMOS/Gemini-S spectra; (11,12,13,14) age, [Fe/H], E(B-V), distance from VISCACHA CMD isochrone fitting; (15,16,17,18) ( $\mu_{\alpha}\cdot \cos(\delta)$, $\mu_{\delta}$ ) PMs from Gaia eDR3 with their respective dispersions $\sigma_{\mu_{\alpha}}$ and $\sigma_{\mu_{\delta}}$}; (18) cluster tidal radius (r$_t$).
    \label{tab:results}
    \centering
    \footnotesize
    \begin{tabular}{lrrcccc}
    \hline
    \noalign{\smallskip}
    \multicolumn{1}{c}{Cluster} &
    \multicolumn{1}{c}{$\alpha_{\rm J2000}$} &
    \multicolumn{1}{c}{$\delta_{\rm J2000}$} &
    \multicolumn{1}{c}{$\alpha_{rel}$} &
    \multicolumn{1}{c}{$\delta_{rel}$} &
    \multicolumn{1}{c}{$a$} &
    \multicolumn{1}{c}{2D/3D Class$^*$}  \\
    \multicolumn{1}{c}{} &
    \multicolumn{1}{c}{(hh:mm:ss.s)} &
    \multicolumn{1}{c}{(dd:mm:ss)} &
    \multicolumn{1}{c}{(deg)} &
    \multicolumn{1}{c}{(deg)} &
    \multicolumn{1}{c}{(deg)} &
    \multicolumn{1}{c}{}    \\
    \noalign{\smallskip}
    \multicolumn{1}{c}{(1)} &
    \multicolumn{1}{c}{(2)} &
    \multicolumn{1}{c}{(3)} &
    \multicolumn{1}{c}{(4)} &
    \multicolumn{1}{c}{(5)} &
    \multicolumn{1}{c}{(6)} &
    \multicolumn{1}{c}{(7)} \\
    \noalign{\smallskip}
    \hline
    \noalign{\smallskip}
B\,4            &   00:24:54.3 & $-$73:01:50  & -2.03 & -0.20 &  3.032 & WH/MB  \\
B\,98ne         &   01:00:28.8 & $-$73:52:32  &  0.54 & -1.05 &  2.267 & SB/B  \\
BS\,75          &   00:54:30.9 & $-$74:11:07  &  0.04 & -1.36 &  2.264 & SB/MB  \\
BS\,80          &   00:56:12.6 & $-$74:09:24  &  0.24 & -1.33 &  2.345 & SB/MB  \\
ESO\,51SC9      &   00:58:58.0 & $-$68:54:55  &  0.56 & 3.91  &  5.701 & CB/CB  \\
HW\,20          &   00:44:48.0 & $-$74:21:46  & -0.54 & -1.53 &  2.033 & SB/MB \\
HW\,36          &   00:59:03.7 & $-$73:50:30  &  0.44 & -1.01 &  2.092 & SB/B$^{**}$  \\
HW\,66          &   01:12:05.1 & $-$75:11:51  &  1.24 & -2.37 &  5.158 & SB/B  \\
HW\,79          &   01:22:48.0 & $-$75:00:06  &  1.94 & -2.17 &  5.825 & SB/B  \\
L\,14           &   00:32:41.0 & $-$72:34:53  & -1.50 &  0.25 &  2.628 & WH/CB  \\
RZ\,107         &   00:58:14.8 & $-$74:36:32  &  0.37 & -1.78 &  3.195 & SB/CB \\
RZ\,158         &   01:06:45.1 & $-$74:49:58  &  0.92 & -1.99 &  4.201 & SB/B \\
    \noalign{\smallskip}
    \hline \hline
    \noalign{\smallskip}
    \multicolumn{1}{c}{Cluster} &
     \multicolumn{1}{c}{N$_{mem}$/N$_{obs}$} &
    \multicolumn{1}{c}{${\rm RV_{hel}}$} &
    \multicolumn{1}{c}{[Fe/H]$_{\rm CaT}$} &
    \multicolumn{1}{c}{Age} &
    \multicolumn{1}{c}{[Fe/H]$_{\rm CMD}$} &
    \multicolumn{1}{c}{E(B-V)} \\
    \multicolumn{1}{c}{} &
    \multicolumn{1}{c}{} &
    \multicolumn{1}{c}{(${\rm km\ s^{-1}}$)} &
    \multicolumn{1}{c}{(dex)} &
    \multicolumn{1}{c}{(Gyr)} &
    \multicolumn{1}{c}{(dex)} &
    \multicolumn{1}{c}{(mag)}    \\
    \noalign{\smallskip}
    \multicolumn{1}{c}{(cont.)} &
    \multicolumn{1}{c}{(8)} &
    \multicolumn{1}{c}{(9)} &
    \multicolumn{1}{c}{(10)} &
    \multicolumn{1}{c}{(11)} &
    \multicolumn{1}{c}{(12)} &
    \multicolumn{1}{c}{(13)}    \\
    \noalign{\smallskip}
    \hline
    \noalign{\smallskip}
B\,4       & 4/29 & $200.4 \pm 3.0(6.0)$ & $-0.99\pm0.05(0.11)$  &   $2.3\pm0.1$   &   $-0.75\pm0.17$ &   $0.10\pm0.05$ \\ 
B\,98ne    & 5/35 & $186.7 \pm 3.1(7.0)$ & $-0.96\pm0.05(0.11)$  &   $3.5\pm0.5$   &   $-0.81\pm0.16$ &   $0.09\pm0.04$ \\ 
BS\,75     & 2/38 & $196.0 \pm 7.3(10.3)$ & $-0.99\pm0.10(0.14)$ &   $2.4\pm0.2$   &   $-0.84\pm0.11$ &   $0.10\pm0.04$ \\ 
BS\,80     & 4/29 & $160.1 \pm 4.5(8.9)$ & $-1.00\pm0.08(0.17)$  &   $3.0\pm0.5$   &   $-0.81\pm0.17$ &   $0.08\pm0.04$ \\ 
ESO\,51SC9 & 7/17 & $142.8 \pm 0.5(3.8)$ & $-1.08\pm0.02(0.13)$  &   $4.1\pm0.6$   &   $-0.86\pm0.22$ &   $0.02\pm0.03$ \\ 
HW\,20     & 3/40 & $161.8 \pm 5.2(9.0)$ & $-0.74\pm0.09(0.15)$  &   $1.26\pm0.17$ &   $-0.66\pm0.17$ &   $0.08\pm0.05$ \\ 
HW\,36     & 3/39 & $154.4 \pm 3.7(6.5)$ & $-0.84\pm0.14(0.23)$  &   $2.8\pm0.7$   &   $-0.84\pm0.16$ &   $0.05\pm0.06$\\ 
HW\,66     & 8/25 & $182.1 \pm 1.1(3.0)$ & $-1.16\pm0.04(0.11)$  &   $4.0\pm0.5$   &   $-1.06\pm0.22$ &   $0.16\pm0.04$ \\ 
HW\,79     & 17/31 & $176.2 \pm 0.9(3.7)$ & $-1.26\pm0.03(0.12)$ &   $4.9\pm0.2$   &   $-1.04\pm0.12$ &   $0.08\pm0.02$ \\ 
L\,14      & 4/20 & $111.7 \pm 2.8(5.6)$ & $-0.96\pm0.07(0.11)$  &   $2.9\pm0.2$   &   $-0.83\pm0.11$ &   $0.03\pm0.03$ \\ 
RZ\,107    & 6/37 & $189.2 \pm 1.2(3.0)$ & $-0.92\pm0.04(0.11)$  &   $2.9\pm0.2$   &   $-0.83\pm0.10$ &   $0.06\pm0.04$  \\ 
RZ\,158    & 4/33 & $160.7 \pm 4.5(9.0)$ & $-1.11\pm0.05(0.11)$  &   $4.7\pm1.4$   &   $-1.11\pm0.19$ &   $0.08\pm0.05$ \\ 
    \noalign{\smallskip}
    \hline
    \hline
    \noalign{\smallskip}
    \multicolumn{1}{c}{Cluster} &
    \multicolumn{1}{c}{d} &
    \multicolumn{1}{c}{$\mu_{\alpha}\cdot \cos(\delta)$} &
    \multicolumn{1}{c}{$\sigma_{\mu_{\alpha}}$} &
    \multicolumn{1}{c}{$\mu_{\delta}$} &
    \multicolumn{1}{c}{$\sigma_{\mu_{\delta}}$} &
    \multicolumn{1}{c}{(r$_t$)} \\
    \multicolumn{1}{c}{} &
    \multicolumn{1}{c}{(kpc)} &
    \multicolumn{1}{c}{(${\rm mas\ yr^{-1}}$)} &
    \multicolumn{1}{c}{(${\rm mas\ yr^{-1}}$)} &
    \multicolumn{1}{c}{(${\rm mas\ yr^{-1}}$)} &
    \multicolumn{1}{c}{(${\rm mas\ yr^{-1}}$)} &
      \multicolumn{1}{c}{(arcsec)} \\
    \noalign{\smallskip}
    \multicolumn{1}{c}{(cont.)} &
    \multicolumn{1}{c}{(14)} &
    \multicolumn{1}{c}{(15)} &
    \multicolumn{1}{c}{(16)} &
    \multicolumn{1}{c}{(17)} &
    \multicolumn{1}{c}{(18)} &
    \multicolumn{1}{c}{(19)} \\
    \noalign{\smallskip}
    \hline
    \noalign{\smallskip}
B\,4           &   $61.4\pm3.1$ &   $0.36\pm0.07$  & 0.05   &  $-1.18\pm0.08$ & 0.07  & 86 $\pm$ 18  \\
B\,98ne        &   $55.2\pm2.5$ &   $0.99\pm0.04$  & 0.29   &  $-1.30\pm0.04$ & 0.14  & 98 $\pm$ 23   \\
BS\,75         &   $60.5\pm2.5$ &   $0.66\pm0.05$  & 0.26   &  $-1.31\pm0.04$ & 0.11  & 82 $\pm$ 14   \\
BS\,80         &   $58.9\pm3.3$ &   $0.54\pm0.08$  & 0.52   &  $-1.19\pm0.06$ & 0.16  & 72 $\pm$ 22   \\
ESO\,51SC9     &   $66.1\pm3.0$ &   $0.45\pm0.10$  & 0.16   &  $-1.30\pm0.09$ & 0.22  & 139$\pm$ 9   \\
HW\,20         &   $58.9\pm4.3$ &   $0.55\pm0.07$  & 0.29   &  $-1.24\pm0.06$ & 0.02  & 35$\pm$  12  \\
HW\,36         &   $58.1\pm4.3$ &   $0.76\pm0.23$  & 0.00   &  $-1.37\pm0.25$ & 0.00  & 27$\pm$  6   \\
HW\,66         &   $57.5\pm2.4$ &   $0.83\pm0.08$  & 0.40   &  $-0.99\pm0.08$ & 0.23  & 82$\pm$  3   \\
HW\,79         &   $56.5\pm1.1$ &   $1.19\pm0.03$  & 0.23   &  $-1.42\pm0.03$ & 0.25  & 133$\pm$ 34    \\
L\,14          &   $69.8\pm2.3$ &   $0.40\pm0.09$  & 0.04   &  $-1.21\pm0.09$ & 0.00  & 122$\pm$ 11  \\
RZ\,107        &   $67.9\pm3.4$ &   $0.50\pm0.07$  & 0.21   &  $-1.14\pm0.06$ & 0.19  & 55$\pm$  12  \\
RZ\,158        &   $56.0\pm4.4$ &   $0.84\pm0.07$  & 0.24   &  $-1.20\pm0.07$ & 0.16  & 68$\pm$  11   \\
    \noalign{\smallskip}
    \hline \hline    
    \end{tabular}
    \\Notes: $^*$ WH: West Halo, MB: Main Body, SB: Southern Bridge, CB: Counter-Bridge, B: Bridge. $^{**}$ Only one spectroscopic member star of cluster HW\,36 has PMs from Gaia, therefore its movement, and therefore its 3D classification, must be considered with caution. \\
\end{table*}

\begin{table*}
     \caption{Literature parameters for our cluster sample}
    \label{tab:literature}
    \centering
    \footnotesize
    \begin{tabular}{lccccc}
    \hline
    \noalign{\smallskip}
    \multicolumn{1}{c}{Cluster} &
    \multicolumn{1}{c}{[Fe/H]} &
    \multicolumn{1}{c}{Age} &
    \multicolumn{1}{c}{E(B-V)} &
    \multicolumn{1}{c}{d} &
    \multicolumn{1}{c}{Ref.} \\
    \multicolumn{1}{c}{} &
    \multicolumn{1}{c}{} &
    \multicolumn{1}{c}{(Gyr)} &
    \multicolumn{1}{c}{(mag)} &
    \multicolumn{1}{c}{(kpc)} &
    \multicolumn{1}{c}{} \\
    \noalign{\smallskip}
    \multicolumn{1}{c}{(1)} &
    \multicolumn{1}{c}{(2)} &
    \multicolumn{1}{c}{(3)} &
    \multicolumn{1}{c}{(4)} &
    \multicolumn{1}{c}{(5)} &
    \multicolumn{1}{c}{(6)} \\
    \noalign{\smallskip}
    \hline
    \noalign{\smallskip}
BS\,75          & -0.48$\pm$0.26          & 2.51                         &  0.03$\pm$0.01         & 59.70 & \citet{perren+17} \\
                & --                      & 1.26                         &  0.05                  & --    & \citet{glatt+10}  \\
                & --                      & 1.78                         &  0.00                  & --    & \citet{maia+14} \\
\hline
RZ\,107         & --                      & 1.37$^{+2.08}_{-0.01}$       &  --                    & --    & \citet{rz05} \\
\hline
HW\,79          & -0.88$\pm$0.65          & 6.31                         & 0.01$\pm$0.01          & 63.68 & \citet{perren+17} \\
                & -1.30/-1.40$\pm$0.2     & 5.00$\pm$1.30/4.30$\pm$1.20  & 0.06/0.07              & --    & \citet{piatti+11} \\
                & --                      & 4.10$\pm$0.50                & --                     & --    & \citet{parisi+14} \\
\hline
B\,98ne     & --                      & 1.40$^{+5.77}_{-0.35}$       & --                     & --    & \citet{rz05} \\
\hline
RZ\,158         & --                      & 7.27$^{+2.73}_{-0.18}$       & --                     & --            & \citet{rz05} \\
                & -0.90$^{+0.43}_{-0.39}$ & 4.80$^{+1.60}_{-1.30}$       & 0.06                   & 54.70$\pm$3.5 &  \citetalias{bica+22}\\
\hline
HW\,20          & --                      & 5.69$^{+4.32}_{-0.41}$       & --                     & --    & \citet{rz05}\\
                & -0.55$^{+0.13}_{-0.10}$ & 1.10$^{+0.08}_{-0.14}$       & 0.07$^{+0.02}_{-0.01}$ & 62.20$^{+2.50}_{-1.20}$ & \citetalias{maia+19}\\
\hline 
HW\,36          &                         & 1.38$^{+4.90}_{-0.03}$       & --                     & --    & \citet{rz05} \\
                & --                      & 1.32$^{+0.72}_{-0.0.30}$     & 0.09                   & --    & \citet{gatto+21} \\
\hline
BS\,80          &  -0.88$\pm$0.65         & 3.98                         & 0.020$\pm$0.009        & 64.27 & \citet{perren+17} \\
                &    --                   & 1.00                         & 0.02                   & --    & \citet{glatt+10} \\
                &    --                   & 2.82                         & 0.00                   & --    & \citet{maia+14} \\
\hline
HW\,66          &  -0.88$\pm$0.65         & 2.82                         & 0.02$\pm$0.01          & 64.27 & \citet{perren+17} \\
                & -1.30/-1.35$\pm$0.20    & 4.00$\pm$0.90/3.50$\pm$1.00  & 0.05                   & --    & \citet{piatti+11} \\
                & --                      & 3.40$\pm$0.40                & --                     & --    & \citet{parisi+14} \\
\hline
ESO\,51SC9      &                         & 5.20$\pm$0.40                & --                     & --    & \citet{parisi+14} \\
                & -1.00$\pm$0.15          & 7.00$\pm$1.30                & --                     & --    & \citet{piatti12}\\

\hline
B\,4            &  -1.19$\pm$0.24         & 3.8$\pm$0.6                  & 0.05$\pm$0.04          & 66.60$\pm$3.70  & \citet{dias+16} \\
                & -0.98$^{+0.13}_{-0.12}$ & 2.75$^{+0.05}_{-0.25}$       & 0.096                  & 57.00$^{+8.00}_{-2.00}$ & \citetalias{saroon+23}. \\
\hline
L\,14           & --                      & 1.26                         & 0.01                   & --    &  \citet{glatt+10}\\
                & --                      & 6.29$^{+0.81}_{-0.59}$       & --                     & --    & \citet{rz05} \\  
                & --                      & 2.40$^{+0.69}_{-0.16}$       & 0.13                   & --    & \citet{gatto+21} \\
                & -1.14$\pm$0.11          & 2.8$\pm$0.4                  & 0.03$\pm$0.02          & 70.60$\pm$1.60 & \citet{dias+16} \\
                & -1.00$\pm$0.10          & 3.20$^{+0.20}_{-0.40}$       & 0.04                   & 63.00$^{+4.00}_{-2.00}$  & \citetalias{saroon+23}. \\
    \noalign{\smallskip}
    \hline
    \end{tabular}
    \\Note: The ages included in this table from \citet{rz05} correspond to the extinction corrected values, STARBURST model and Z=0.001.
\end{table*}

\section{Discussions}
\label{sec:discussions}

The structure of the SMC presents a main body and some external arms or substructures as seen in the projected distribution of star clusters \citep[e.g.][and others]{dias+14,dias+16,parisi+22}. When the third dimension is added to the analysis, a more complete and complex view of the SMC outskirts appears. As pointed out in \citetalias{dias+21} the SMC seems to have three branches of a Bridge pointing to and moving towards the LMC on the Eastern side and a shell of star clusters moving outwards in the Western region from the North all the way down to the Southern regions, with some clusters on the background of the bridge as well moving together with the Counter-Bridge. This observational evidence is consistent with the model predictions by \citet{zivick+21} describing the SMC with an inner cylindrical rotation and an outer tidal expansion, as well as with partial evidence showing foreground and background stellar populations by e.g. \citet[][]{nidever+13,omkumar+21} and by outward proper motions by Gaia, VMC, HST \citep[e.g.][]{zivick+18,niederhofer+18,piatti+21}. We have detected the alignment of these structures with the LMC-SMC direction combined with the outward motion for the first time in a self-consistent analysis of full 3D positions and motions of selected SMC star clusters in two opposite regions, the Northern Bridge \citepalias{dias+21} and West Halo \citepalias{dias+22}. In this work, we add the Southern bridge region to the analysis to check whether it follows or not the Northern Bridge. In this section we discuss the structure, kinematics and age metallicity relation (AMR) of the Southern Bridge, as traced by star clusters, in comparison with the Northern Bridge. In order to perform that comparison, we added clusters from the literature analysed with similar methods to the ones applied here to our sample, specifically, the clusters analysed in \citetalias{dias+21} and those having CaT metallicities from the works of \citet{parisi+09,parisi+15,parisi+22}.  \\

A first indication of the movement of the clusters in our sample are the VPDs (Fig. \ref{fig:vpd}). 
As can be seen, considering projected distances, the two clusters catalogued by \citet{dias+14,dias+16} as West Halo (B\,4 and L\,14), the Counter-Bridge cluster (ESO\,51SC9) and three of the nine Southern Bridge clusters (BS\,80, HW\,20 and RZ\,107) are moving in the direction away from the LMC. In the case of the remaining 
six clusters, B\,98ne, HW\,66 and HW\,79 seem to move towards the LMC while the movement of BS\,75 and HW\,36 is difficult to ensure due to the large uncertainties in their 
meand PMs.
However, as shown in \citetalias{dias+21} and \citetalias{dias+22}, including real distances instead of projected distances brings a more realistic picture of the kinematics and structure of the SMC traced by its star cluster system. The 2D classification originally proposed by \citet{dias+14,dias+16} was superseded by a new 3D classification defined in \citetalias{dias+21} and \citetalias{dias+22}, which is based on full phase-space analysis. 
Clusters located within the SMC tidal radius (r$_{\rm SMC}$) are considered to belong to the Main Body. The outer clusters (r>r$_{\rm SMC}$) are classified as Bridge or Counter-Bridge according to their 3D positions and velocity directions: Bridge clusters if they are foreground clusters moving towards the LMC and Counter-Bridge if they are moving in the opposite direction, as predicted by dynamical  models.\\

\begin{figure*}
    \centering
    \includegraphics[width=0.49\textwidth]{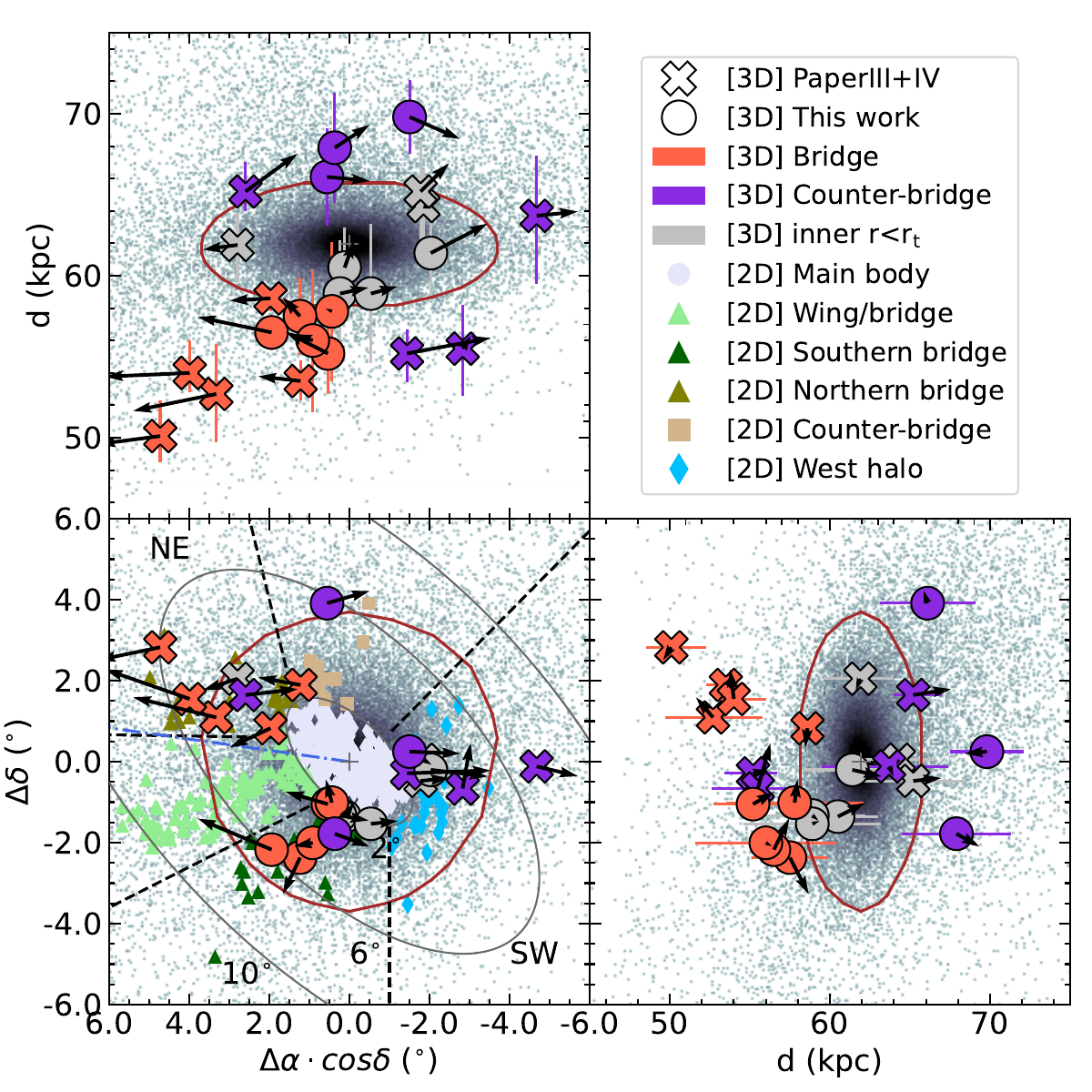}
    \includegraphics[width=0.49\textwidth]{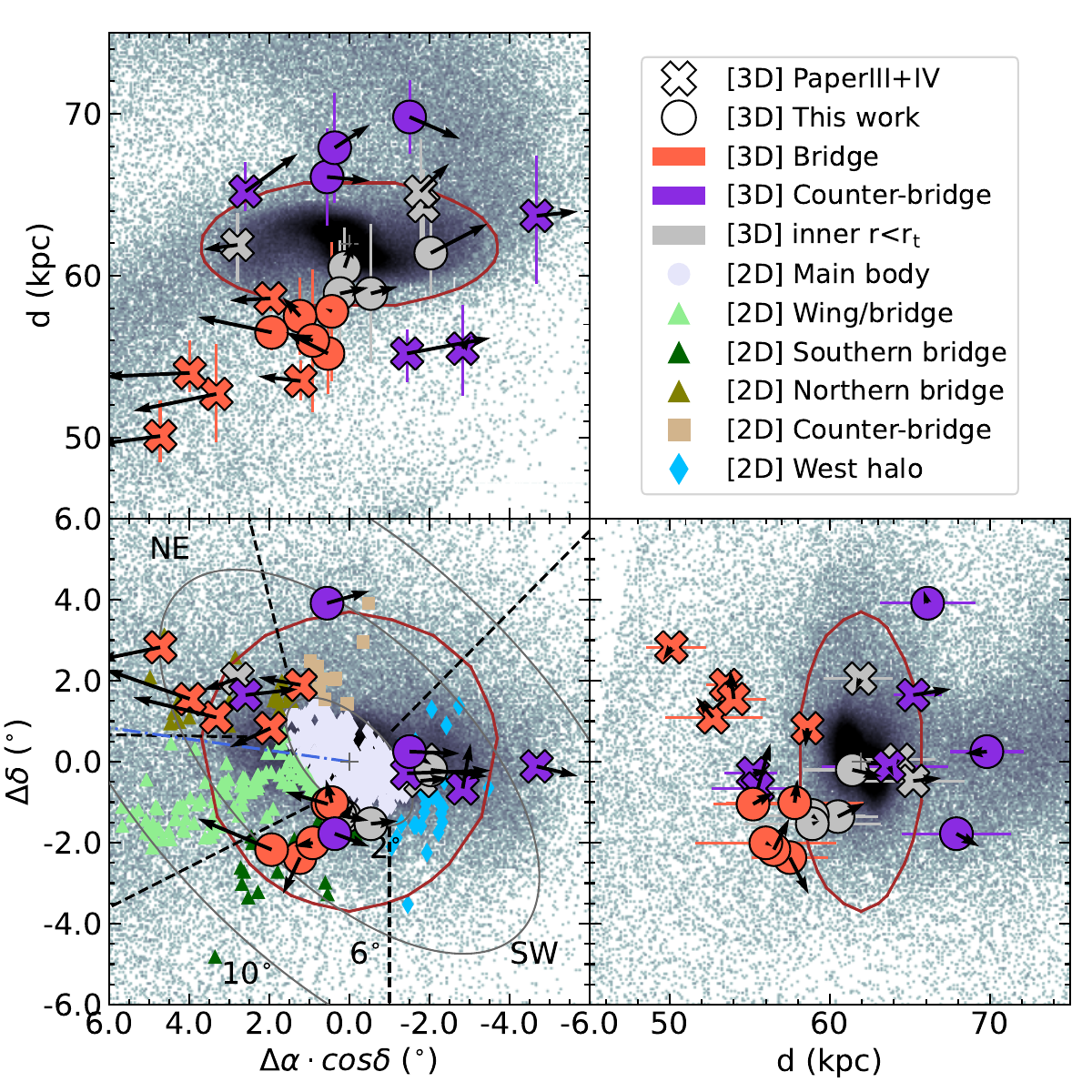}
    \caption{3D distribution of the star clusters analysed in this paper combined with the previous sample. Cross symbols are clusters from the two previous works and the circles are the additions from this work. Colours indicate their classification according to \citetalias{dias+21} whereas the 2D projected regions defined by \citet{dias+14,dias+16,dias+21} have smaller symbols in the bottom left panel only. The brown circles and ellipses are the projections of a sphere of radius 4 kpc around the SMC centre representative of the SMC tidal radius. Blue ellipse represents the break radius from \citet{dias+21}. Left and right sets of panels contain exactly the same observational data, the difference is the background simulations by \citet{diaz+12} and are shown for comparison. Left panels contain stellar particles from the spheroid component and right panels contain the gas particles from the disc component from the same simulation. }
    \label{fig:struct}
    \end{figure*}

We show the 3D distribution of the star clusters in Fig. \ref{fig:struct}. We consider a sphere of 4 kpc as representative of the r$_{\rm SMC}$ for the SMC. As can be seen, the sky regions Southern Bridge, West Halo and Counter-Bridge analysed in this work contain: 5 clusters (HW\,36, B\,98ne, RZ\,158, HW\,66 and HW\,79) following the structure of the Northern Bridge sample analysed in \citetalias{dias+21} and represented in Fig. \ref{fig:struct} with orange crosses; 4 clusters (B\,4, HW\,20, BS\,75 and BS\,80) from the Main Body  with distances similar to the SMC centre; and 3 background cluster (RZ\,107, L\,14 and ESO\,51SC9) which mans that they are consistent with the 3D Counter-Bridge shell moving outwards found in \citetalias{dias+22}. 

That reinforces, despite the low number of clusters, the suggestion of \citet{tatton+21} and \citetalias{dias+22} that the West Halo clusters belong to the Counter-Bridge, and perhaps also to the Main Body. On the other hand, the only  2D Counter-Bridge cluster (ESO\,51SC9) in our sample is confirmed to belong to the Counter-Bridge according to the 3D classification. As can be seen in Fig. \ref{fig:struct}, the five Bridge clusters and the three Counter-Bridge clusters, reclassified in this work, follow the velocity patterns of their neighbours in the corresponding components. Also in Fig. \ref{fig:struct} it is evident that, in the planes that involve the distance, Southern Bridge and Northern Bridge clusters appear to follow the same tail from the SMC, but being the Southern Bridge ones closer to the SMC main body in the line of sight. 

\begin{figure*}
    \centering
    \includegraphics[width=0.49\textwidth]{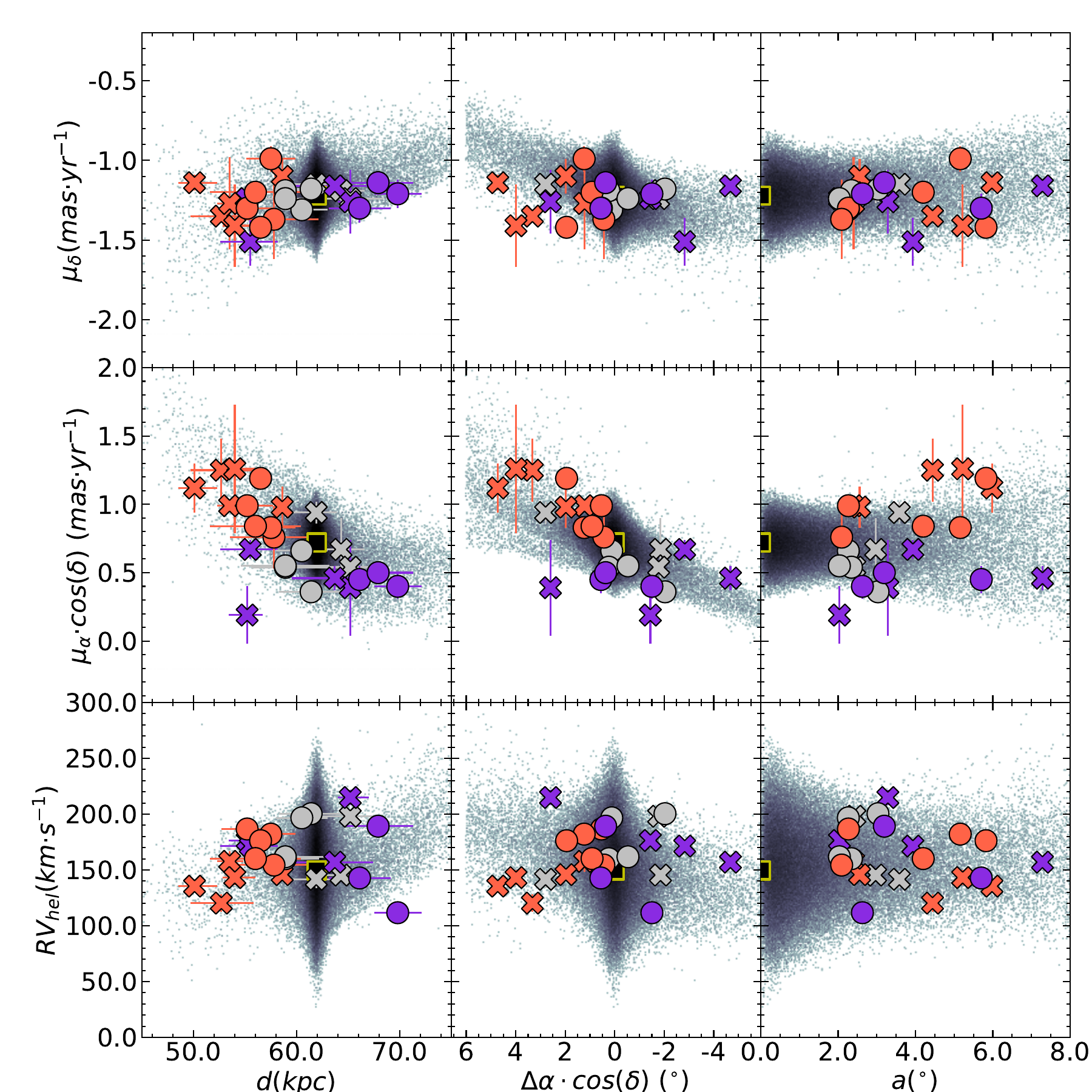}
    \includegraphics[width=0.49\textwidth]{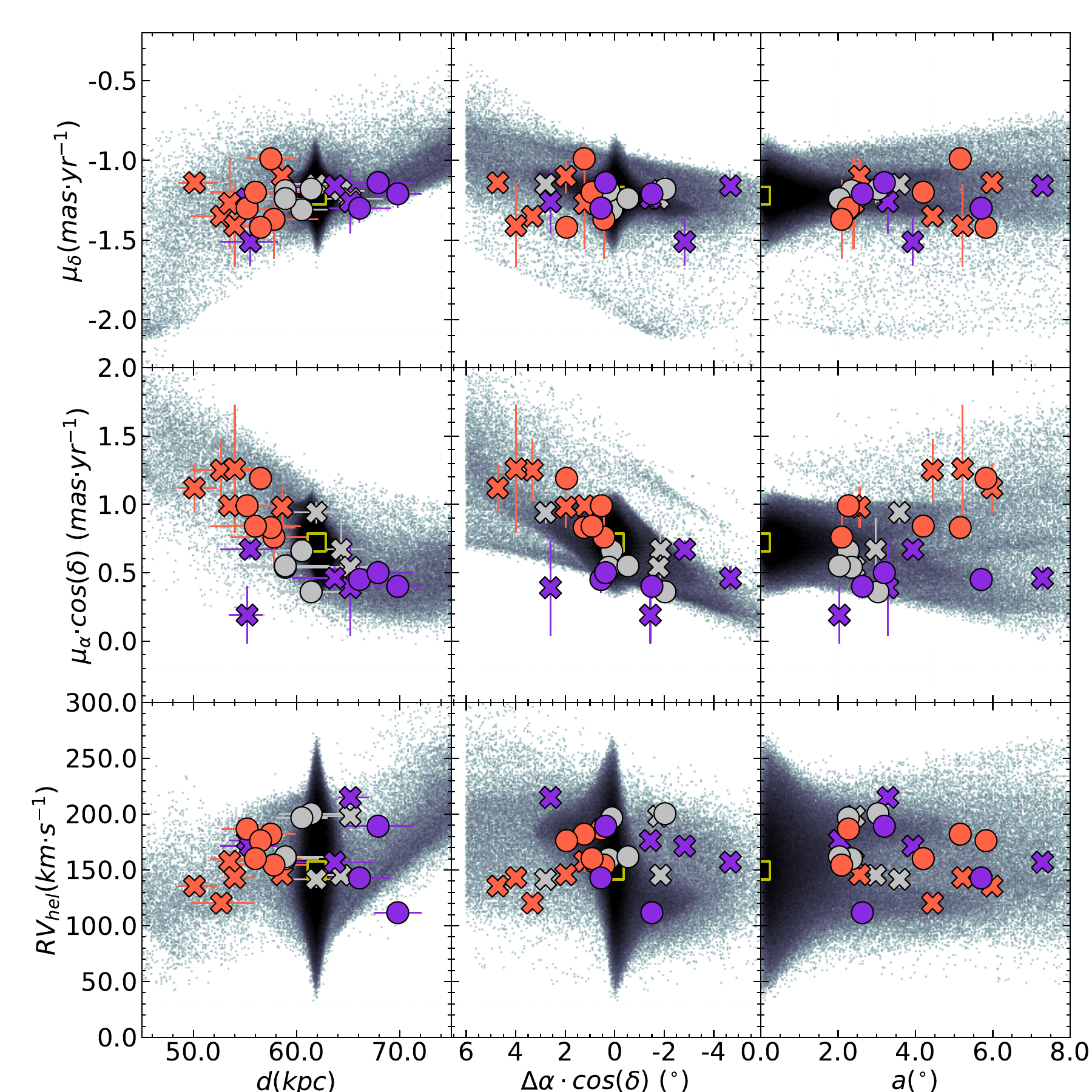}
    \caption{Kinematics of the star clusters analysed in this paper combined with the previous sample from \citetalias{dias+21} and \citetalias{dias+22}. Symbols and colours are the same as defined in Fig. \ref{fig:struct}. Background particles are simulated stellar particles from the spheroid component (left panel) and gas particles (right panel) from \citet{diaz+12} for comparison. }
    \label{fig:kines}
\end{figure*}

The kinematics of the full sample is shown in Fig. \ref{fig:kines} as a direct comparison with the simulated stellar particles from \citet{diaz+12}. The plots reinforce the idea that the five genuine Bridge clusters studied in this work follow the kinematics of the Northern Bridge, also confirming  that the Southern Bridge is moving towards the LMC. Besides, the three clusters reclassified here as Counter-Bridge clusters show  kinematics consistent with the shell moving outwards in the opposite direction to the Bridge, also reinforcing the suggestion of \citetalias{dias+22} that the Counter-Bridge forms a ring in the boundary of the SMC moving away from the SMC.

One aspect important to be mentioned in  Fig. \ref{fig:kines} is that the kinematics become confusing when the parameter $a$ is used instead of the real distances. The deprojected distance $a$ has been defined by \citet{piatti+05} as the semi-major axis of an ellipse centred in the center of the SMC, with axis ratio of b/a $=$ 1/2, and coinciding with the position of the cluster. This is equivalent of stating that real cluster positions are in a circular disc. However, new evidence \citep[e.g.][]{zivick+21} consolidates that the SMC cannot be described simply by a disc. 
Also, in \citetalias{dias+22} we show that the metallicity and age radial gradients can be observed  when the deprojected distance $a$ is used but they vanish when real 3D distances are considered, showing that they are not real. So this parameter is not appropriate to accurately perform any analysis involving distances and positions of the clusters. Both aforementioned VISCACHA's previous works and the present paper clearly demonstrate that  the use of the parameter $a$ as an indicator of distance is dispensable in future analyses. To make the necessary effort to determine real distances, as the VISCACHA collaboration has been systematically doing, is mandatory. \\ 

In general terms, the 3D distribution and 3D kinematics of the star clusters presented in this work are comparable to the predictions from the simulations of \citet{diaz+12}. However, the simulations are not able to reproduce a couple of smaller scale details, for example: (i) there is not evident overdensity of stars to characterize the Northern and Sourthern Bridge, and the gas particles show an overdensity for the young bridge and counter-bridge, which does not match with the intermediate-age clusters discussed here; (ii) the kinematics of the foreground bridge stars and gas shows a clear pattern that is similar in the simulations and data, but this does not seem to be the case for the background counter-bridge clusters. We stress that these are the most recent simulations of the entire Magellanic System orbits around the Milky Way reaching this level of details of small-scale structure of the SMC including structure and kinematics. However this model is more than 10 years old, and since then the LMC mass determination has shown that the LMC is a factor of 10 heavier, favouring the first infall scenario proposed by \citet{bkh07}. The full simulations of \citet{bkh10,besla+12} are focused on the large scale-structure of the Magellanic System and does not provide a SMC small-scale structure comparable with observations, which was the focus of more recent models by \citet{zivick+21} for example, but without the full simulation of the LMC, SMC orbit. We expect that the accumulating evidence from the VISCACHA star clusters can serve as hard constraints for forthcoming full chemo-dynamical simulations that are timely and necessary.

\begin{figure}
    \centering
    \includegraphics[width=\columnwidth]{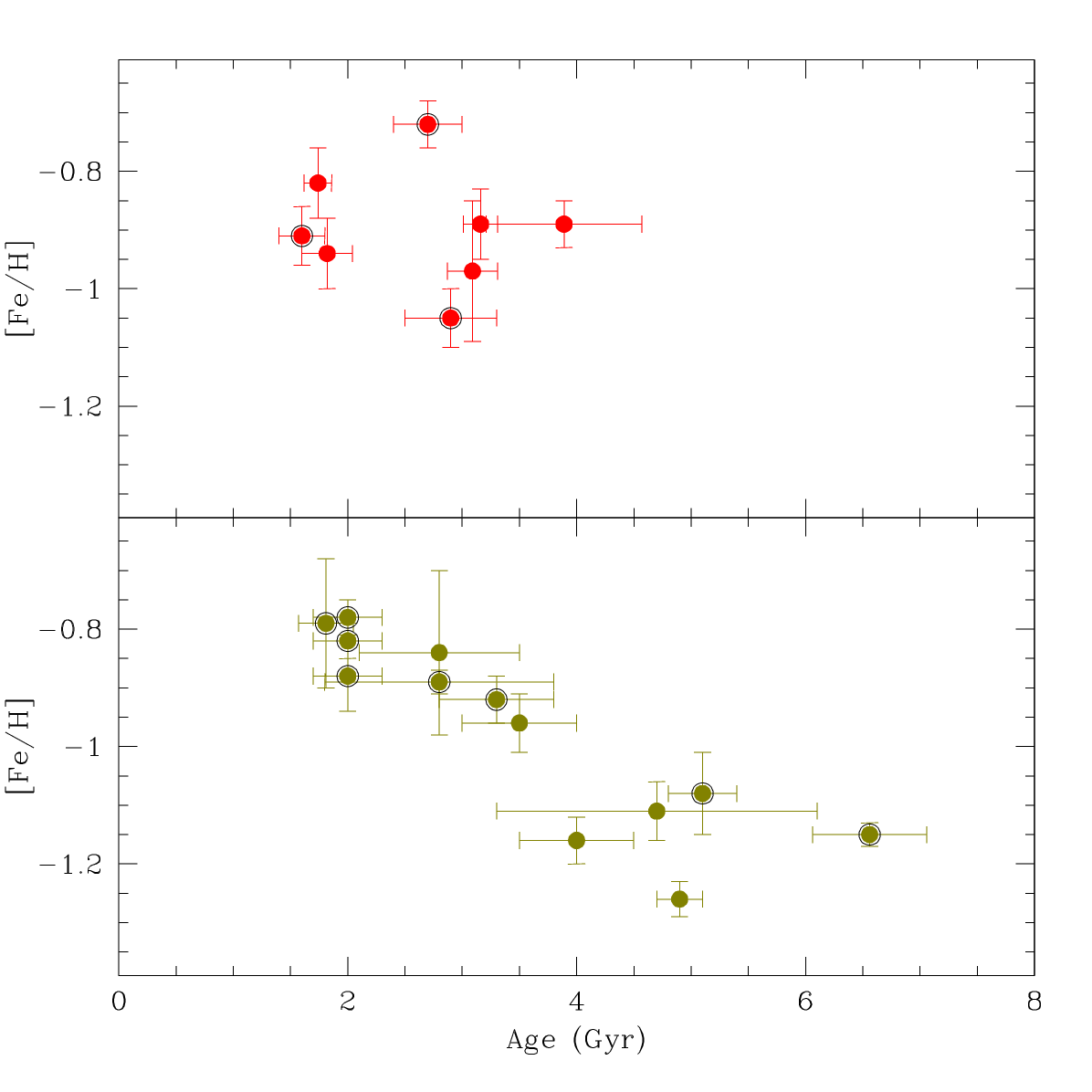}
    \caption{Age metallicity relation of clusters with CaT metallicties. Clusters from the Southern Bridge and the Northern Bridge are shown in the bottom and top panels, respectively. Clusters taken from the literature not  studied by VISCACHA (Table \ref{tab:lit_CaT}), are marked with black open circles. }
    \label{fig:AMR}
\end{figure}

As mentioned above, since the Southern Bridge and Northern Bridge clusters are located in different regions of the sky but share the same kinematics, it is interesting to analyse whether or not they share the same chemical evolution. In Fig. \ref{fig:AMR} we compare the behaviour of the CaT metallicity as a function of age for clusters in both components. The age and metallicity values of the additional CaT sample are listed in table \ref{tab:lit_CaT}. In \citet{parisi+22} the SMC AMR was analysed using all the clusters having CaT at that moment splitting the sample into different subsamples considering the 2D classification. Two of the Northern Bridge clusters used in \citet{parisi+22} belong in fact to the Main Body (B\,168) and to the Counter-Brdige (L\,102) according to the 3D classification \citepalias{dias+21} and therefore we do not consider them here. On the other hand, we add  five Southern Bridge clusters to the eight ones included in \citet{parisi+22}. As can be observed in Fig. \ref{fig:AMR}, this improved sample of clusters corroborates that the Southern Bridge presents a clear AMR: clusters show a  trend in the sense that  the material from which the clusters have been formed between  6.5 and  2 Gyr ago  has been enriched from $-$1.2 to $-$0.8 dex, being the metallicity dispersion at any given age very small. In contrast, the Northern Bridge clusters cover a substantially large range in metallicities ($\sim$0.4 dex) occupying a small range of ages ($\sim$3 Gyr), which reinforces the idea that this region, apparently, does not show a clear chemical enrichment between 4 and 2 Gyr ago, which is the age range covered by the current sample. If we only consider clusters younger than $\sim$4 Gyrs the mean metallicity of the Northern and Southern Bridge regions are $-$0.90 dex and  $-$0.86 dex, respectively.  Both regions present statistically consistent mean metallicities being the intrinsic metallicity dispersion (i.e., uncertainties removed) slightly larger in the Northern Bridge (0.10 dex) than in the Southern Bridge (0.06 dex). This result is expected considering that both populations cover a similar range of metallicities, but it does not reflect the trend and lack of trend shown in Figure \ref{fig:AMR} for Southern Bridge and Northen Bridge, respectively. We note that clusters from Table \ref{tab:lit_CaT} do not have distances and 3D classification, therefore the final AMR from Fig. \ref{fig:AMR} may or may not slightly change when this info is available in the future. Nevertheless, even if we remove these clusters from Fig. \ref{fig:AMR}, the Northern Bridge and Southern bridge sample clusters still show different age distributions.
 
There is a clear lack of clusters in the Northern Bridge more metal$-$poor than $-$1.1  and older than 4 Gyr. We believe that this cannot be due to a selection effect since we have performed a good spatial coverage of that region (see Fig. \ref{fig:distrib}) and we have not applied any selection criteria that favour the youngest and most metal$-$rich clusters. We suggest that this evidence shows that although these two branches of the Bridge present similar kinematic characteristics, compatible with the tidal effects due to the interaction with the LMC, the Northern and Southern part of the SMC have not experienced the same history of chemical enrichment.
\citet[][submitted]{almeida+23} analysed the chemistry of field stars split in a rough estimate of line-of-sight distances and concluded that all field stars analysed in the Main Body, East and West SMC regions share a similar chemical evolution, but the metallicity distribution (MD) is different. They concluded that the foreground Eastern stars have a similar MD as the Main body whereas the background Eastern stars have a similar MD as the Western region. Our new evidence in the present work shows that two branches of the Bridge that are part of the Eastern foreground region present a clear difference in their MD, questioning the generalisation of the conclusions by \citet[][submitted]{almeida+23}.

\begin{table*}
     \caption{Clusters from the literature with CaT metallicities}
    \label{tab:lit_CaT}
    \centering
    \footnotesize
    \begin{tabular}{lccccc}
    \hline
    \noalign{\smallskip}
    \multicolumn{1}{c}{Cluster} &
    \multicolumn{1}{c}{Age} &
    \multicolumn{1}{c}{Ref.} &
    \multicolumn{1}{c}{[Fe/H]} &
    \multicolumn{1}{c}{Ref.} &
    \multicolumn{1}{c}{2D Class$^*$} \\
    \multicolumn{1}{c}{} &
    \multicolumn{1}{c}{(Gyr)} &
    \multicolumn{1}{c}{} &
    \multicolumn{1}{c}{} &
    \multicolumn{1}{c}{} &
    \multicolumn{1}{c}{} \\
    \noalign{\smallskip}
    \multicolumn{1}{c}{(1)} &
    \multicolumn{1}{c}{(2)} &
    \multicolumn{1}{c}{(3)} &
    \multicolumn{1}{c}{(4)} &
    \multicolumn{1}{c}{(5)} &
    \multicolumn{1}{c}{(6)} \\
    \noalign{\smallskip}
    \hline
    \noalign{\smallskip}
HW\,84   & 1.60 $\pm$ 0.20 & 1 & -0.91 $\pm$ 0.05 &  6 & NB\\
HW\,67   & 2.70 $\pm$ 0.30 & 1 & -0.72 $\pm$ 0.04 &  7 & NB\\
L\,108   & 2.90 $\pm$ 0.40 & 1 & -1.05 $\pm$ 0.05 &  6 & NB\\
K\,44    & 2.00 $\pm$ 0.30 & 2 & -0.78 $\pm$ 0.03 &  8 & SB\\
L\,116   & 2.80 $\pm$ 1.00 & 3 & -0.89 $\pm$ 0.02 &  8 & SB\\
NGC\,339 & 6.56 $\pm$ 0.50 & 4 & -1.15 $\pm$ 0.02 &  8 & SB\\
HW\,47   & 3.30 $\pm$ 0.50 & 1 & -0.92 $\pm$ 0.04 &  6 & SB\\
L\,58    & 1.81 $\pm$ 0.24 & 4 & -0.79 $\pm$ 0.11 &  7 & SB\\
L\,106   & 2.00 $\pm$ 0.30 & 1 & -0.88 $\pm$ 0.06 &  6 & SB\\
L\,112   & 5.10 $\pm$ 0.30 & 1 & -1.08 $\pm$ 0.07 &  7 & SB\\
L\,111   & 2.00 $\pm$ 0.30 & 1 & -0.82 $\pm$ 0.03 &  6 & SB\\
    \noalign{\smallskip}
    \hline
    \end{tabular}
    \\References: (1) \citet{parisi+14}; (2) \citet{gatto+21}; (3) \citet{piatti+01}; (4) \citet{lagioia+19}, (5) \citetalias{oliveira+23}; (6) \citet{parisi+09}; (7) \citet{parisi+15}; (8) \citet{parisi+22}. 
    $^*$ The 3D classification for this cluster sample is not available. 
\end{table*}

\section{Summary and Conclusions}
\label{sec:conclusions}

We present an analysis for twelve SMC star clusters located in the projected regions Southern Bridge, West Halo and Counter-Bridge using images from the VISCACHA Survey and its spectroscopic follow-up. Images in the $V$ and $I$ filters and spectra in the region of the CaT lines were obtained with SAM/SOAR and GMOS/Gemini-S, respectively. We derived ages, interstellar reddening, distances, radial velocities and metallicities for this cluster sample and complemented these parameters  with proper motions from the Gaia eDR3 catalogue \citep{gaia+21a}. All this information allowed us to derive the phase-space vectors for our cluster sample from which we analysed the 3D structure and kinematics of the SMC cluster system, complementing our similar previous works \citepalias{dias+21,dias+22}, in which we study the kinematics of the regions defined in 2D as Northern Bridge and West Halo, respectively. The conclusions of the present work can be summarised as follows:

\begin{itemize}
    \item Four clusters belong to the SMC Main Body, five to the Bridge and three to the Counter-Bridge.
    
    \item The Southern Bridge is moving toward the LMC and shares the kinematics of the Northern Bridge.

    \item The three clusters reclassified as Counter-Bridge in 3D (ESO\,51SC9, L\,14 and RZ\,107)  follow the kinematics of this component, reinforcing the discoveries of \citetalias{dias+22}.

    \item The Northern Bridge and the Southern Bridge seem to not have suffered the same chemical enrichment history. While the Southern Bridge presents a clear AMR, clusters of the Northern Bridge cover a wide metallicity range (between $\sim$ $-$0.7 and $\sim$ $-$1.1), having a spread of $\sim$ 0.4 dex,  when all ages are considered. However, clusters younger than $\sim$ 4 Gyrs have similar mean metallicities values ($-$0.90 $\pm$ 0.10 dex and $-$0.86 $\pm$ 0.06 dex for the Northern and Southern Bridge, respectively).

    \item There is a clear lack of Northern Bridge clusters older than 4 Gyr and more metal$-$poor than $[Fe/H]\sim-1.1$\,dex in comparison with the extended sample of older and more metal-poor clusters found in the Southern Bridge region.
    
\end{itemize}

A next necessary step is to provide full phase$-$space information for clusters classified in 2D as Wing/Bridge, analyse the 3D structure of the cluster system in that tidal branch and compare its kinematics with the one of the other two branches of the Bridge. This would help to reach a more accurate understanding of the tidal nature of the Bridge as a whole and its fine structure in particular. The results of these analyses constitute important constraints to the dynamical models that attempt to understand the origin of the Bridge and the consequences of the SMC-LMC tidal interactions. The SMC outskirts remains a very interesting science case and there is still a lot to be learned from the SMC dynamical and chemical evolution history.\\

\section*{Acknowledgements}

MCP and BD thank Sergio Vasquez for providing his script to measure EW of CaT lines.
MA and BD thank German Gimeno for useful discussions on the GMOS/Gemini-S data reduction process.
BD thanks Jonathan Diaz for kindly providing their simulation results from \cite{diaz+12} for direct comparison with our observational results.
BD thanks the Gemini SOS personnel for the pre-image data reduction and acknowledges support by ANID-FONDECYT iniciación grant No. 11221366.
F.F.S.M. acknowledges financial support from Conselho Nacional de Desenvolvimento Científico e Tecnológico - CNPq (proc. 404482/2021-0) and from FAPERJ (proc. E-26/201.386/2022 and E-26/211.475/2021).
This research was partially supported by the Argentinian institution SECYT (Universidad Nacional de Córdoba) and Consejo Nacional de Investigaciones Científicas y Técnicas de la República Argentina, Agencia Nacional de Promoción Científica y Tecnológica.
J.G.F-T acknowledges support provided by Agencia Nacional de Investigaci\'on y Desarrollo de
Chile (ANID) under the Proyecto Fondecyt Iniciaci\'on 2022 Agreement No. 11220340, and from the Joint Committee ESO-Government of
Chile 2021 under the Agreement No. ORP 023/2021.\\
D.M. gratefully acknowledges support from the ANID BASAL projects ACE210002 and FB210003, from Fondecyt Project No. 1220724, and from CNPq Brasil Project 350104/2022-0.
We appreciate the detailed evaluation carried out by the referee and her/his helpful comments.\\
B.P.L.F. acknowledges the financial support from CNPq (140642/2021-8).
Based on observations obtained at the Southern Astrophysical Research (SOAR) telescope, which is a joint project of the Minist\'{e}rio da Ci\^{e}ncia, Tecnologia e Inova\c{c}\~{o}es (MCTI/LNA) do Brasil, the US National Science Foundation’s NOIRLab, the University of North Carolina at Chapel Hill (UNC), and Michigan State University (MSU).\\
Based on observations obtained at the international Gemini Observatory, a program of NSF’s NOIRLab, which is managed by the Association of Universities for Research in Astronomy (AURA) under a cooperative agreement with the National Science Foundation. on behalf of the Gemini Observatory partnership: the National Science Foundation (United States), National Research Council (Canada), Agencia Nacional de Investigaci\'{o}n y Desarrollo (Chile), Ministerio de Ciencia, Tecnolog\'{i}a e Innovaci\'{o}n (Argentina), Minist\'{e}rio da Ci\^{e}ncia, Tecnologia, Inova\c{c}\~{o}es e Comunica\c{c}\~{o}es (Brazil), and Korea Astronomy and Space Science Institute (Republic of Korea).\\
This work presents results from the European Space Agency (ESA) space mission Gaia. Gaia data are being processed by the Gaia Data Processing and Analysis Consortium (DPAC). Funding for the DPAC is provided by national institutions, in particular the institutions participating in the Gaia MultiLateral Agreement (MLA). The Gaia mission website is https://www.cosmos.esa.int/gaia. The Gaia archive website is https://archives.esac.esa.int/gaia.

\section*{Data Availability}
 

The data underlying this article are available in the NOIRLab Astro Data Archive (\url{https://astroarchive.noirlab.edu/}) or upon request to the authors.



\bibliographystyle{mnras}
\bibliography{bibliography} 




\appendix

\section{Figures}
\label{app:figures}
\begin{figure*}
    \centering
    \includegraphics[width=0.32\textwidth]{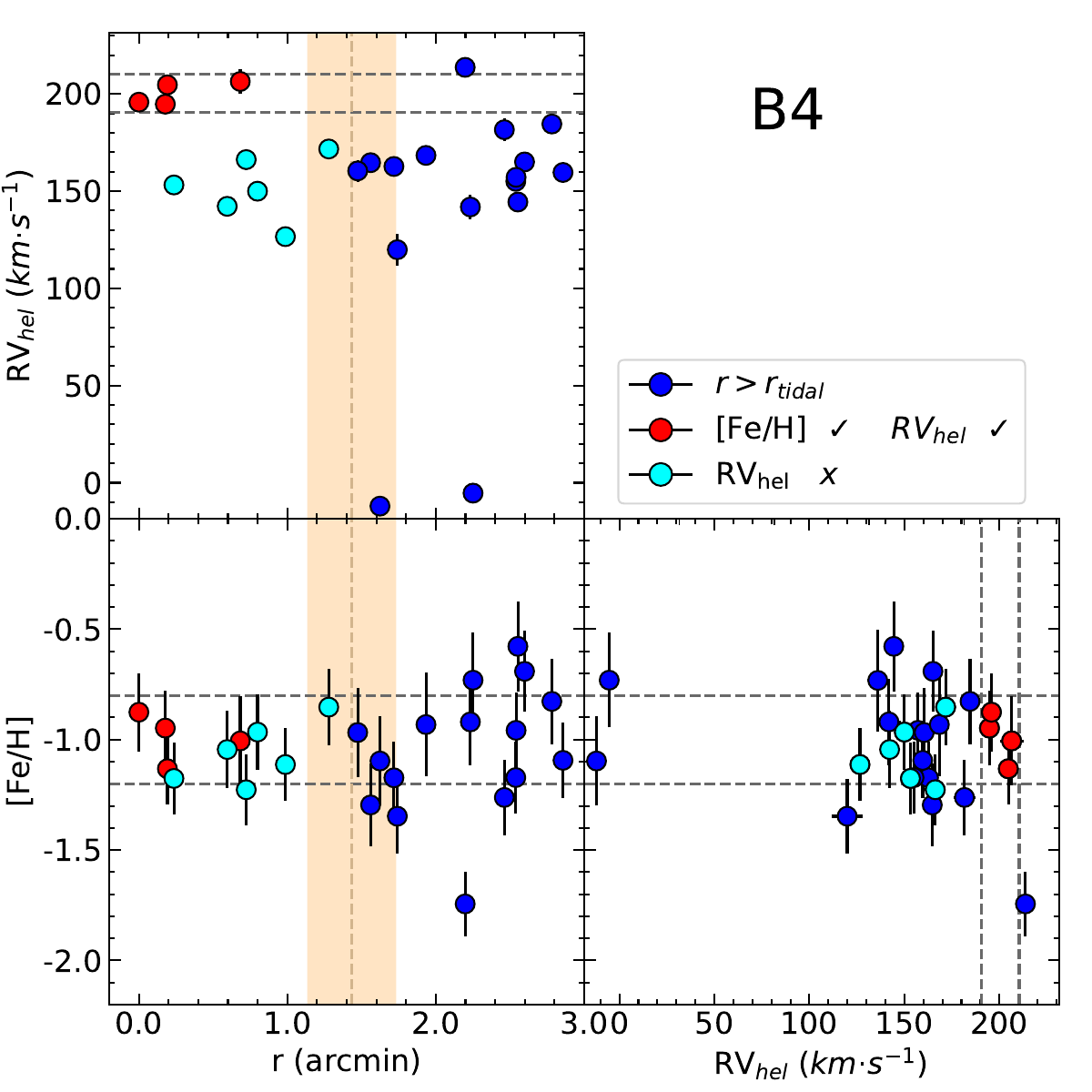}
    \includegraphics[width=0.32\textwidth]{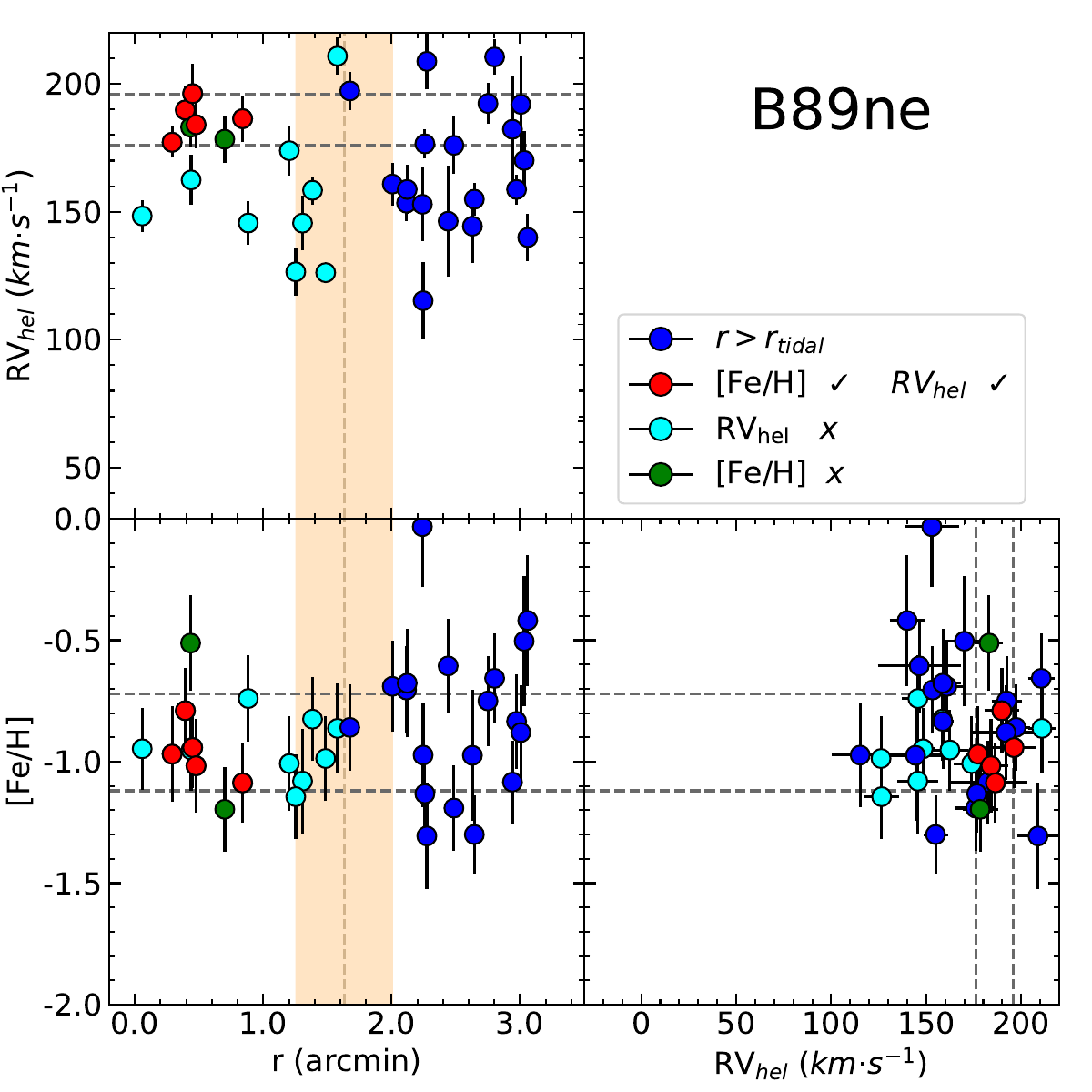}
    \includegraphics[width=0.32\textwidth]{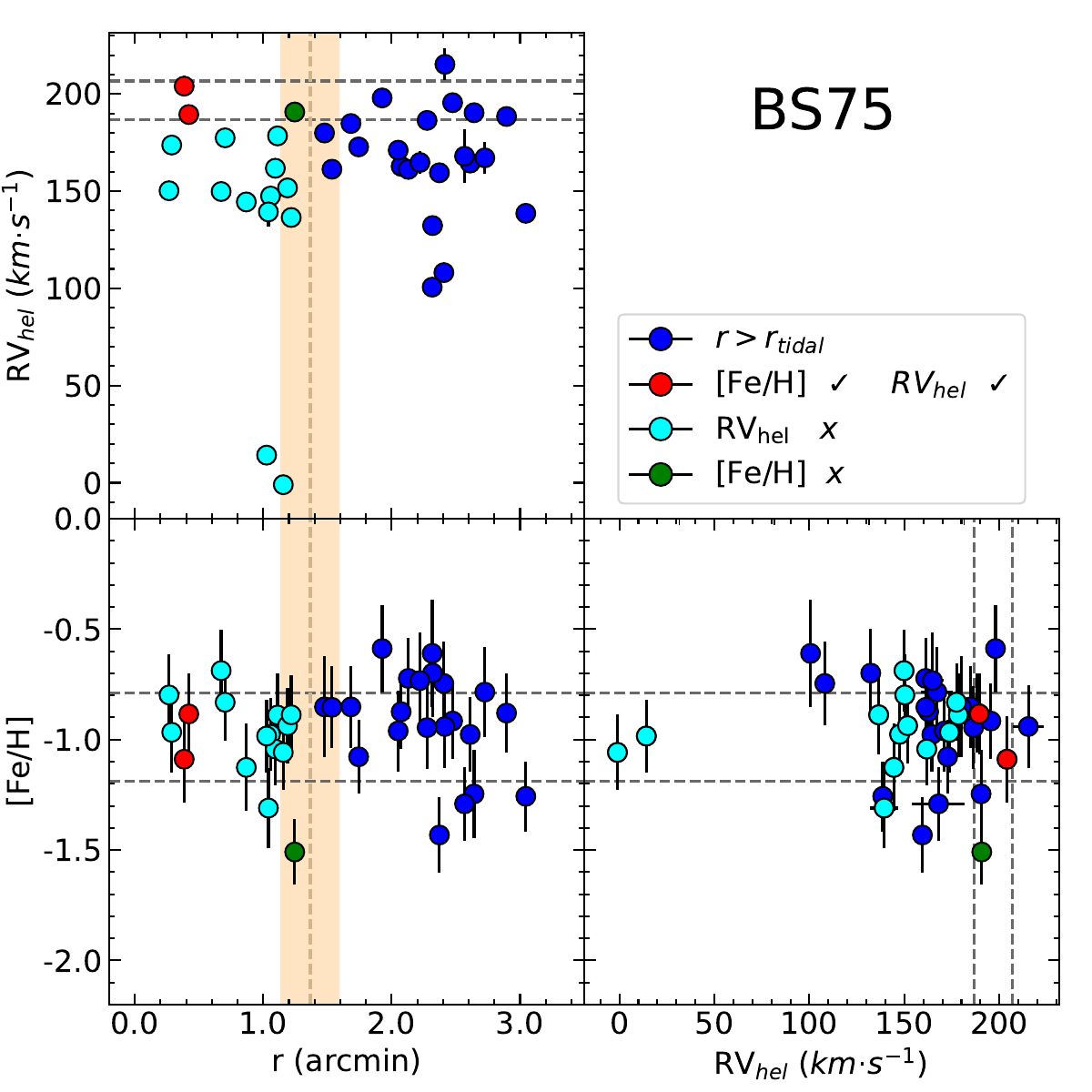}
    \includegraphics[width=0.32\textwidth]{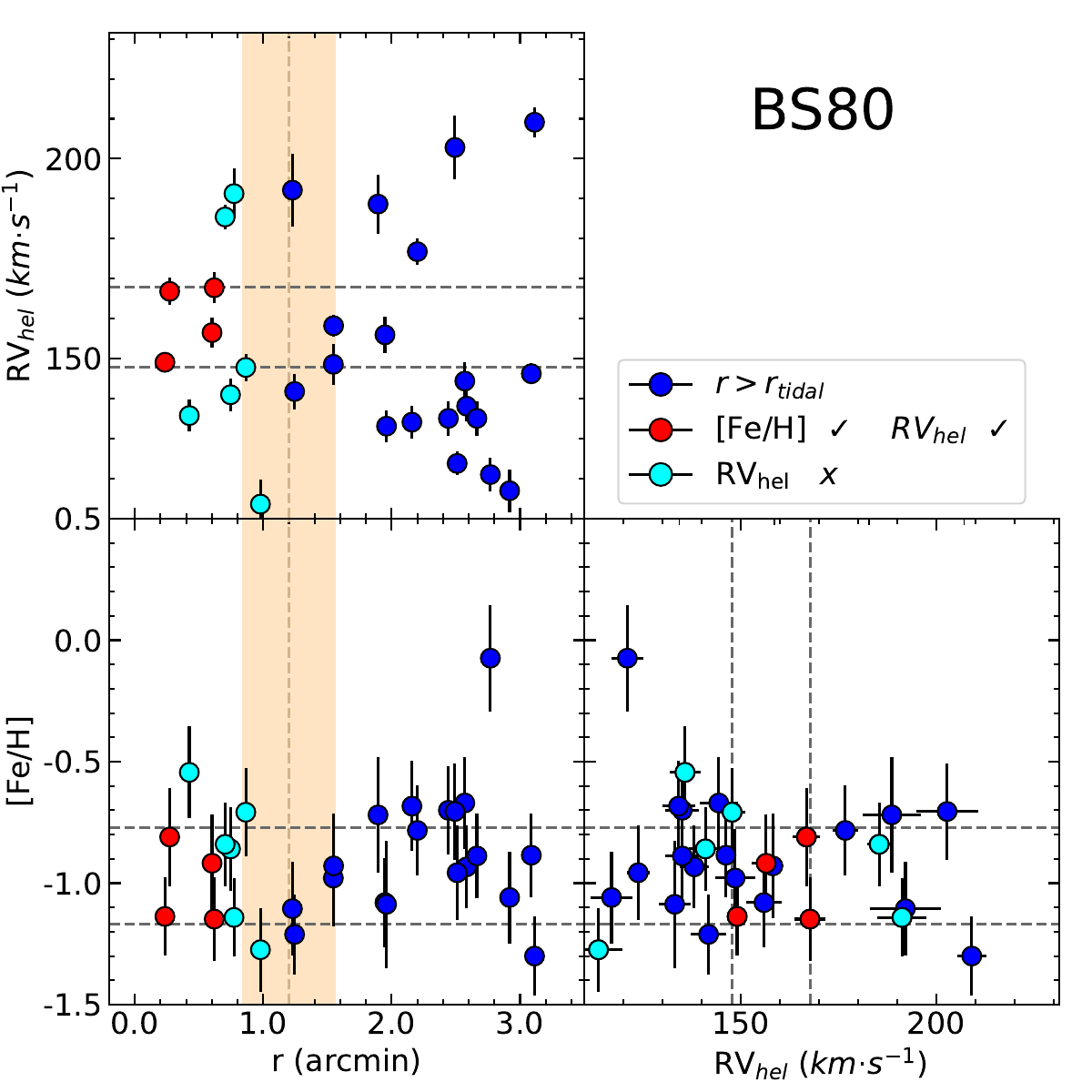}
    \includegraphics[width=0.32\textwidth]{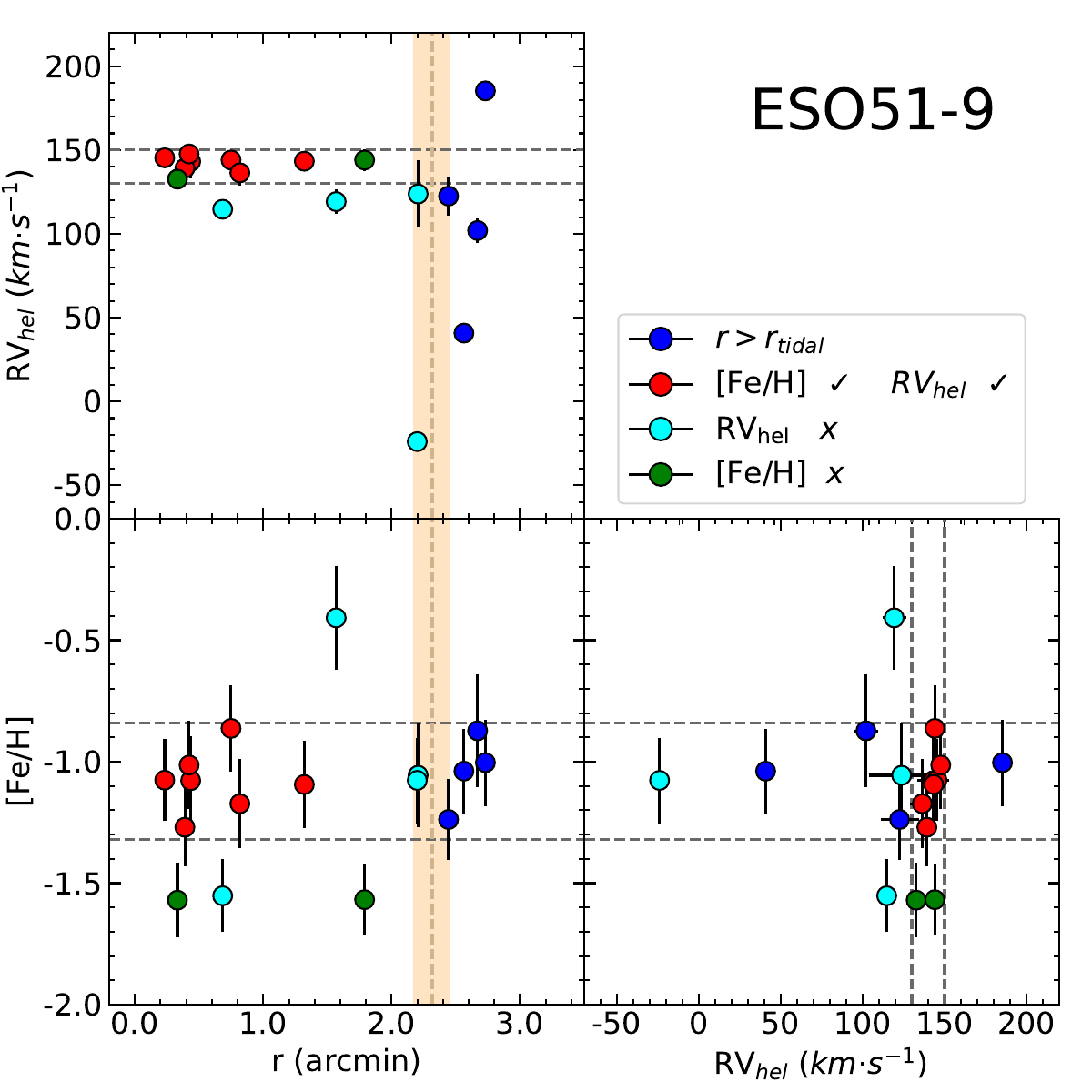}
   \includegraphics[width=0.32\textwidth]{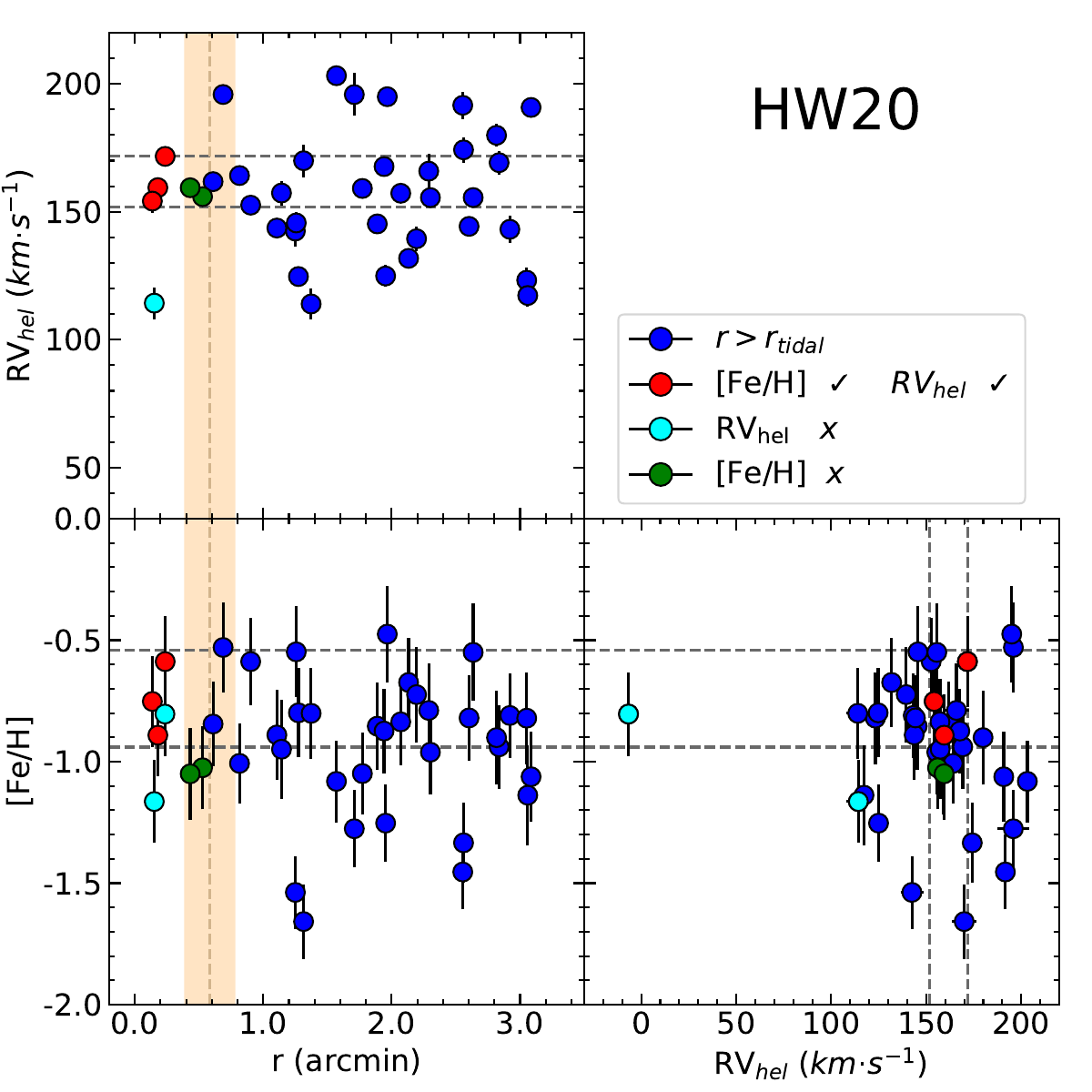}
    \includegraphics[width=0.32\textwidth]{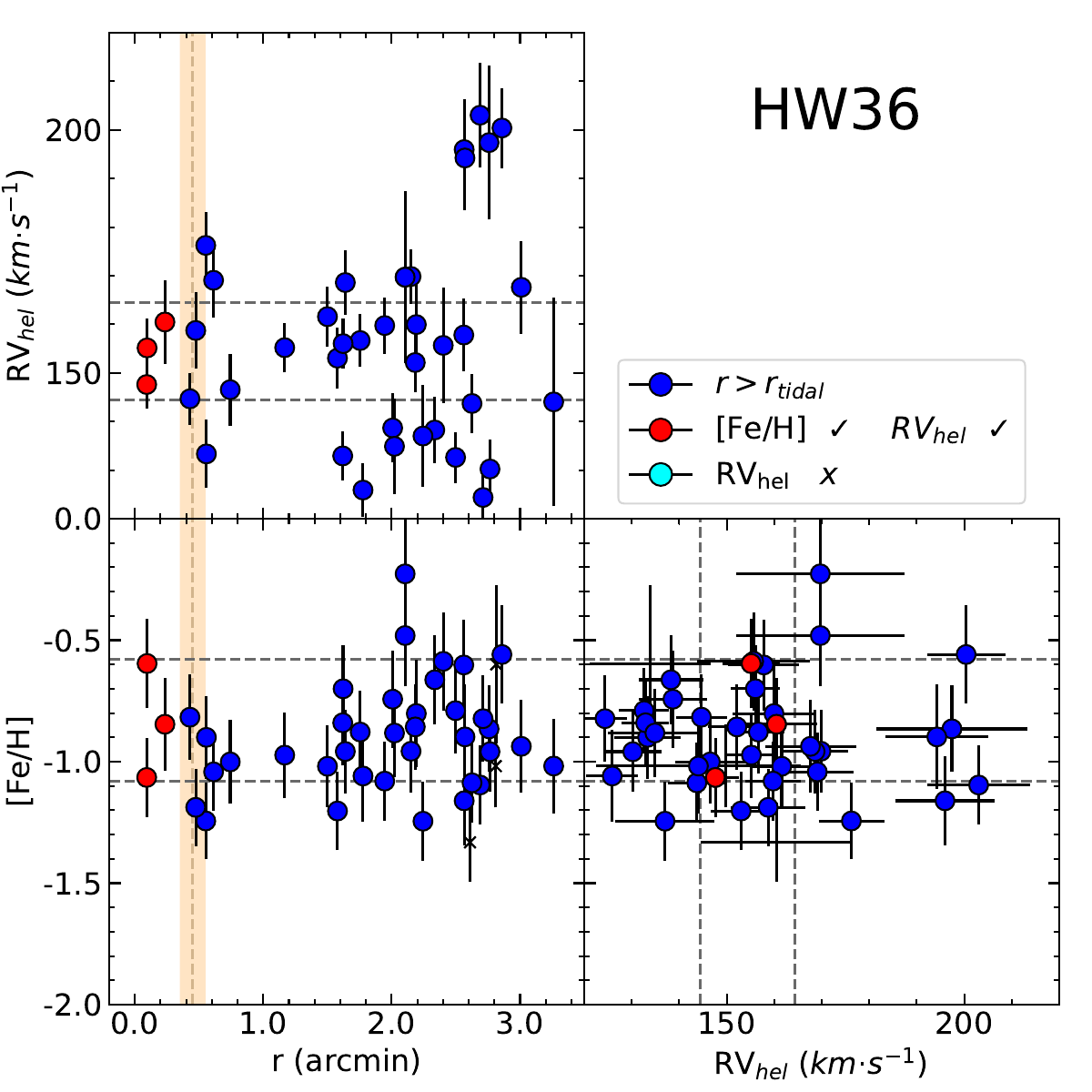}
    \includegraphics[width=0.32\textwidth]{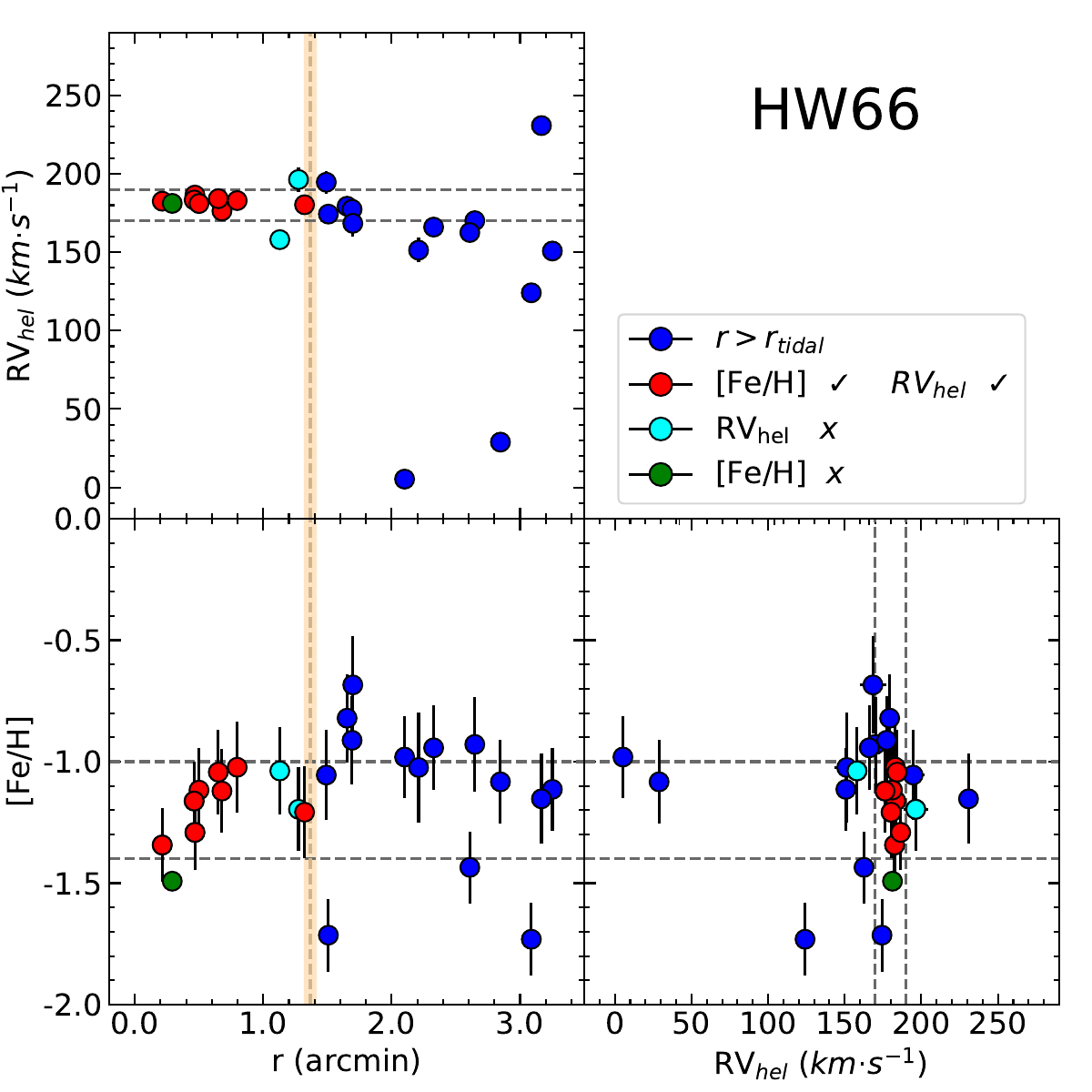}
    \includegraphics[width=0.32\textwidth]{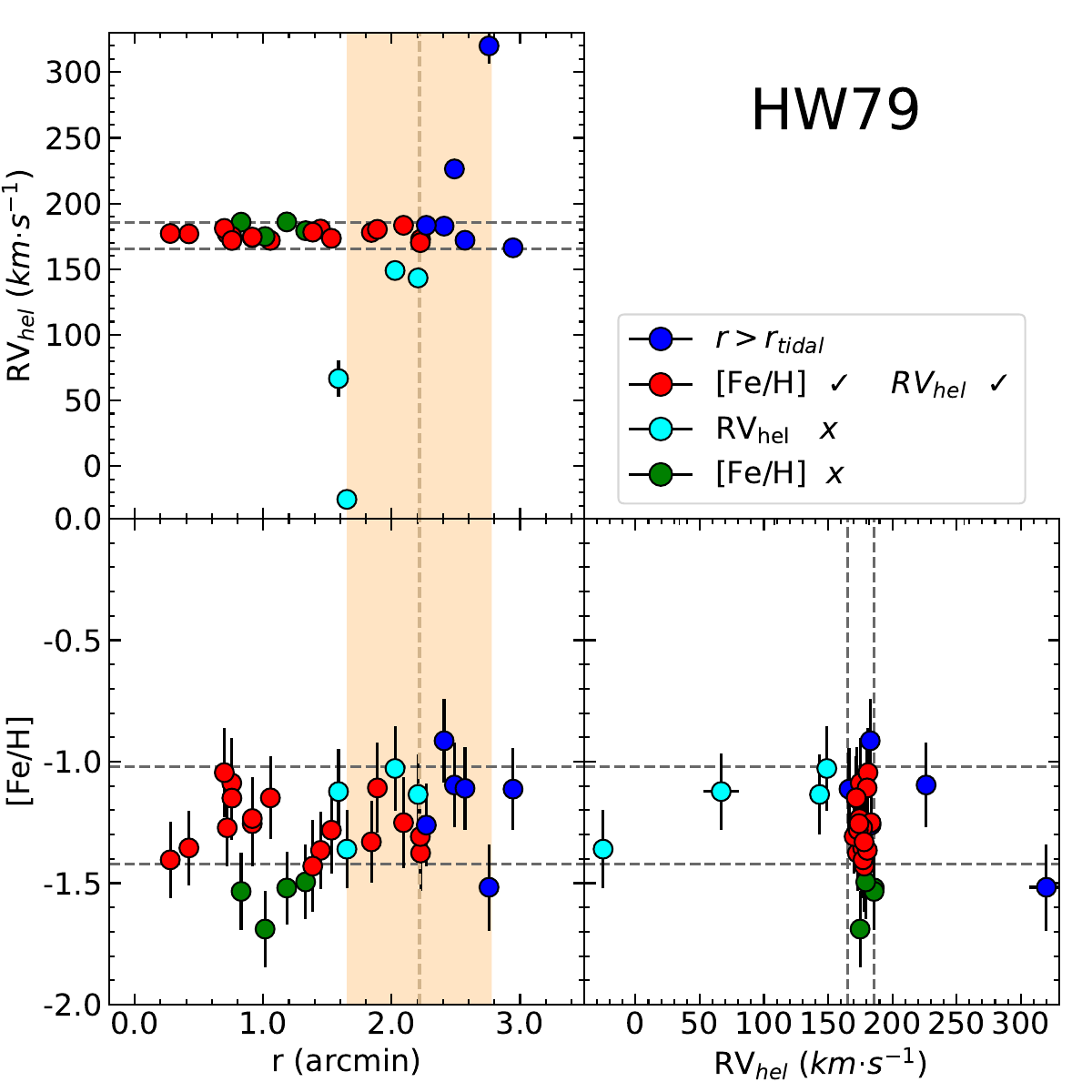}
    \includegraphics[width=0.32\textwidth]{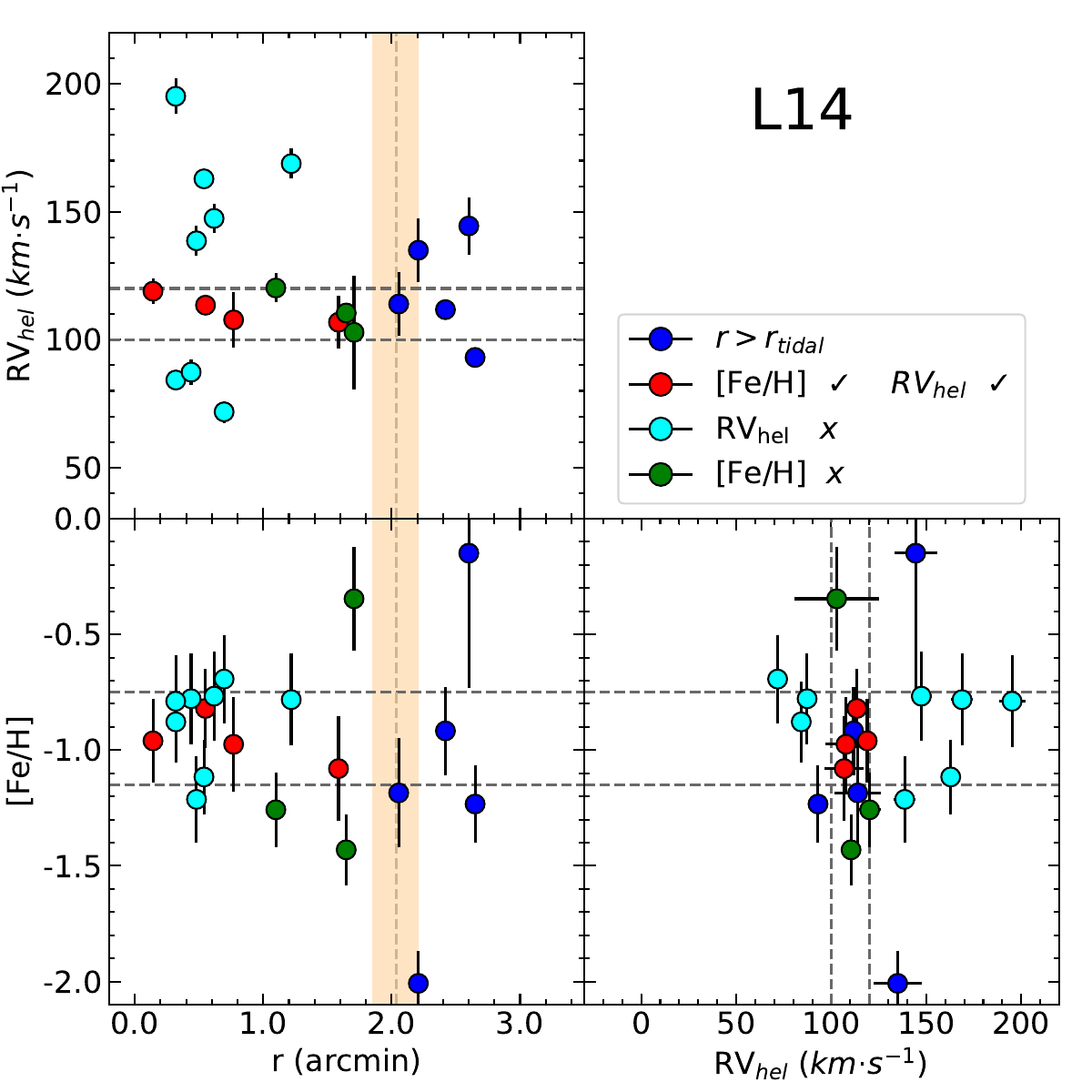}
    \includegraphics[width=0.32\textwidth]{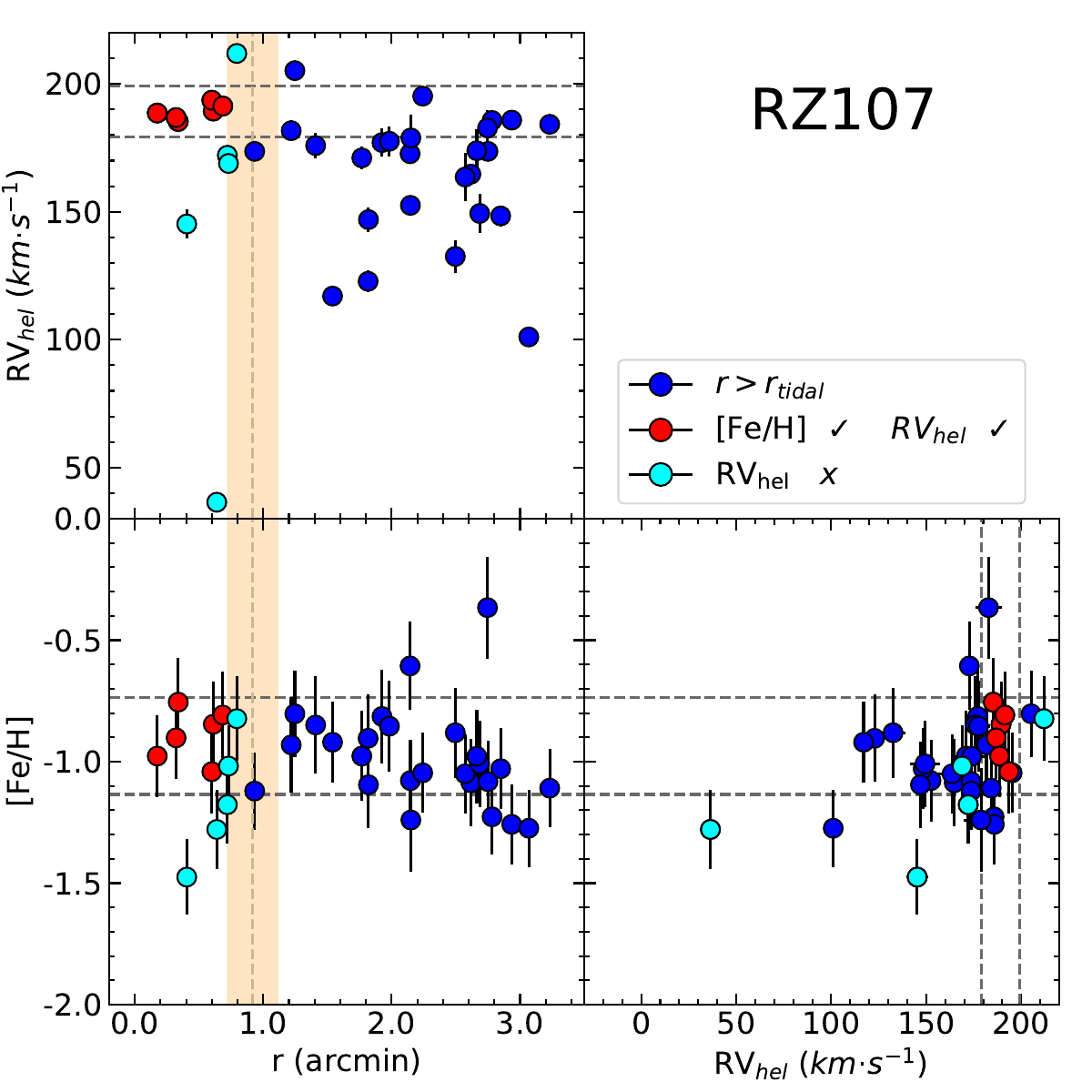}
    \includegraphics[width=0.32\textwidth]{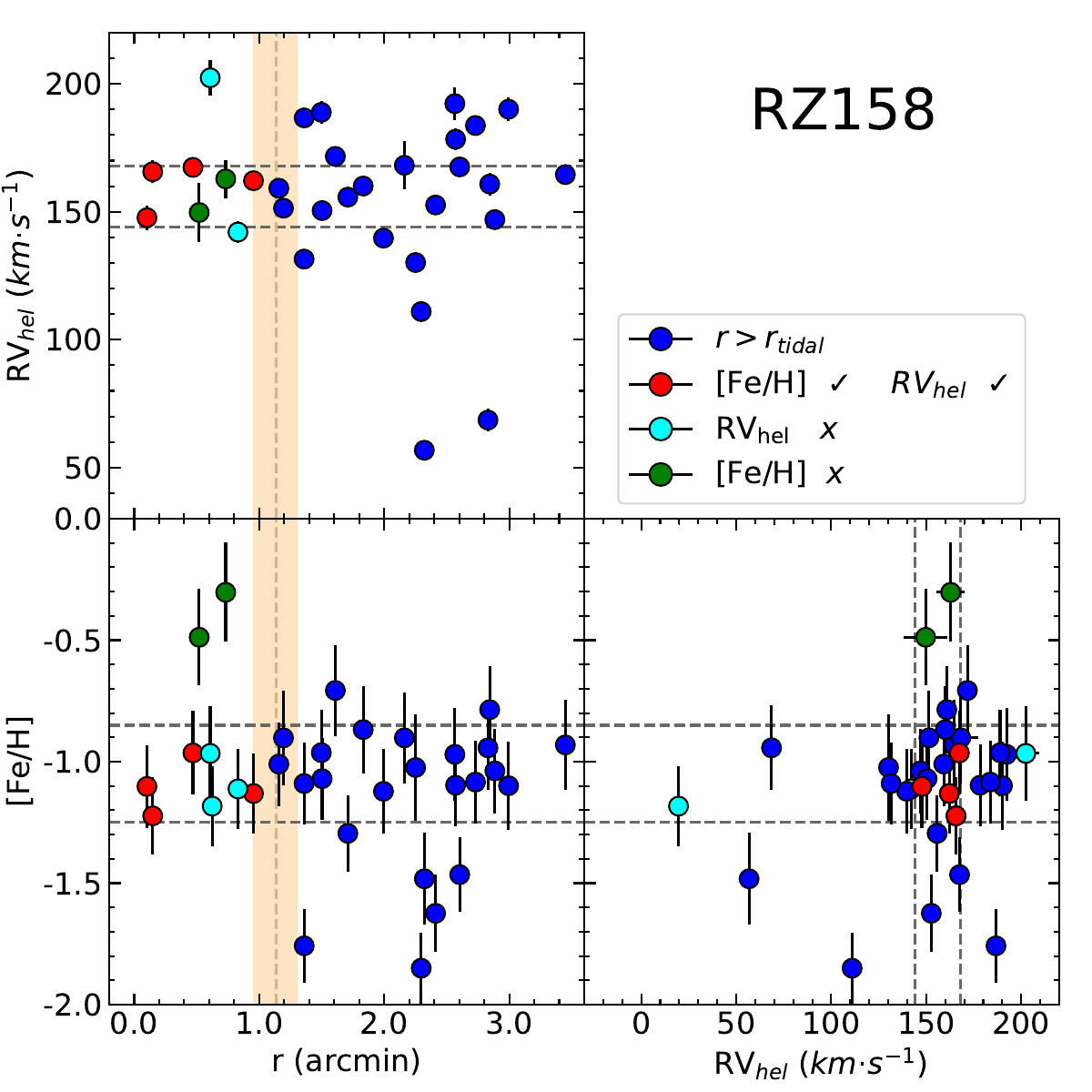}
    \caption{[Fe/H], RVs and distance to the cluster centre of the observed stars. Vertical lines represent our cuts in distance (the corresponding cluster r$_t$) and horizontal lines our cuts in metallicity and RV.  }
    \label{fig:memb}
\end{figure*}

\begin{figure*}
    \centering
    \includegraphics[width=0.33\textwidth]{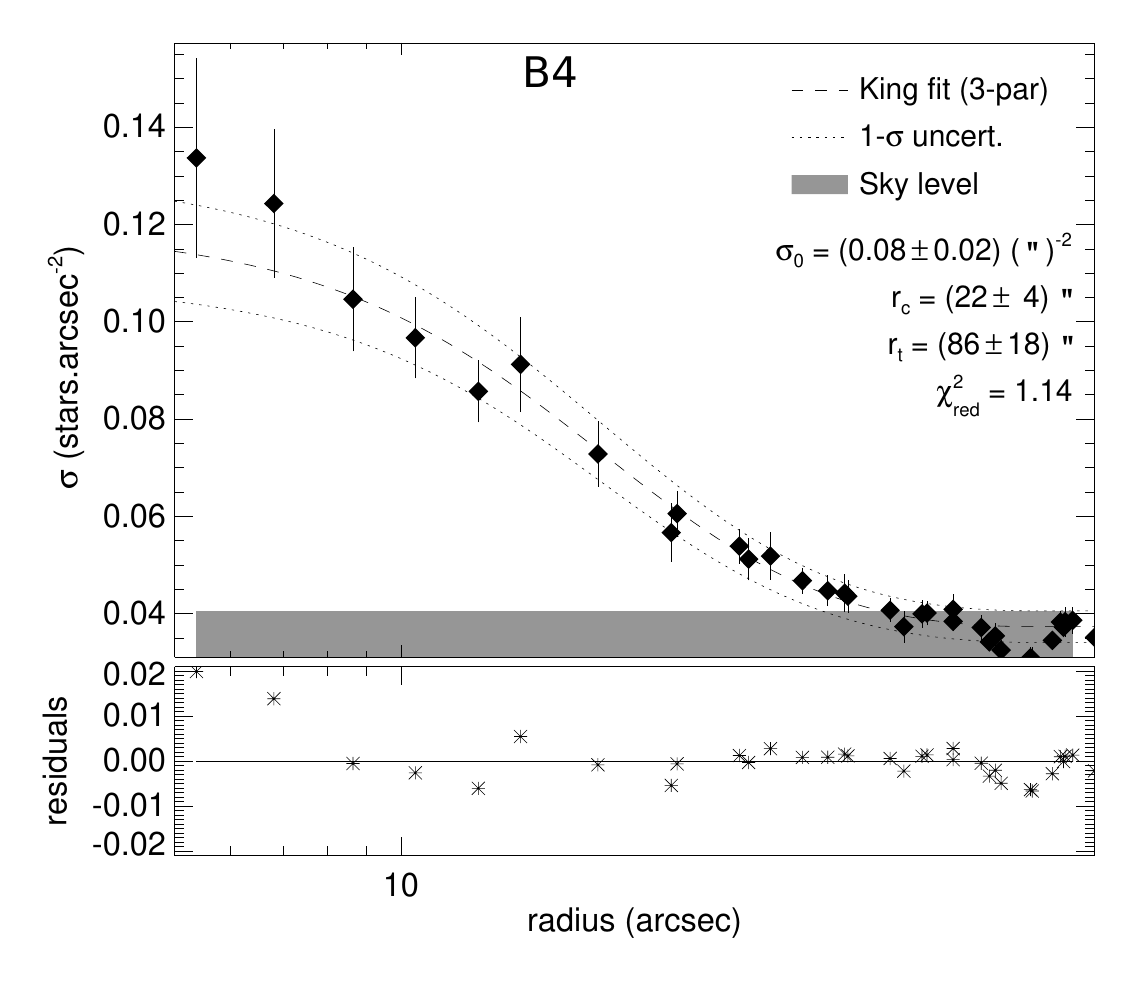}
    \includegraphics[width=0.33\textwidth]{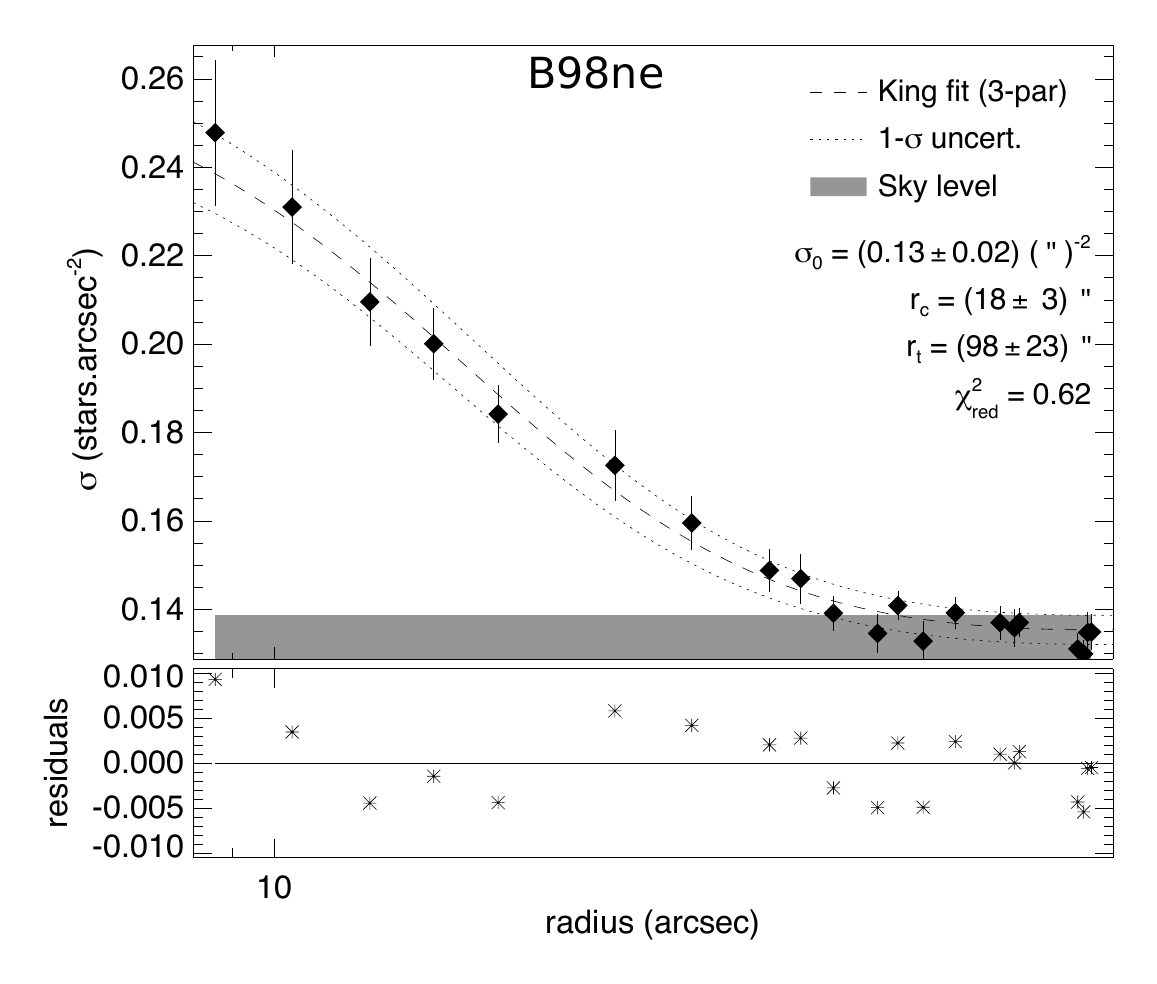}
    \includegraphics[width=0.33\textwidth]{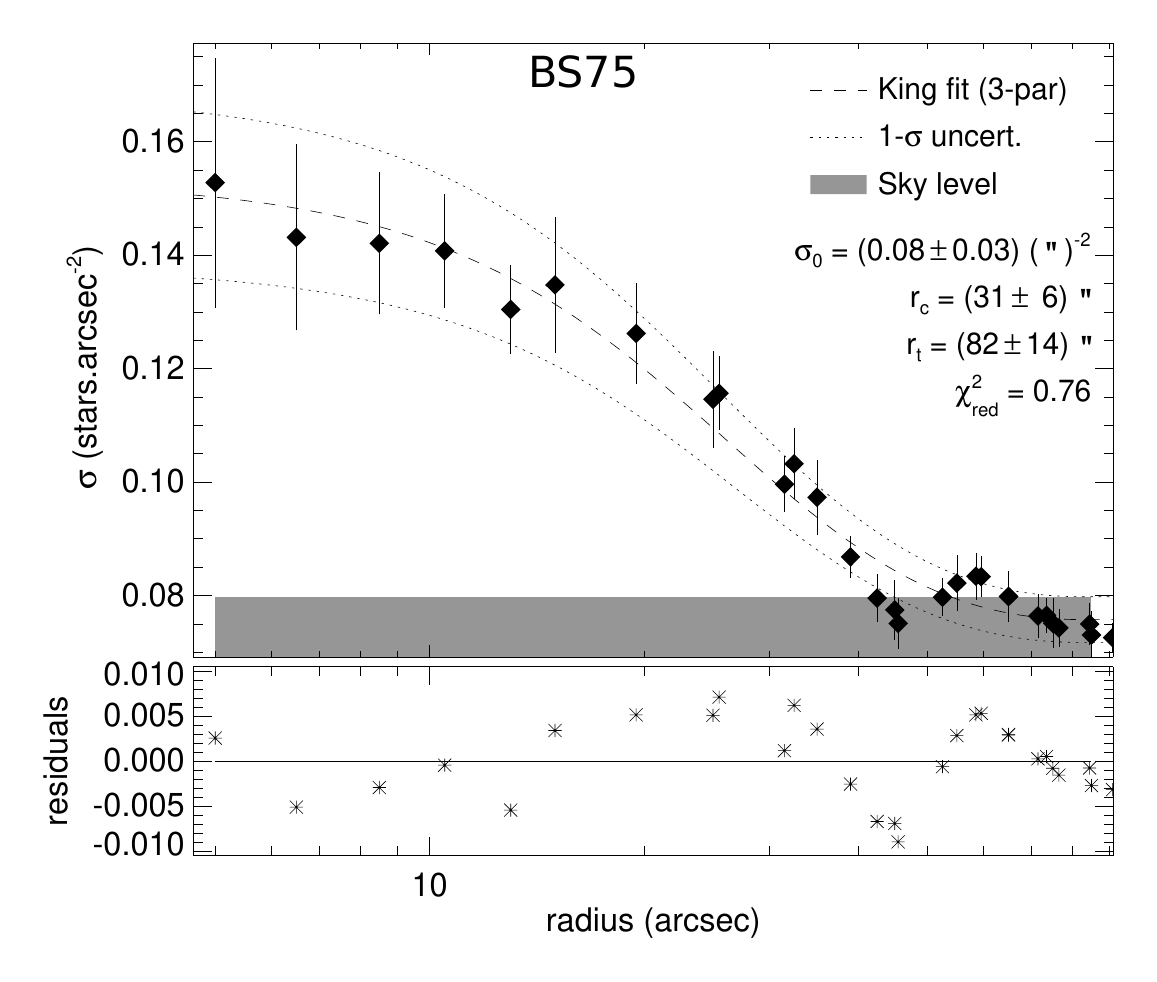}
    \includegraphics[width=0.33\textwidth]{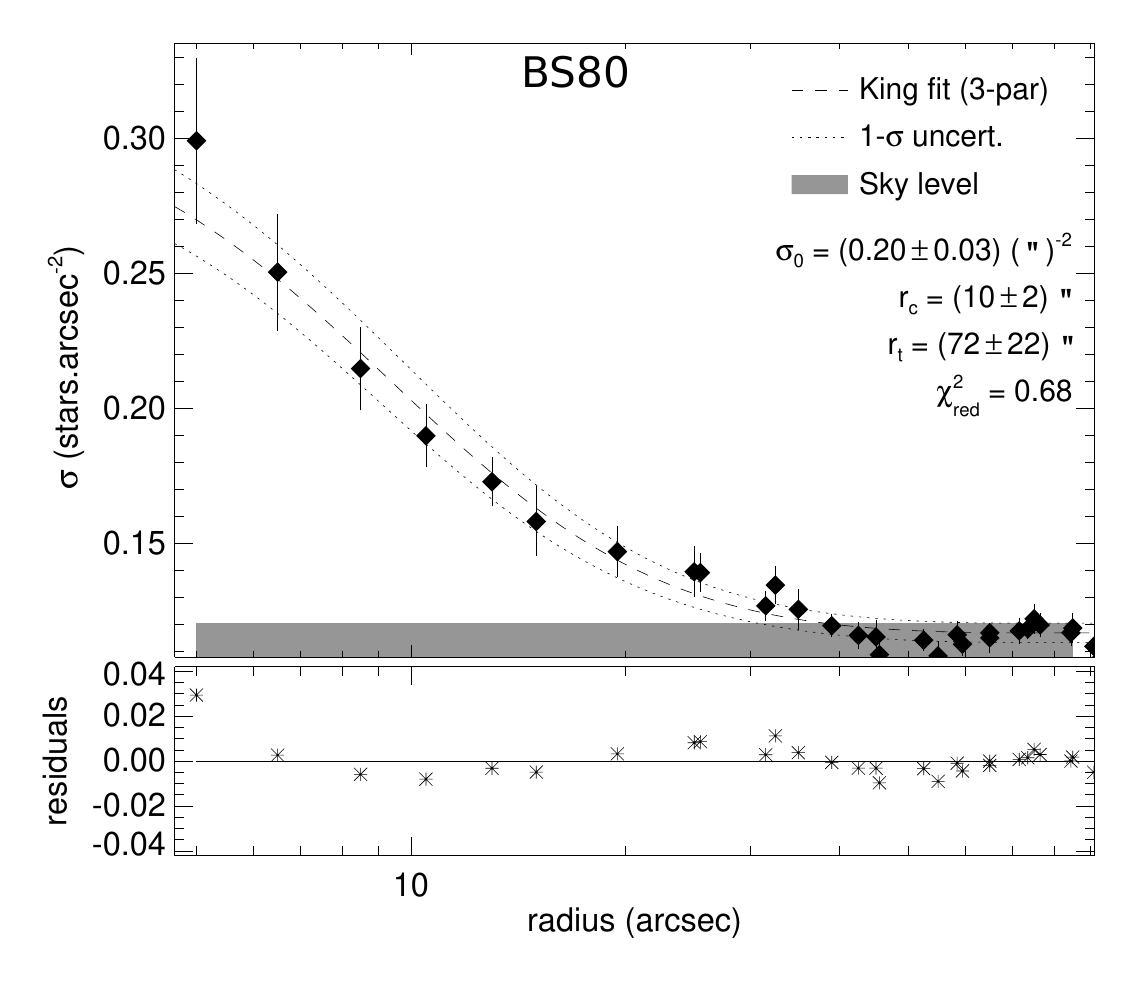}
    \includegraphics[width=0.33\textwidth]{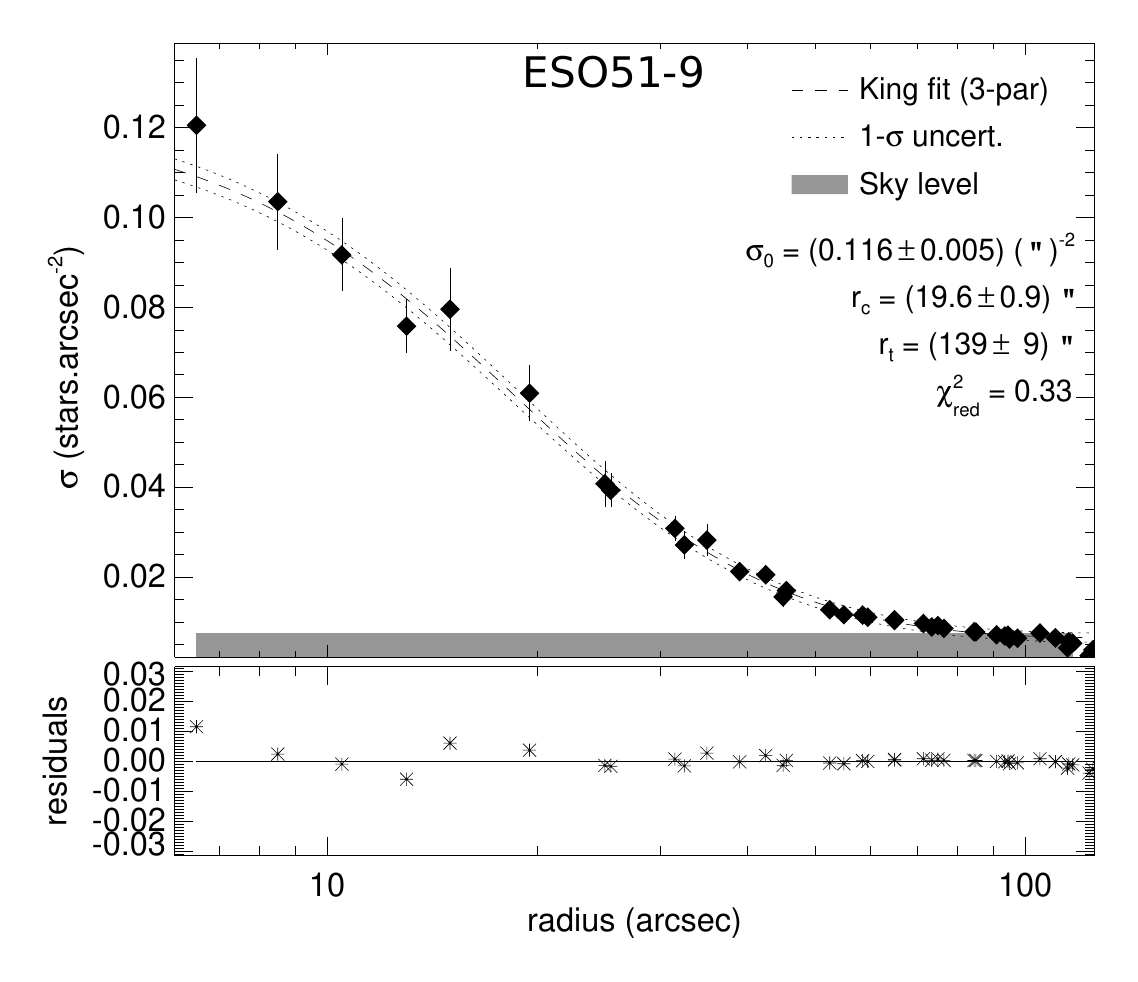}
    \includegraphics[width=0.33\textwidth]{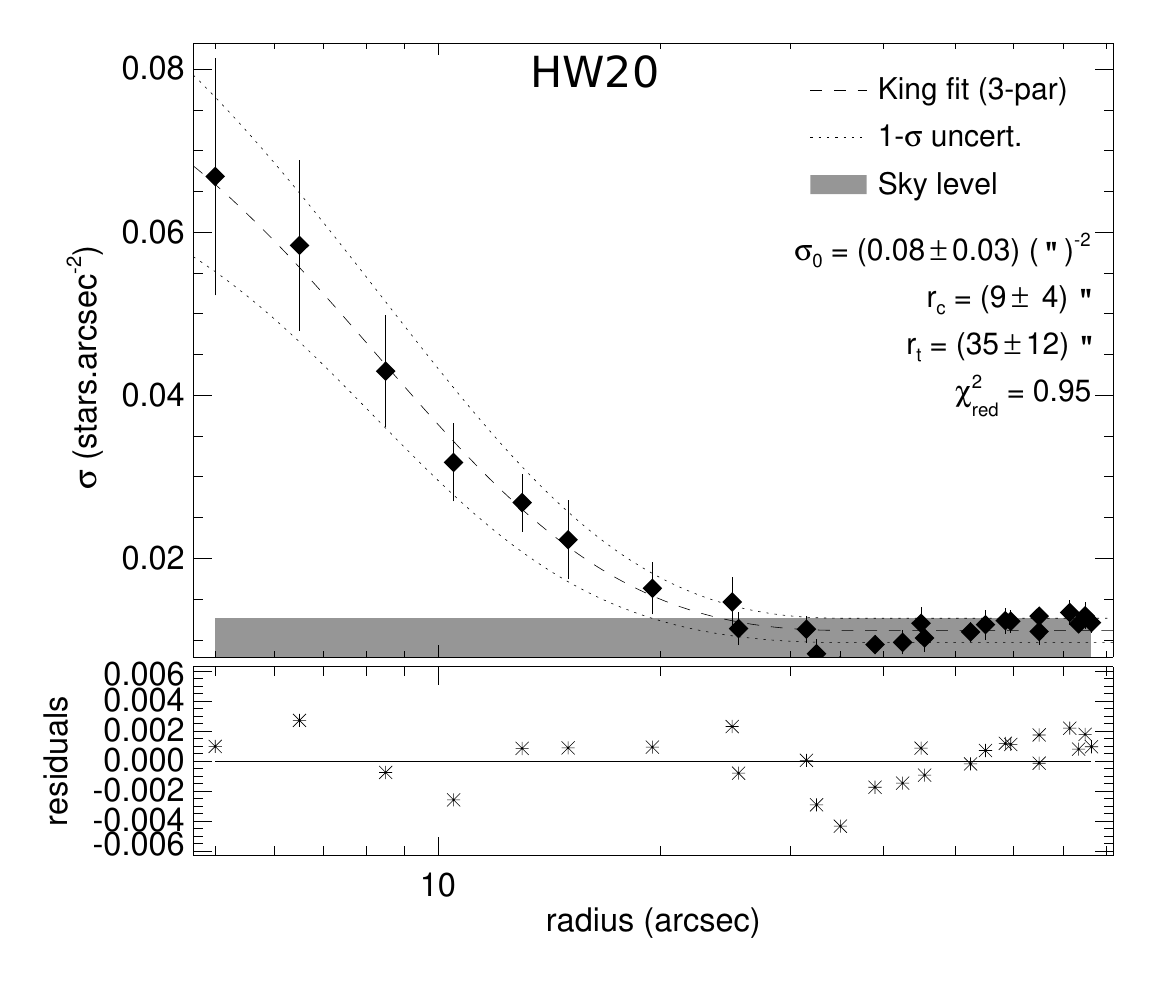}
    \includegraphics[width=0.33\textwidth]{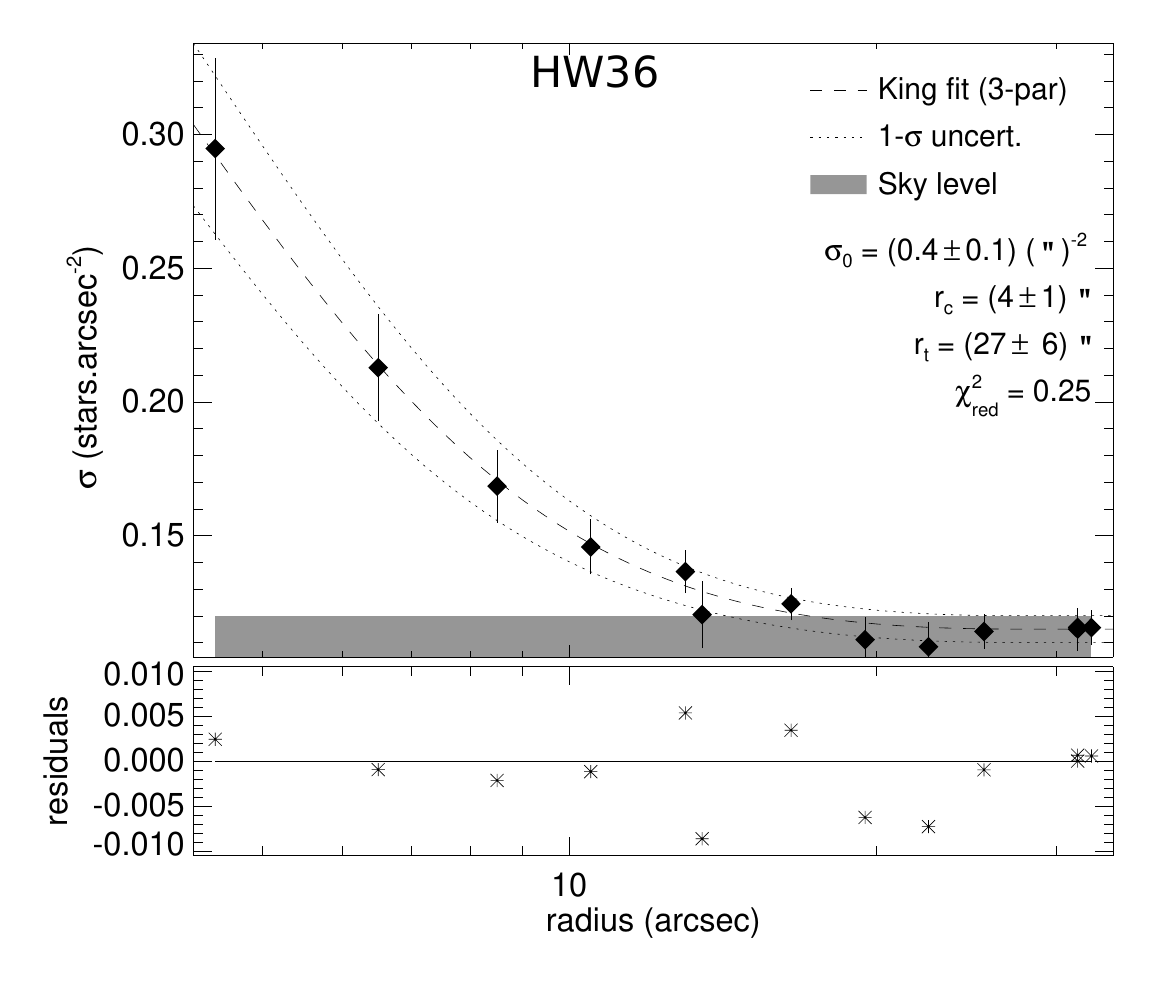}
    \includegraphics[width=0.33\textwidth]{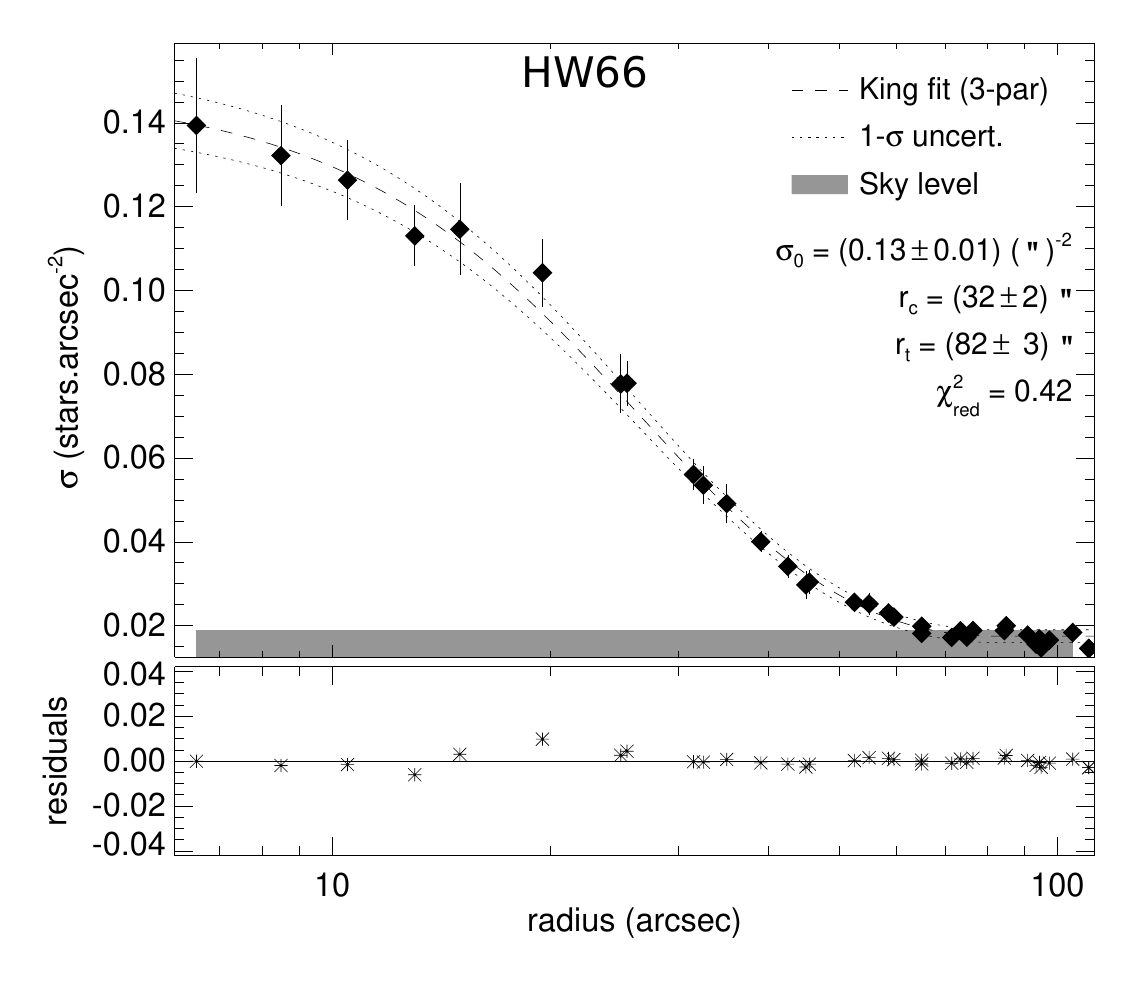}
    \includegraphics[width=0.33\textwidth]{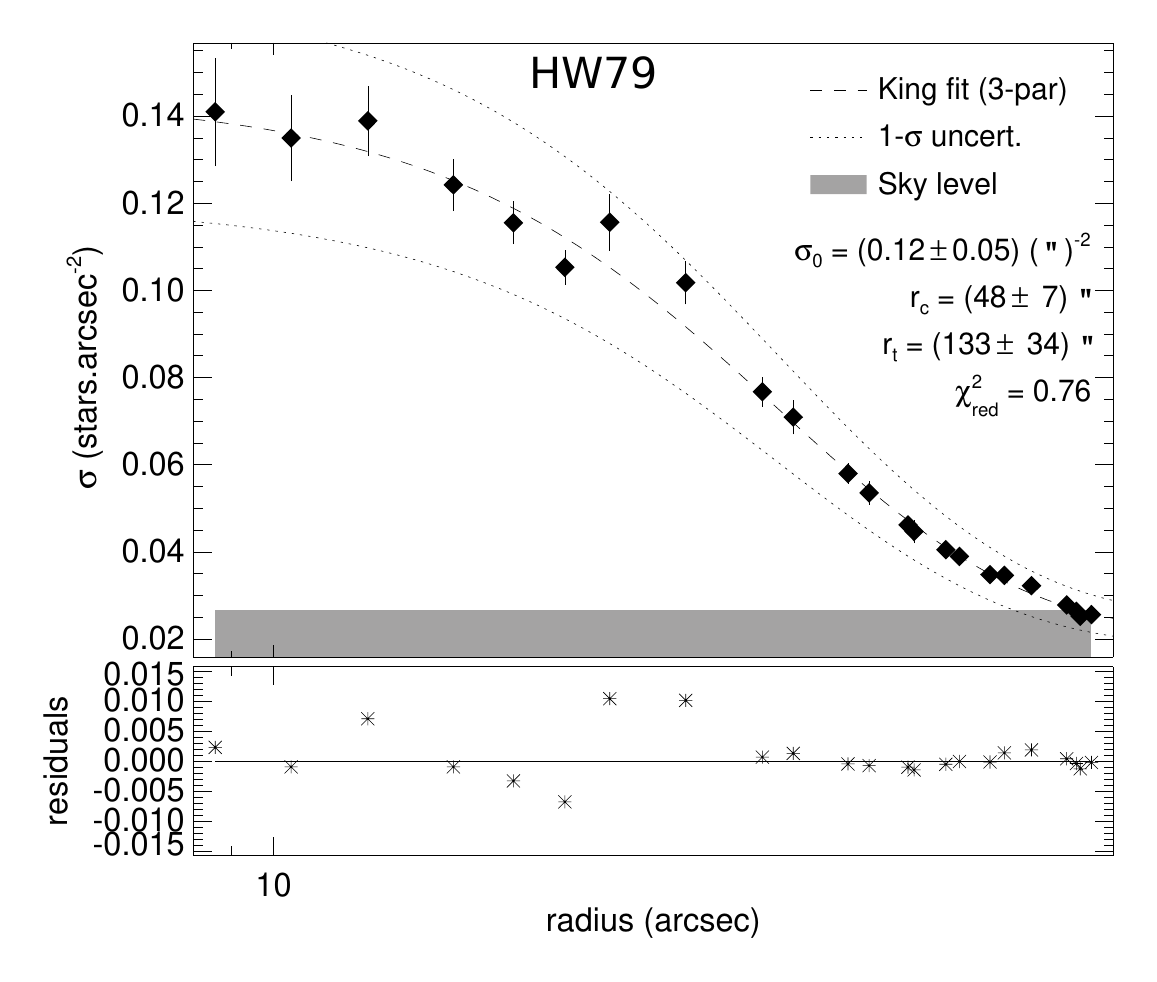}
    \includegraphics[width=0.33\textwidth]{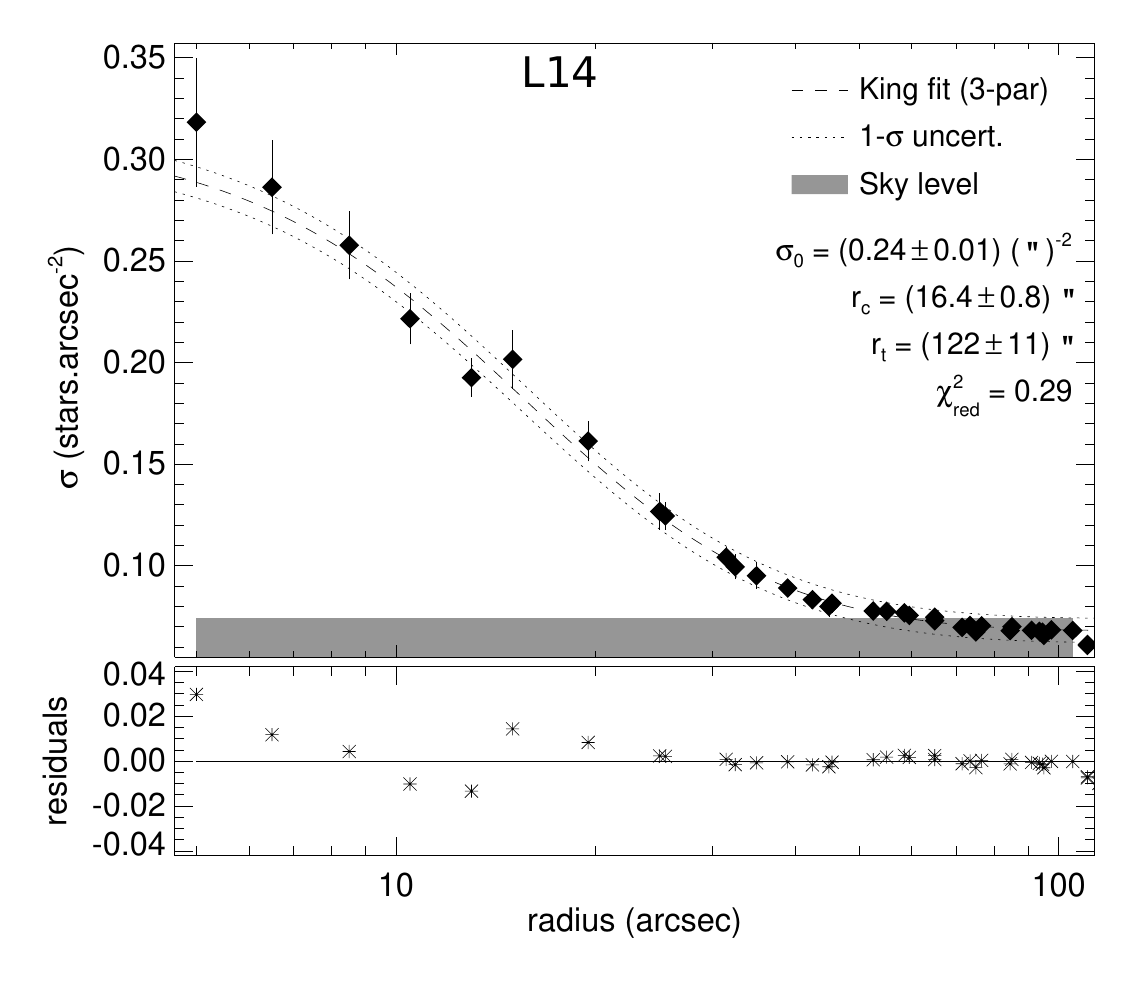}
    \includegraphics[width=0.33\textwidth]{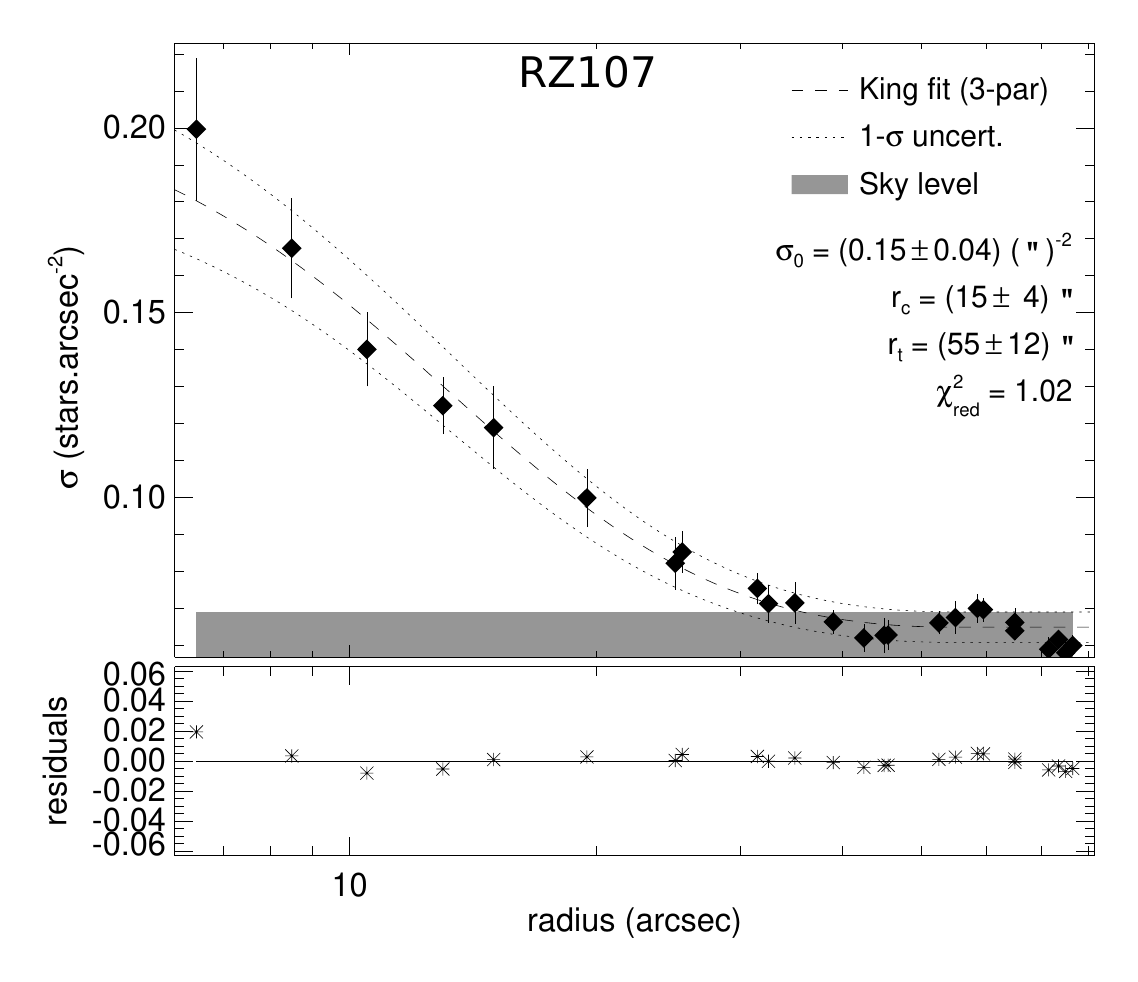}
    \includegraphics[width=0.33\textwidth]{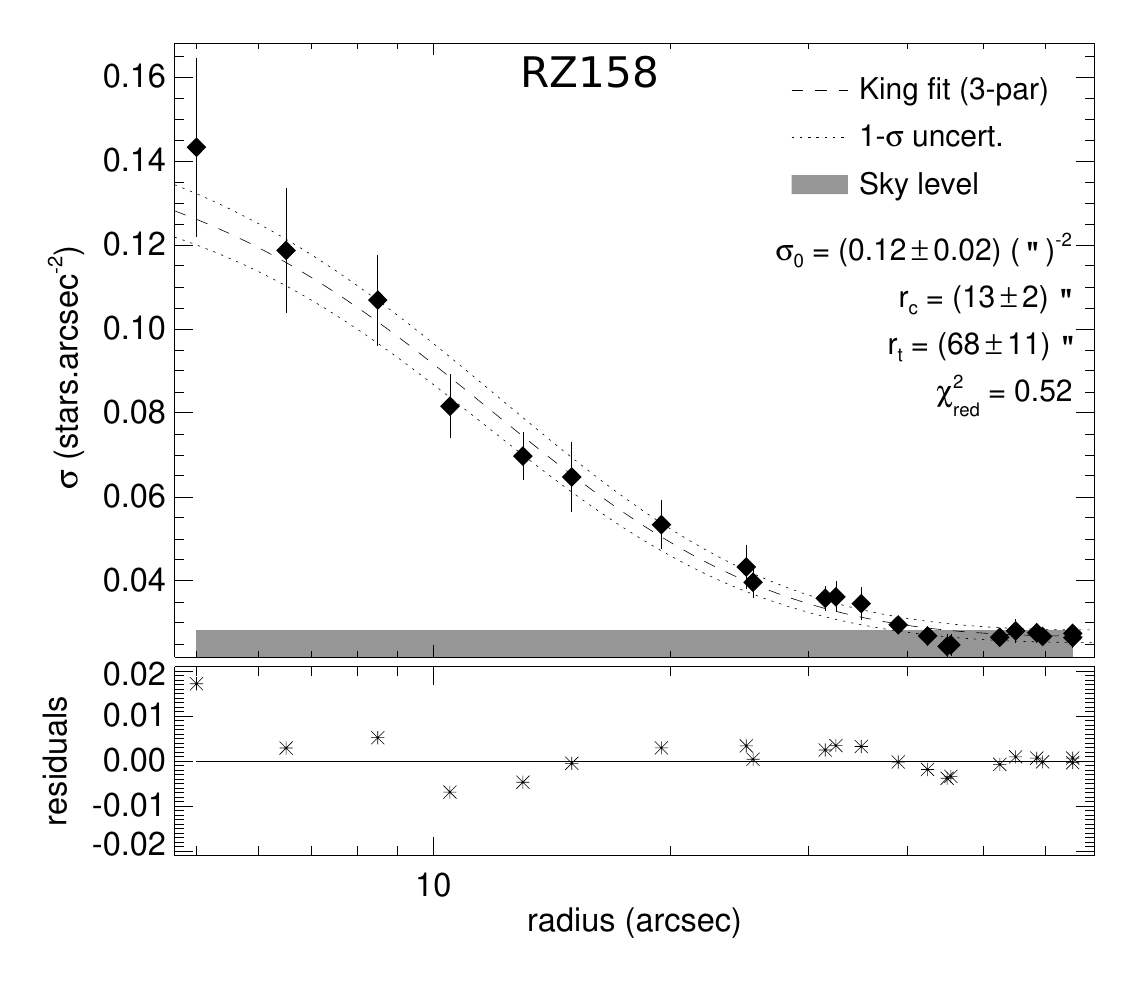}
    \caption{Radial density profiles (diamonds) for clusters of our sample. The fit residuals are shown in the lower sub-panel of each plot.}
     \label{fig:rad_profiles}
\end{figure*}

\begin{figure*}
    \centering
    \includegraphics[width=0.25\textwidth]{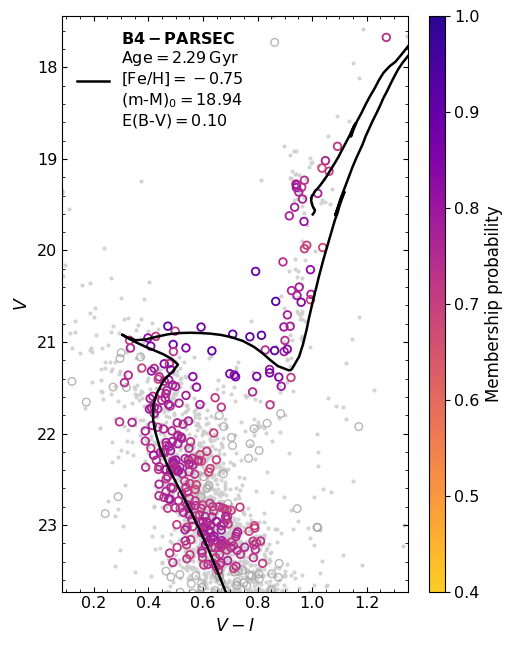}
    \includegraphics[width=0.25\textwidth]{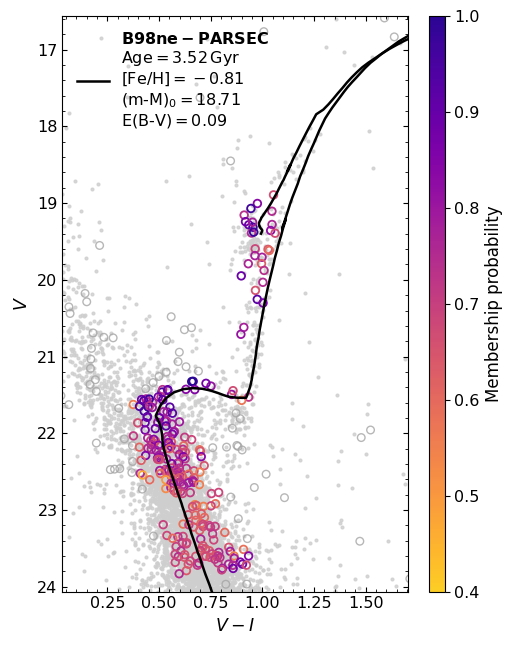}
    \includegraphics[width=0.25\textwidth]{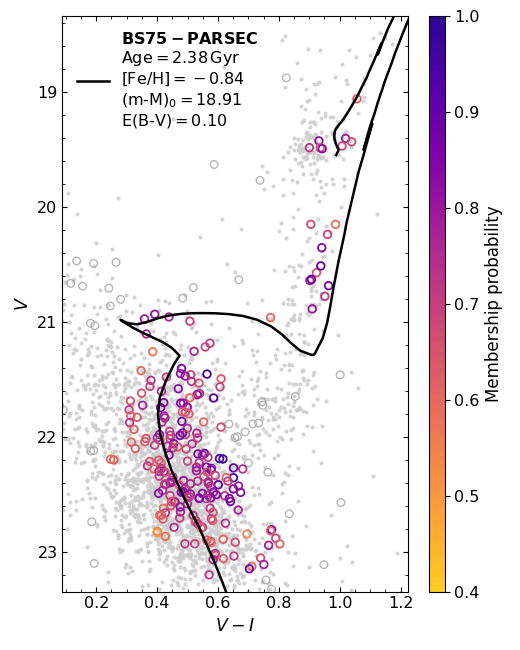}
    \includegraphics[width=0.25\textwidth]{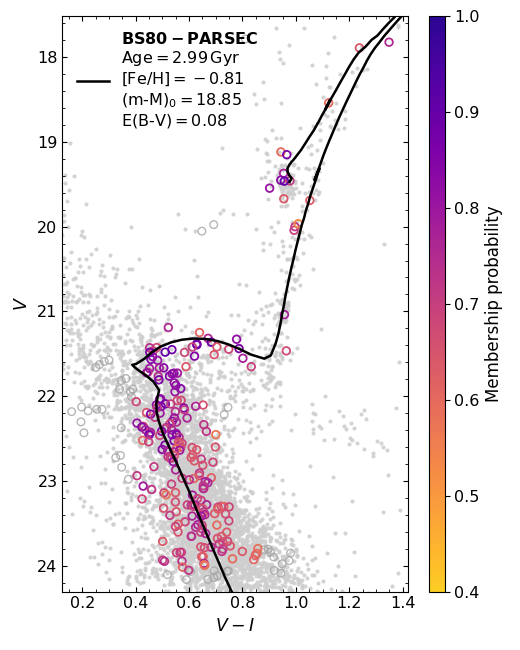}
    \includegraphics[width=0.25\textwidth]{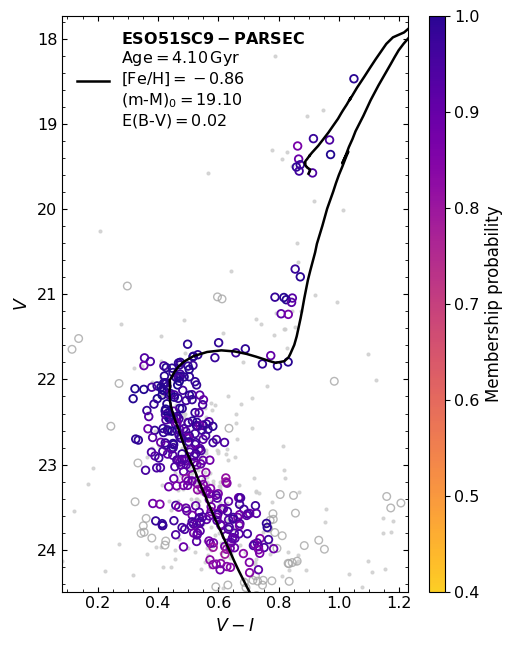}
    \includegraphics[width=0.25\textwidth]{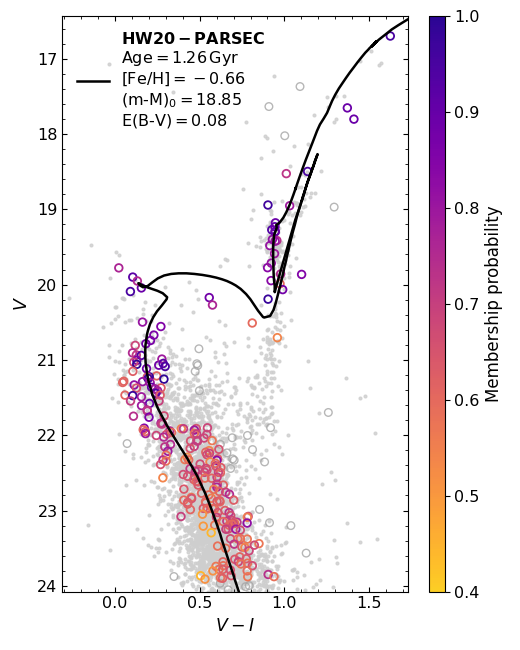}
    \includegraphics[width=0.25\textwidth]{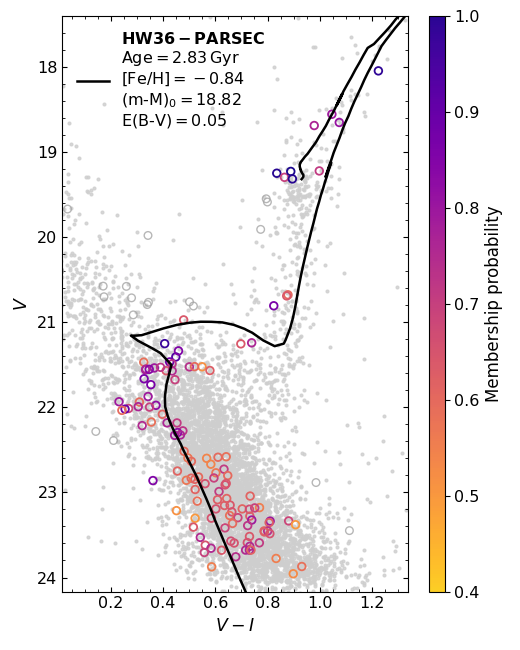}
    \includegraphics[width=0.25\textwidth]{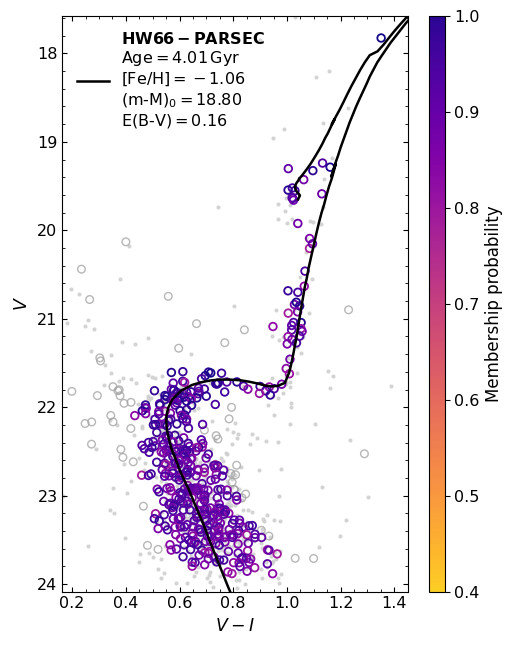}
    \includegraphics[width=0.25\textwidth]{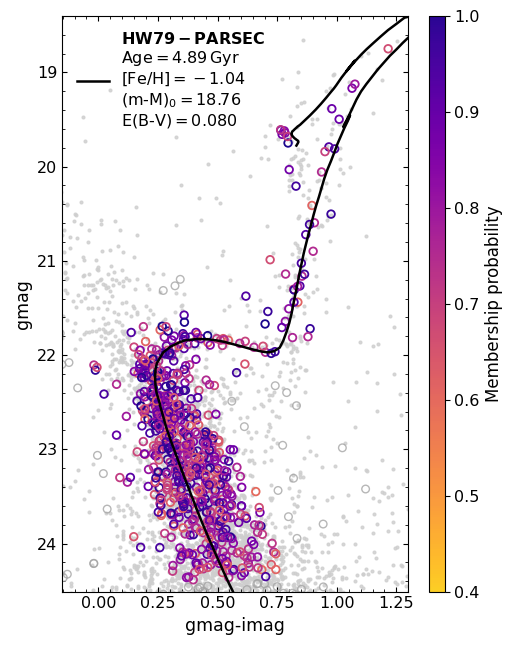}
    \includegraphics[width=0.25\textwidth]{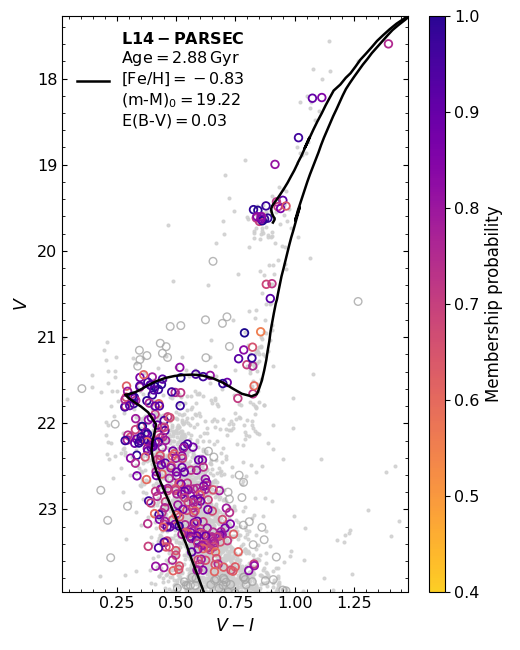}
    \includegraphics[width=0.25\textwidth]{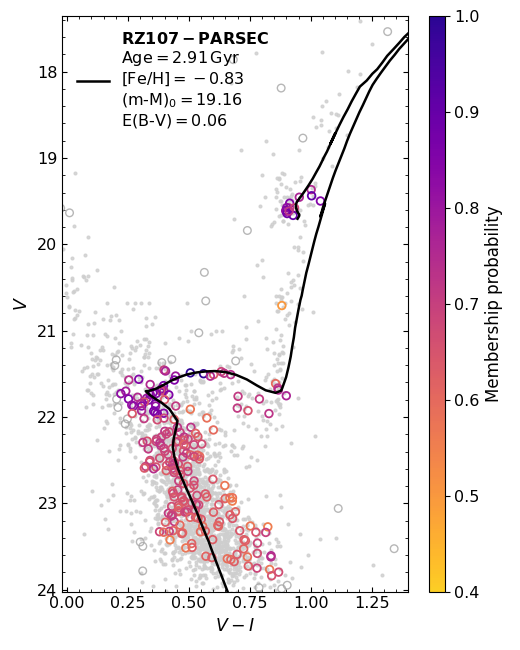}
    \includegraphics[width=0.25\textwidth]{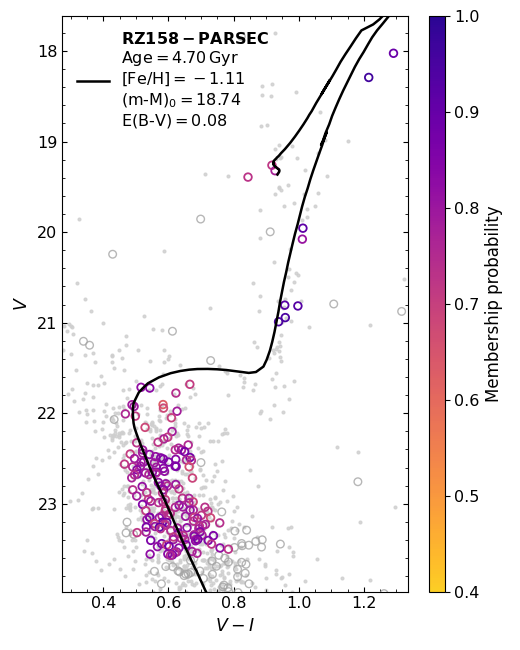}
    \caption{Statistically decontaminated CMDs for our cluster sample. The  best fitting isochrones  using the \texttt{SIRIUS} code are shown by the solid lines. Field and probable cluster member stars are represented by grey dots and circles, respectively. The membership probability is identified with the color code shown in each plot.}
    \label{fig:cmds}
\end{figure*}

\begin{figure*}
    \centering
    \includegraphics[width=0.30\textwidth]{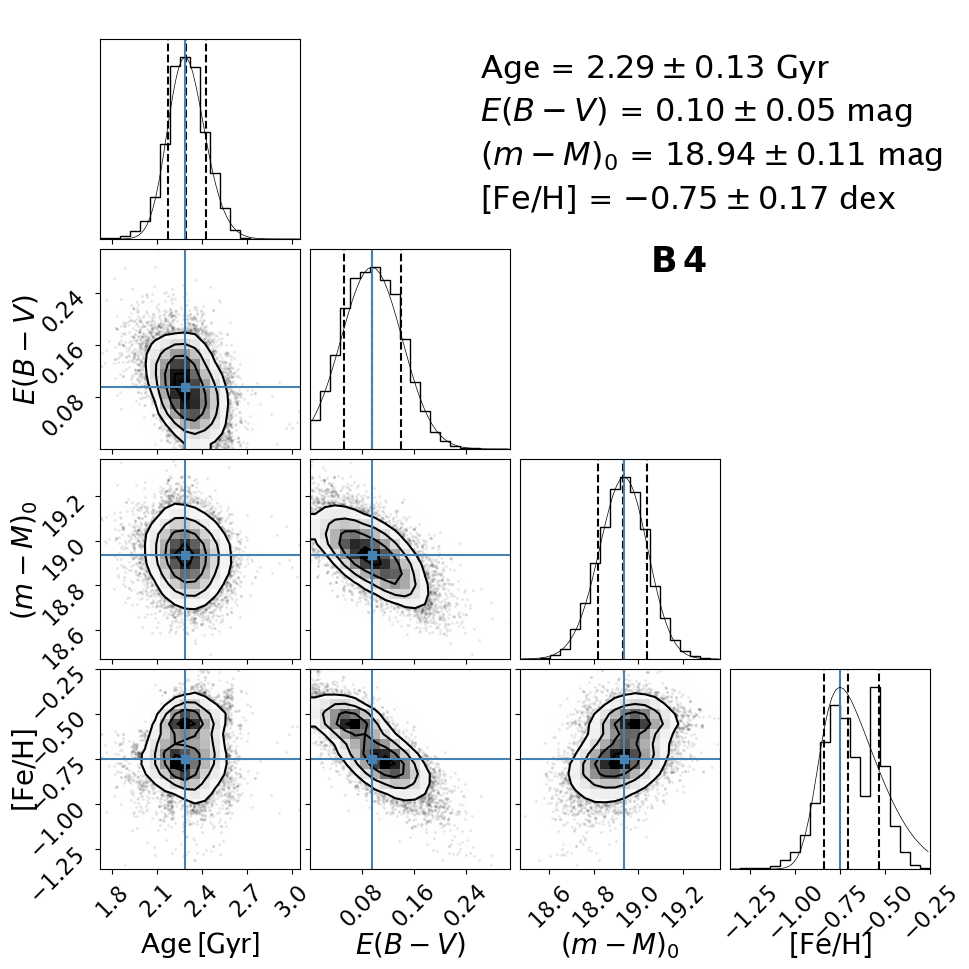}
    \includegraphics[width=0.30\textwidth]{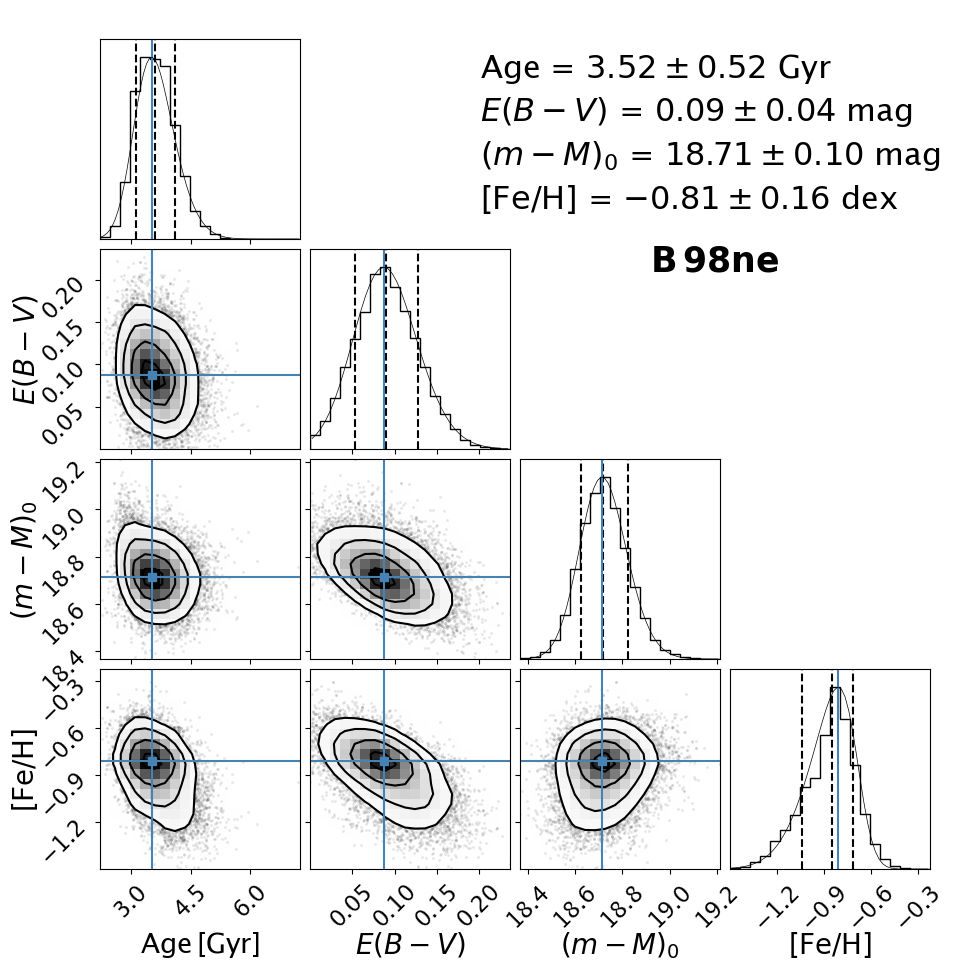}
    \includegraphics[width=0.30\textwidth]{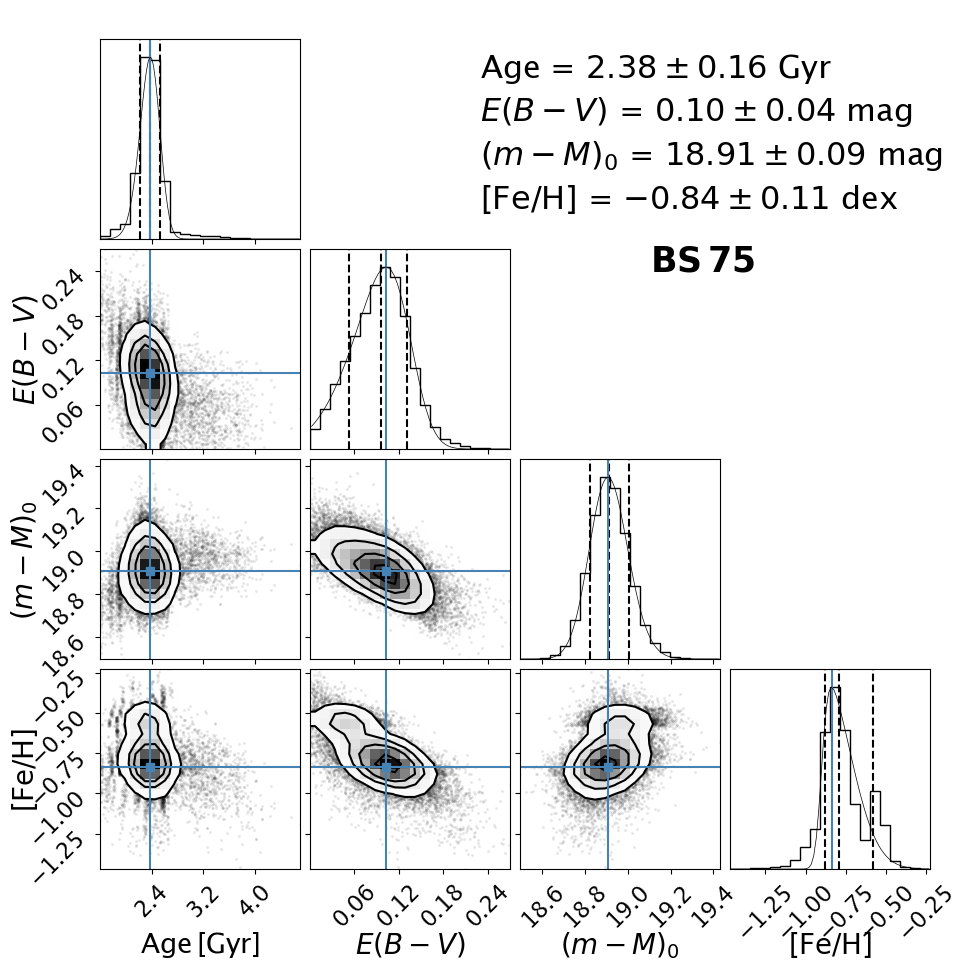}
    \includegraphics[width=0.30\textwidth]{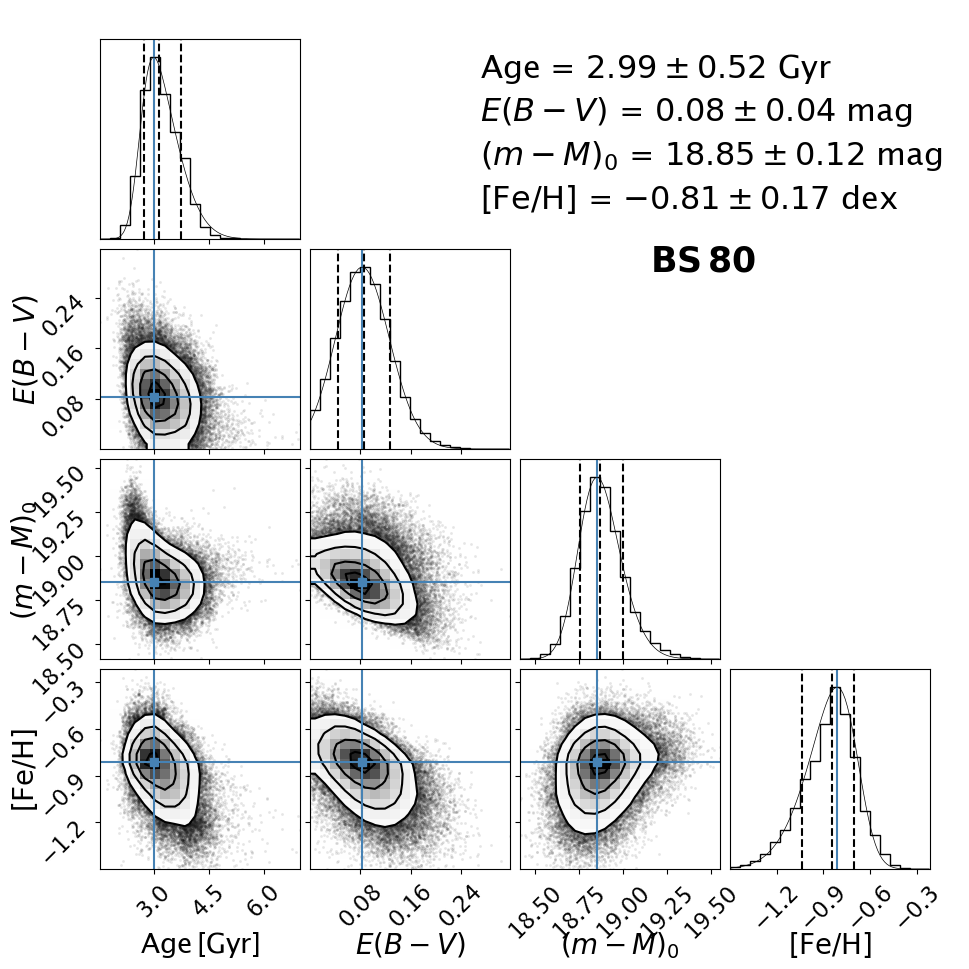}
    \includegraphics[width=0.30\textwidth]{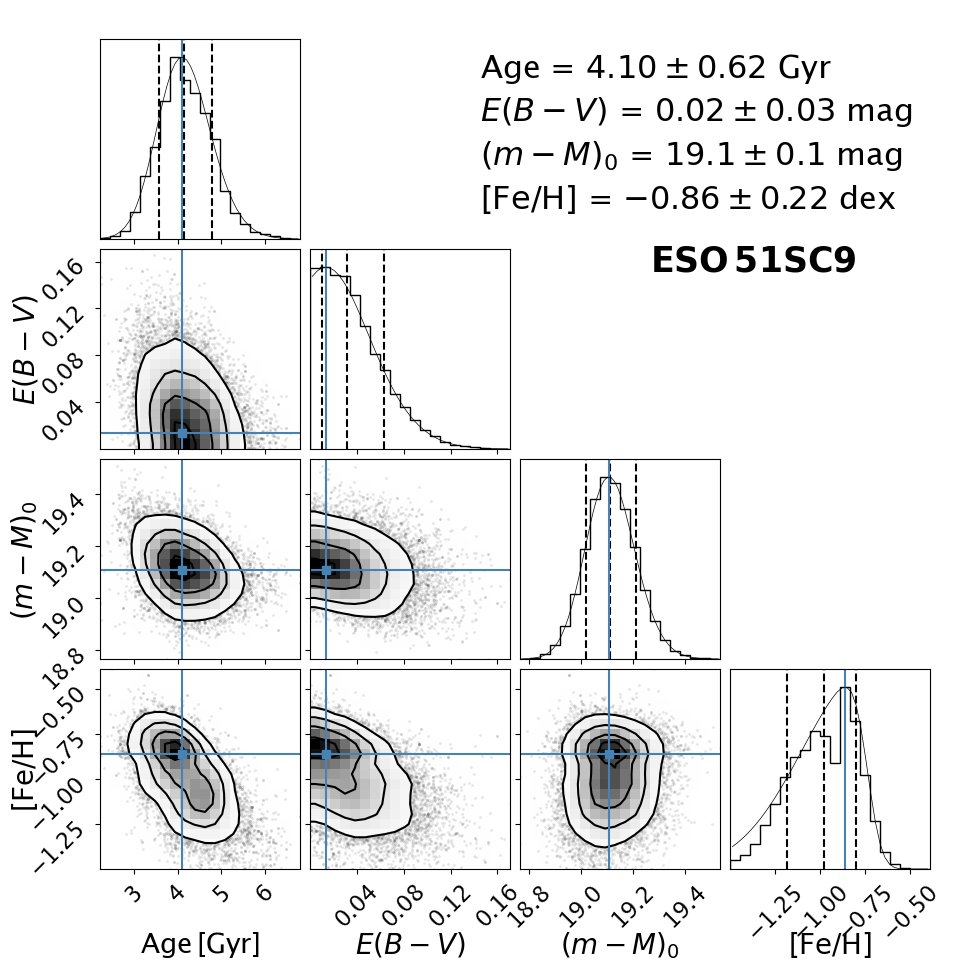}
    \includegraphics[width=0.30\textwidth]{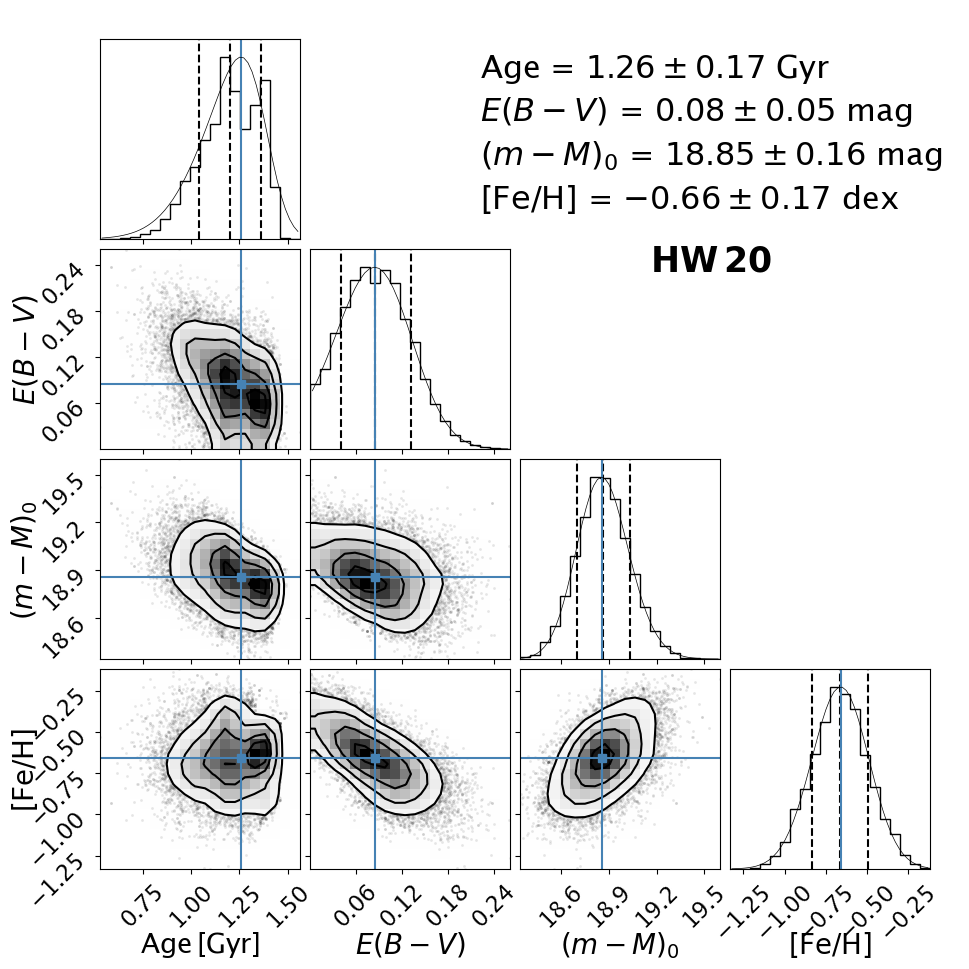}
    \includegraphics[width=0.30\textwidth]{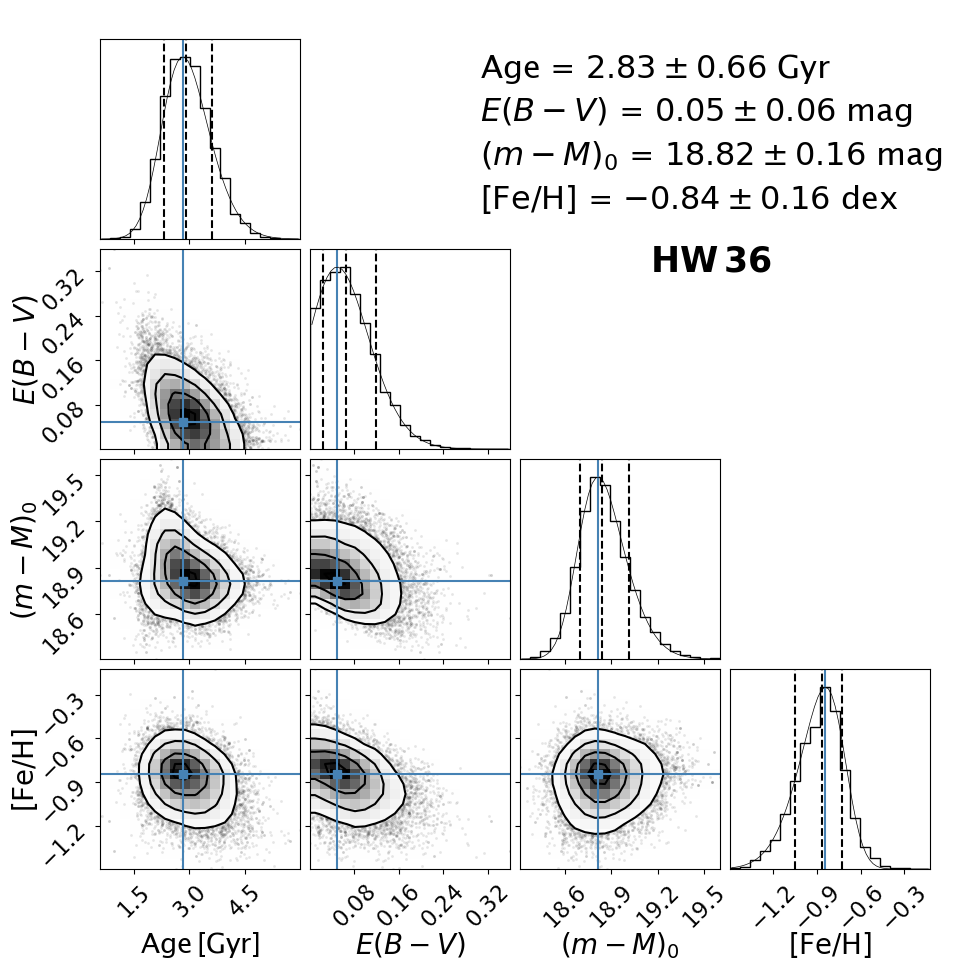}
    \includegraphics[width=0.30\textwidth]{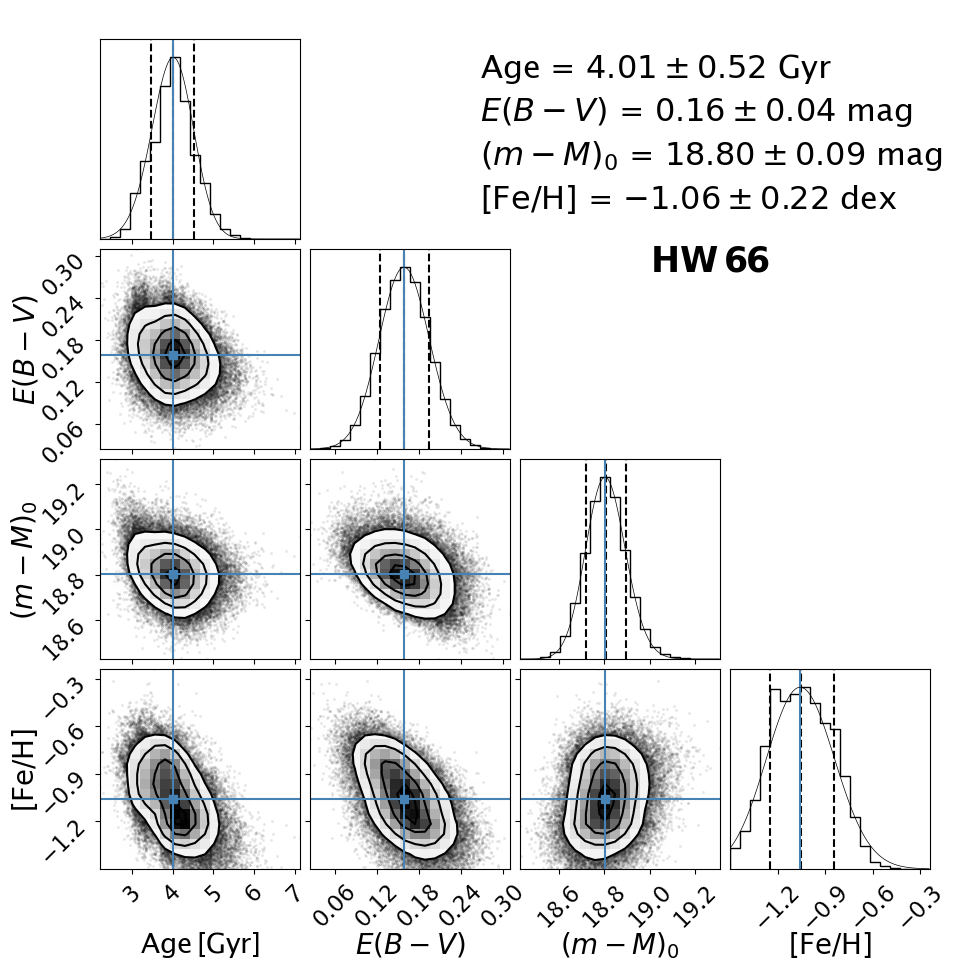}
    \includegraphics[width=0.30\textwidth]{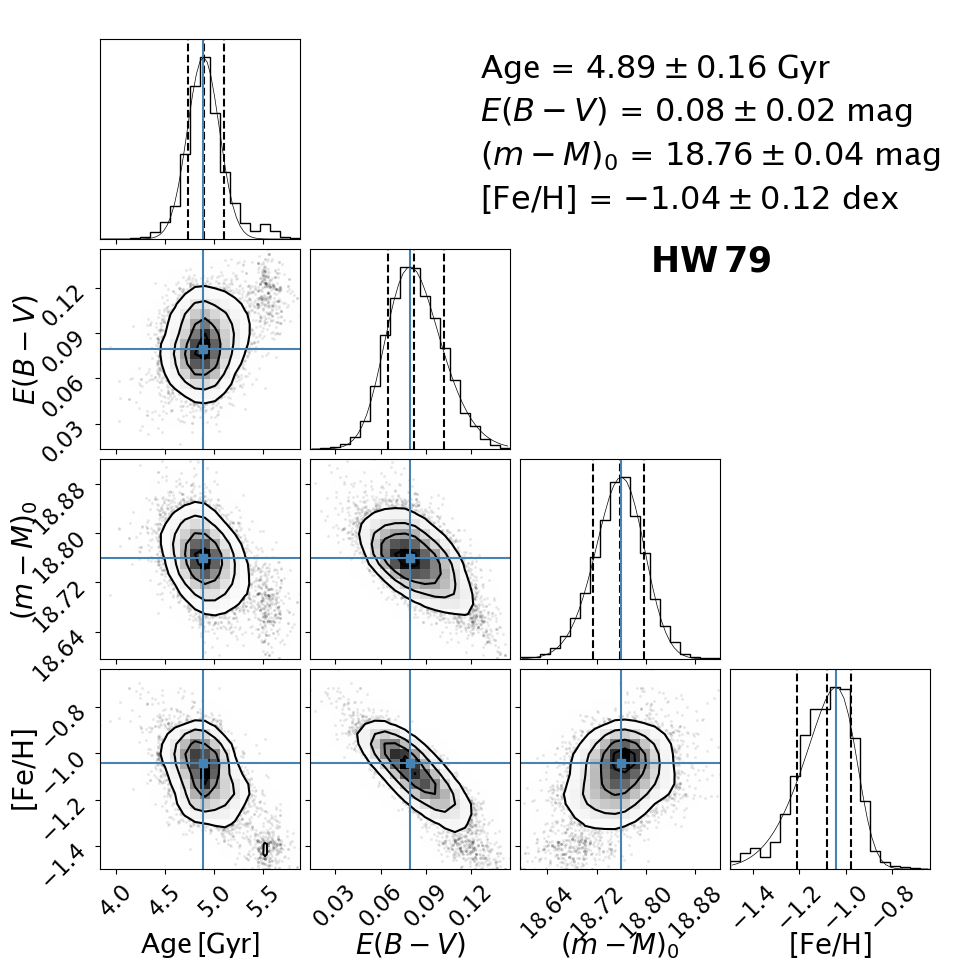}
    \includegraphics[width=0.30\textwidth]{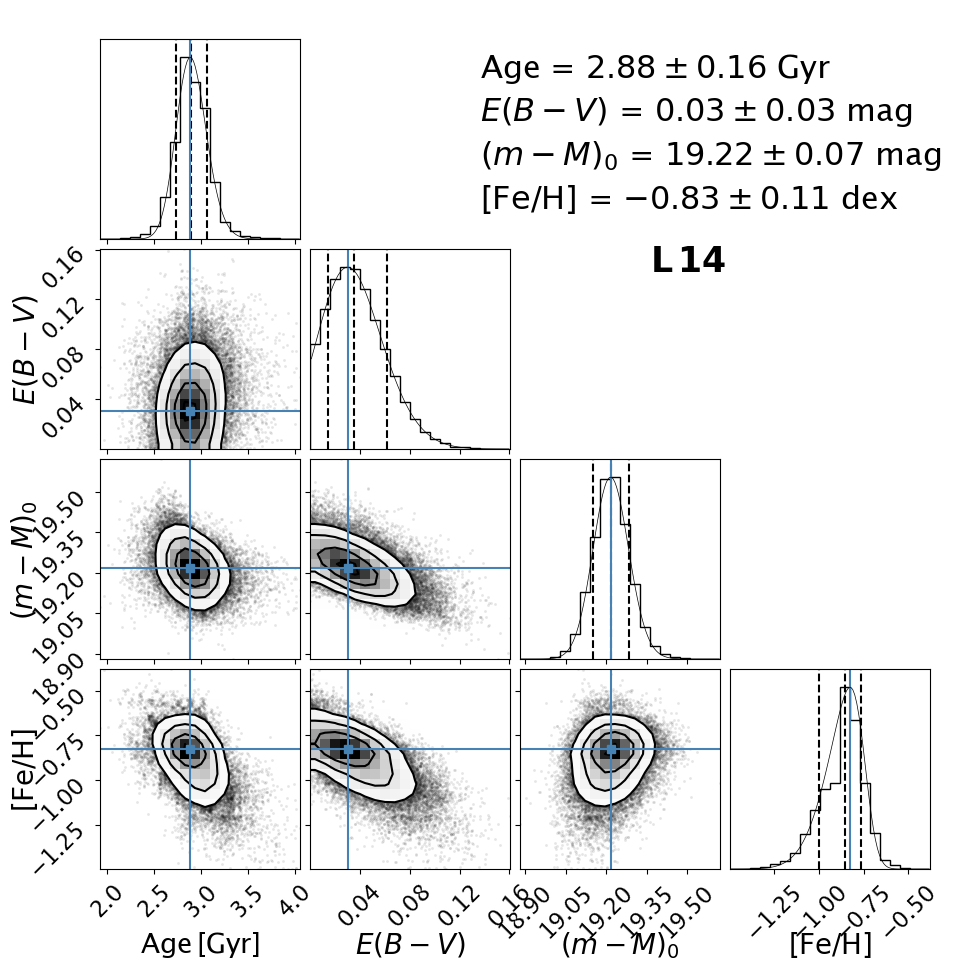}
    \includegraphics[width=0.30\textwidth]{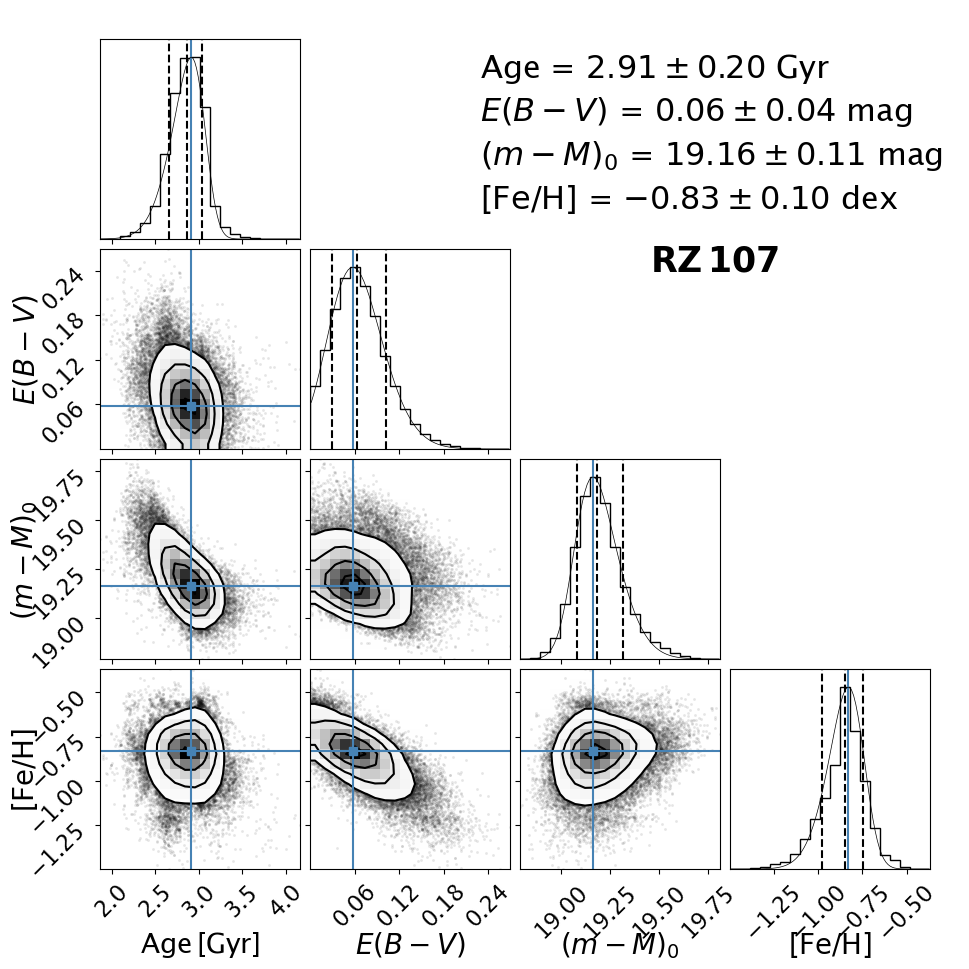}
    \includegraphics[width=0.30\textwidth]{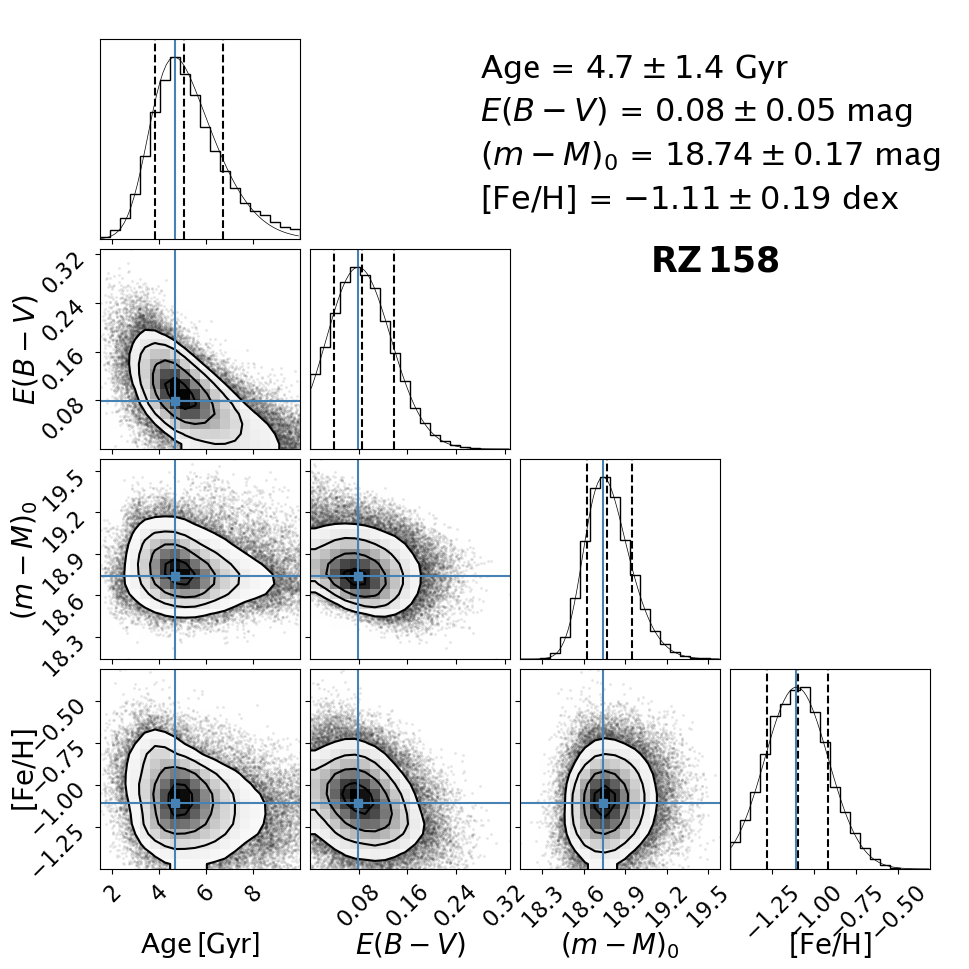}
     \caption{Corner plots for our cluster sample.}
    \label{fig:corner}
\end{figure*}

\begin{figure*}
    \centering
    \includegraphics[width=0.39\textwidth]{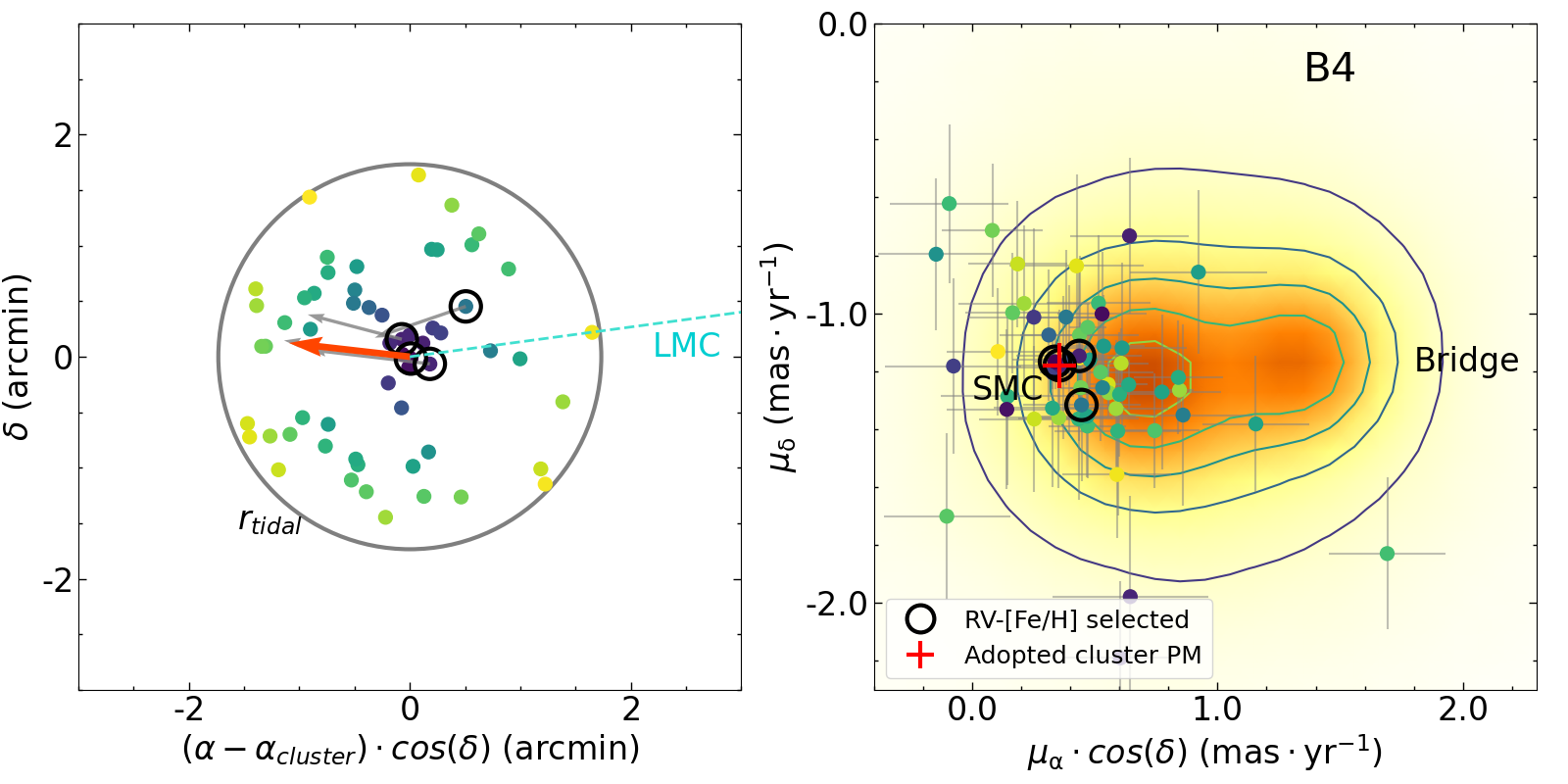} \hspace{0.5cm}
    \includegraphics[width=0.39\textwidth]{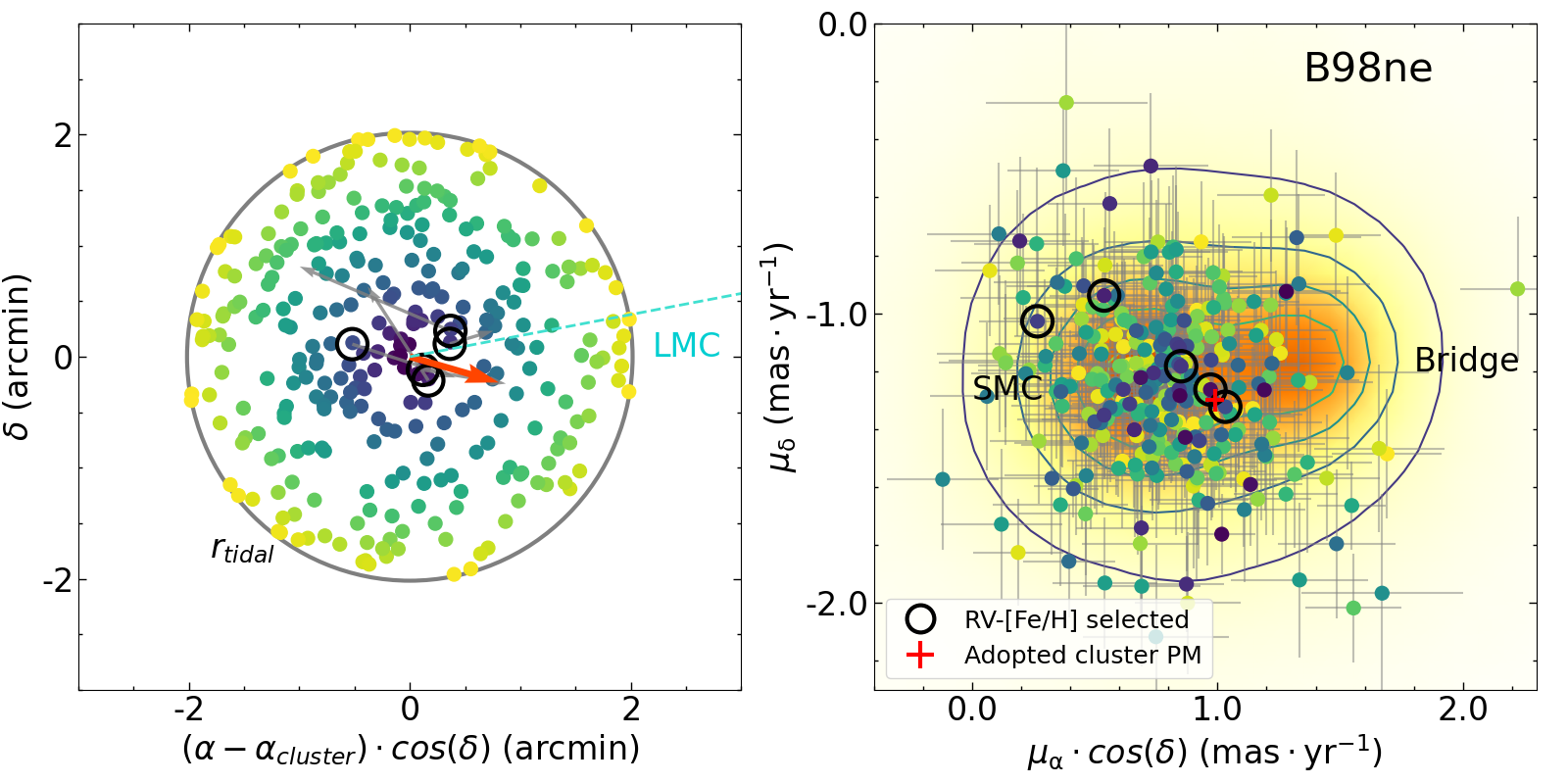}\vspace{0.2cm}
    \includegraphics[width=0.39\textwidth]{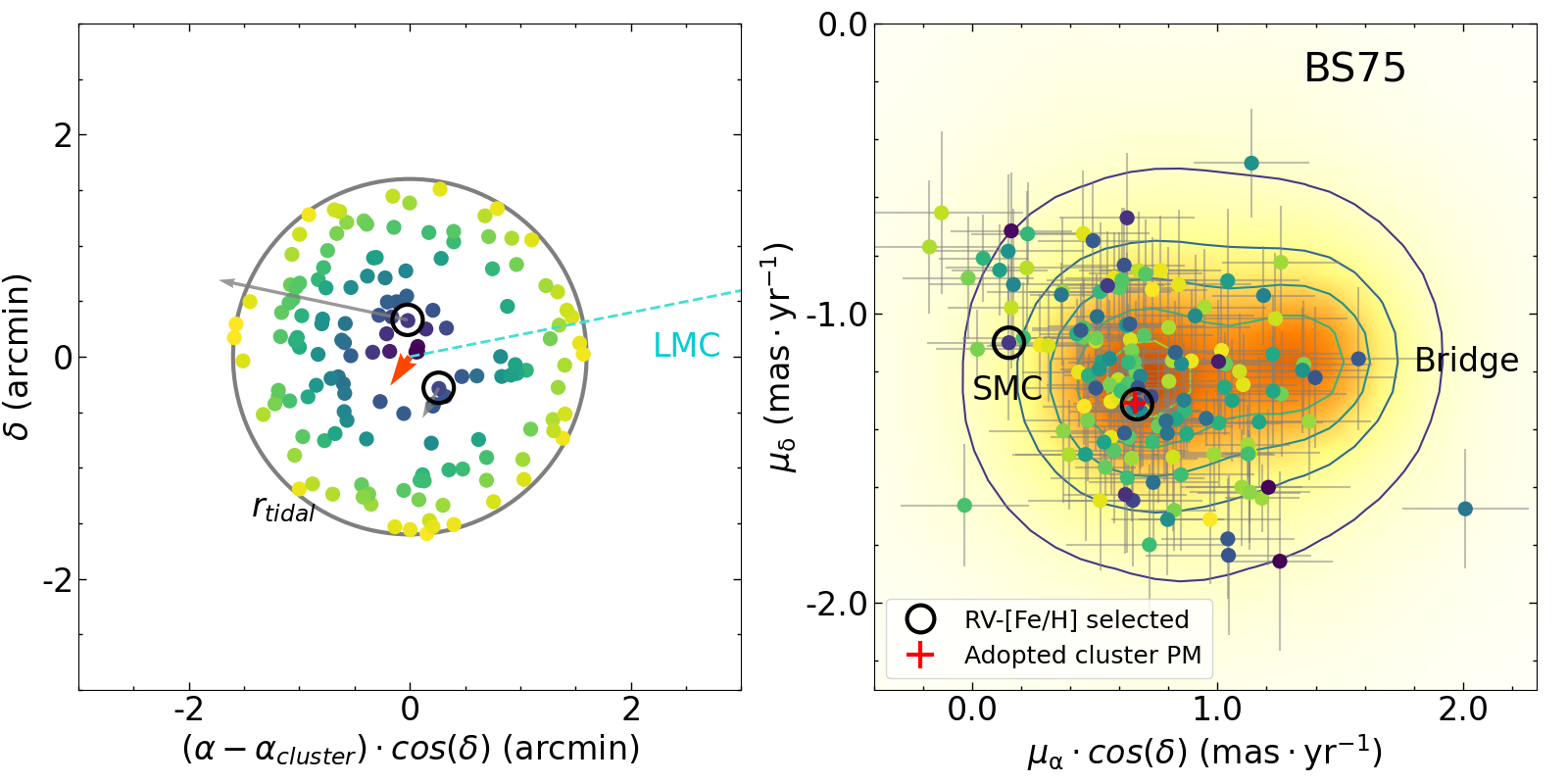}\hspace{0.5cm}
    \includegraphics[width=0.39\textwidth]{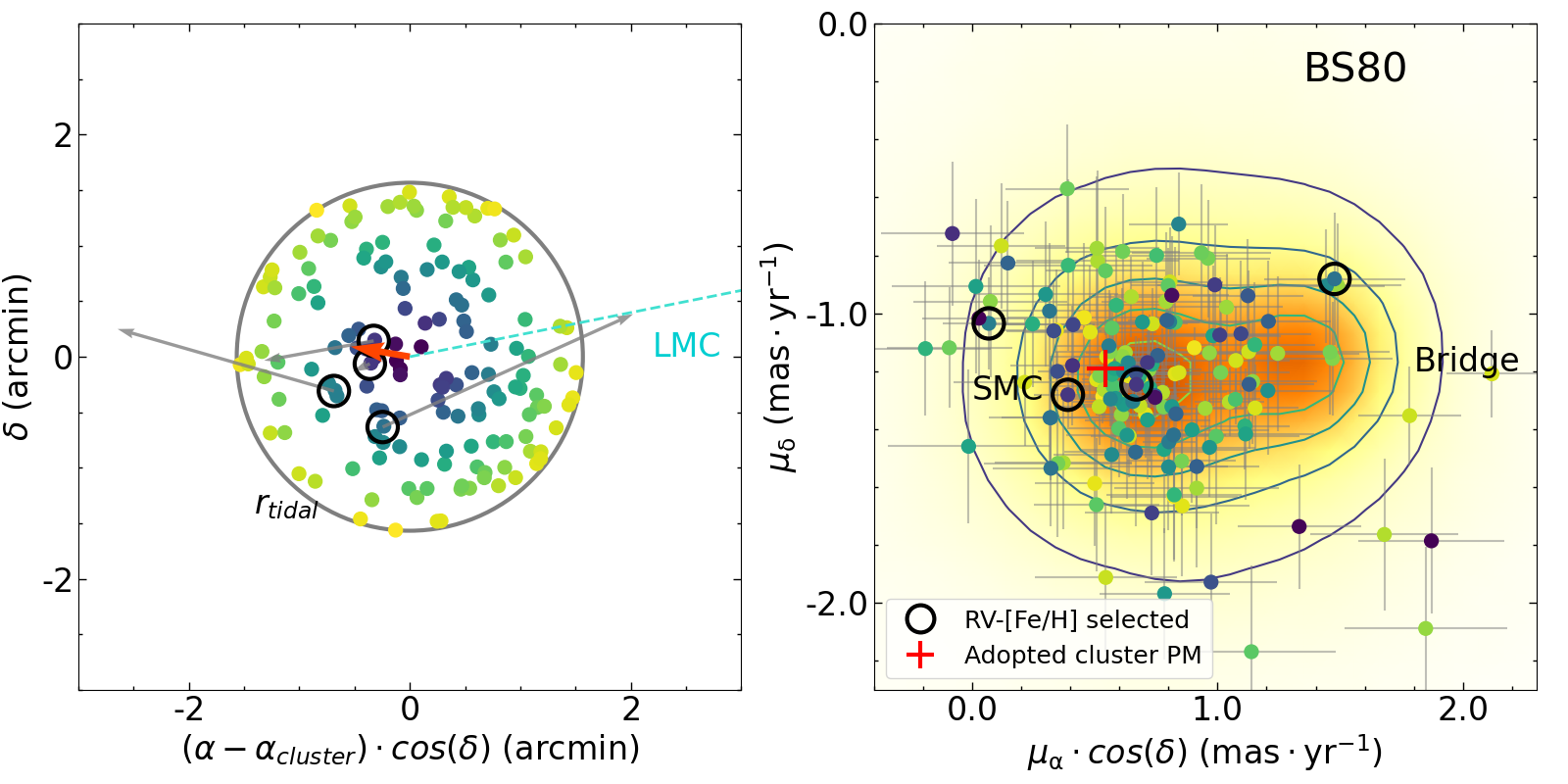}\vspace{0.2cm}
    \includegraphics[width=0.39\textwidth]{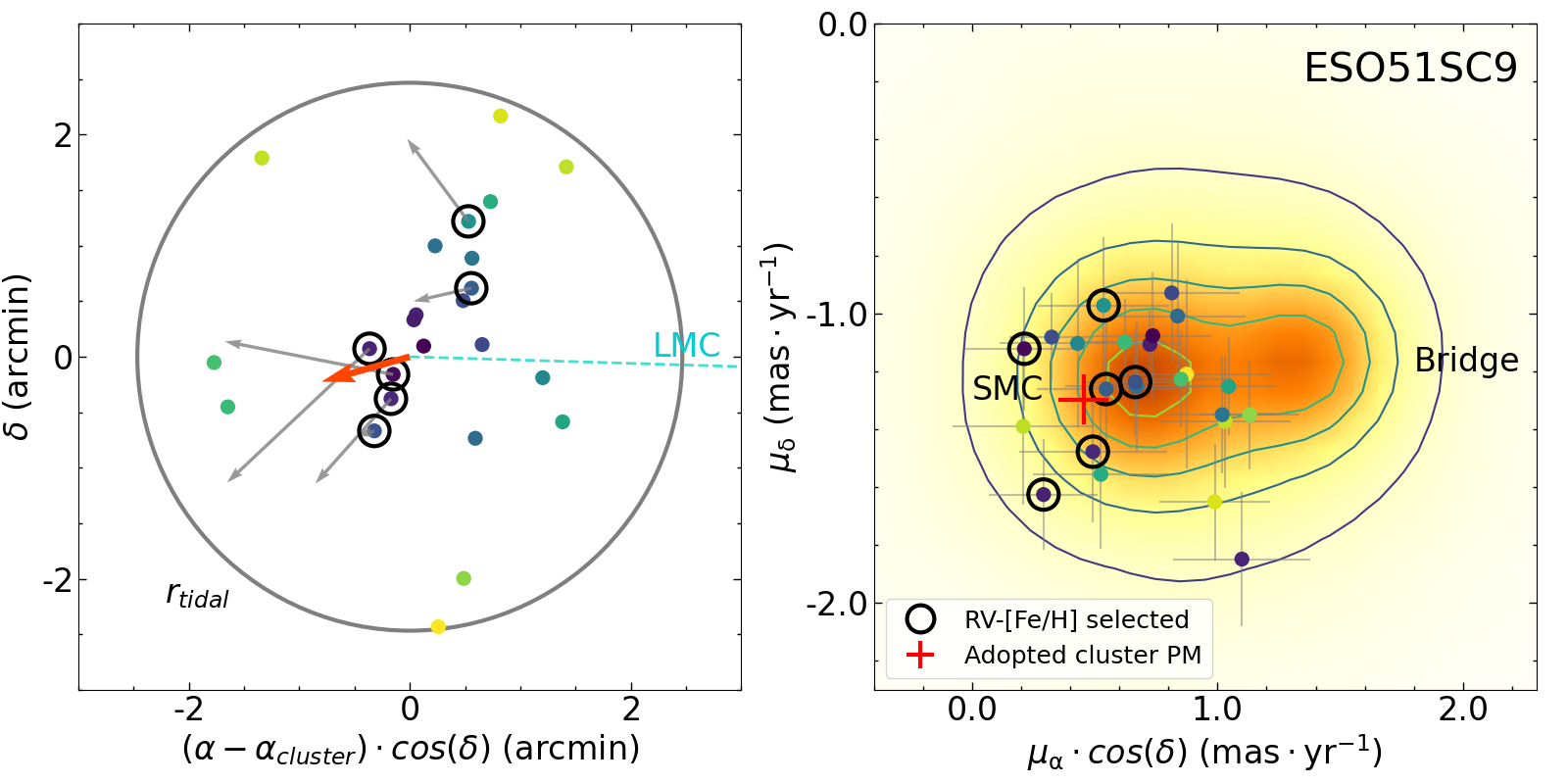}\hspace{0.5cm}
    \includegraphics[width=0.39\textwidth]{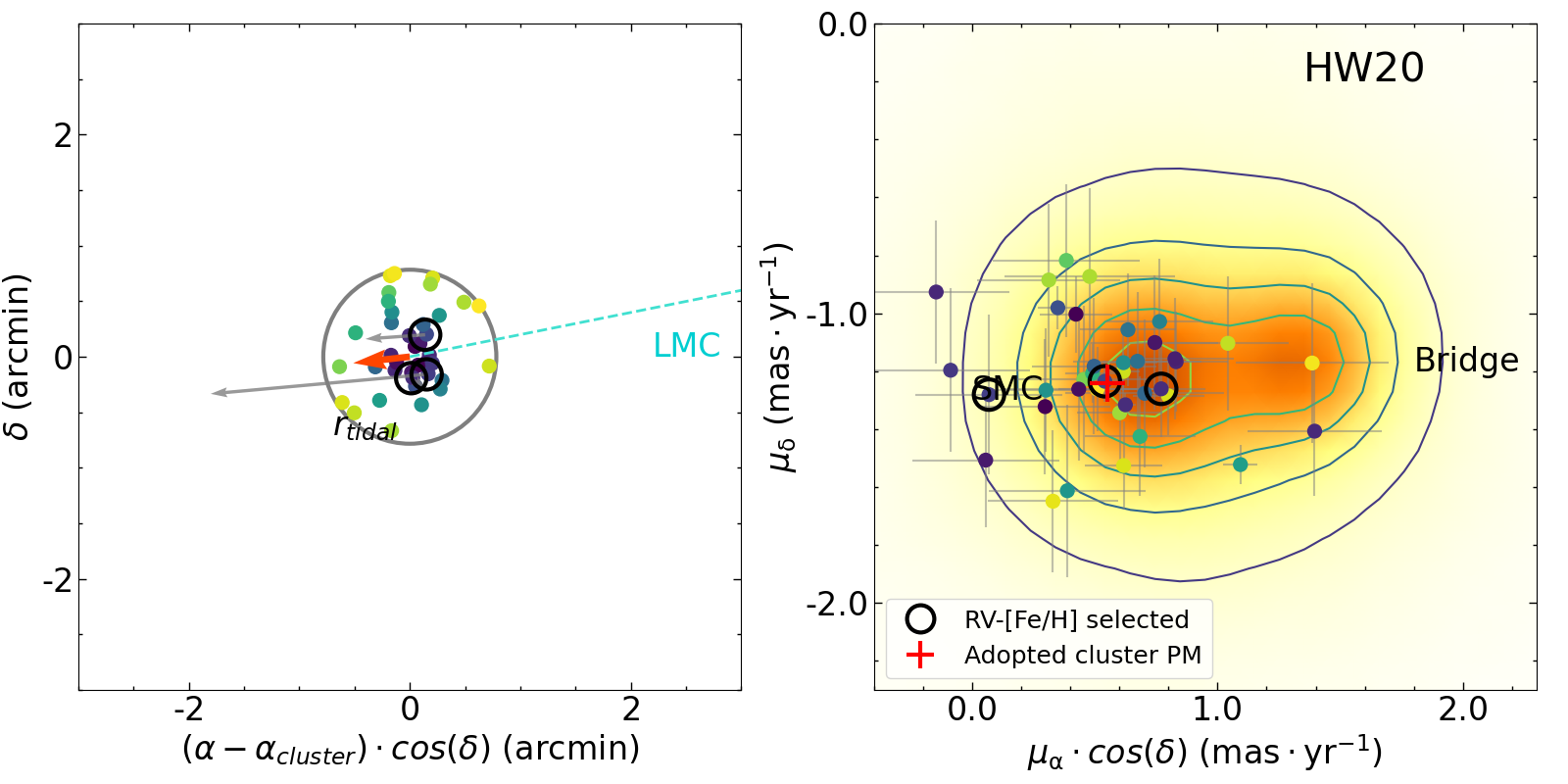}\vspace{0.2cm}
    \includegraphics[width=0.39\textwidth]{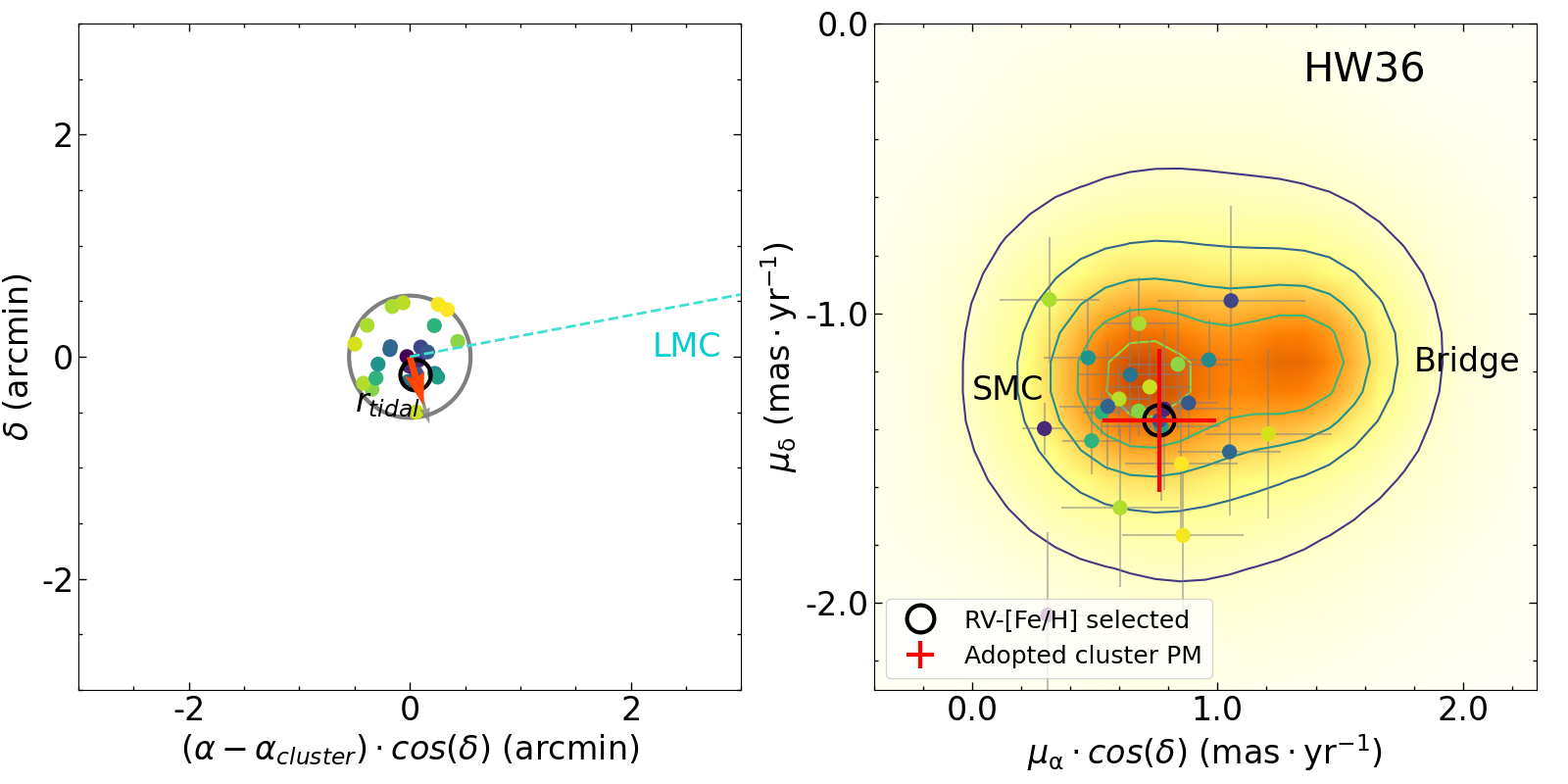}\hspace{0.5cm}
    \includegraphics[width=0.39\textwidth]{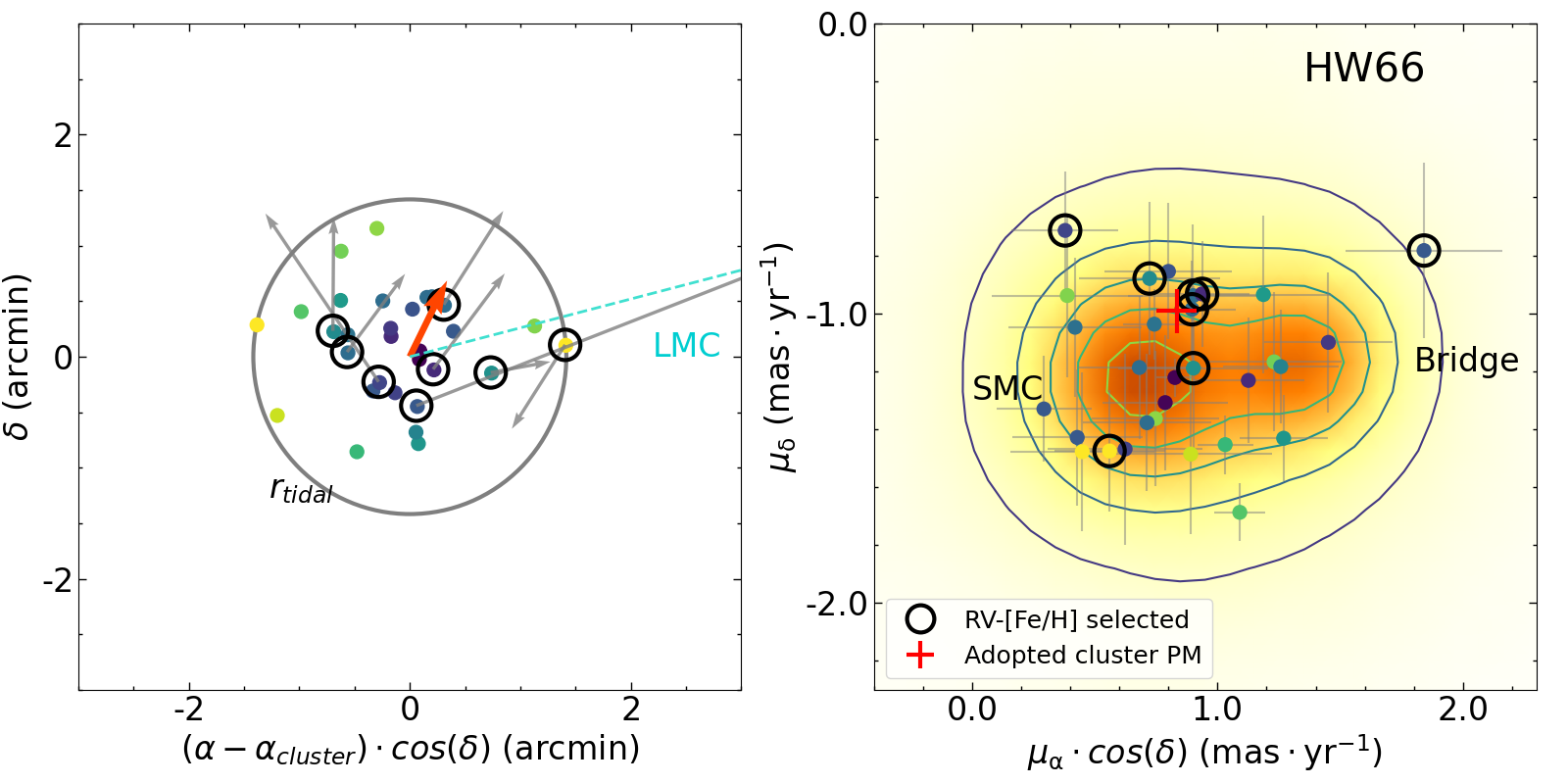}\vspace{0.2cm}
    \includegraphics[width=0.39\textwidth]{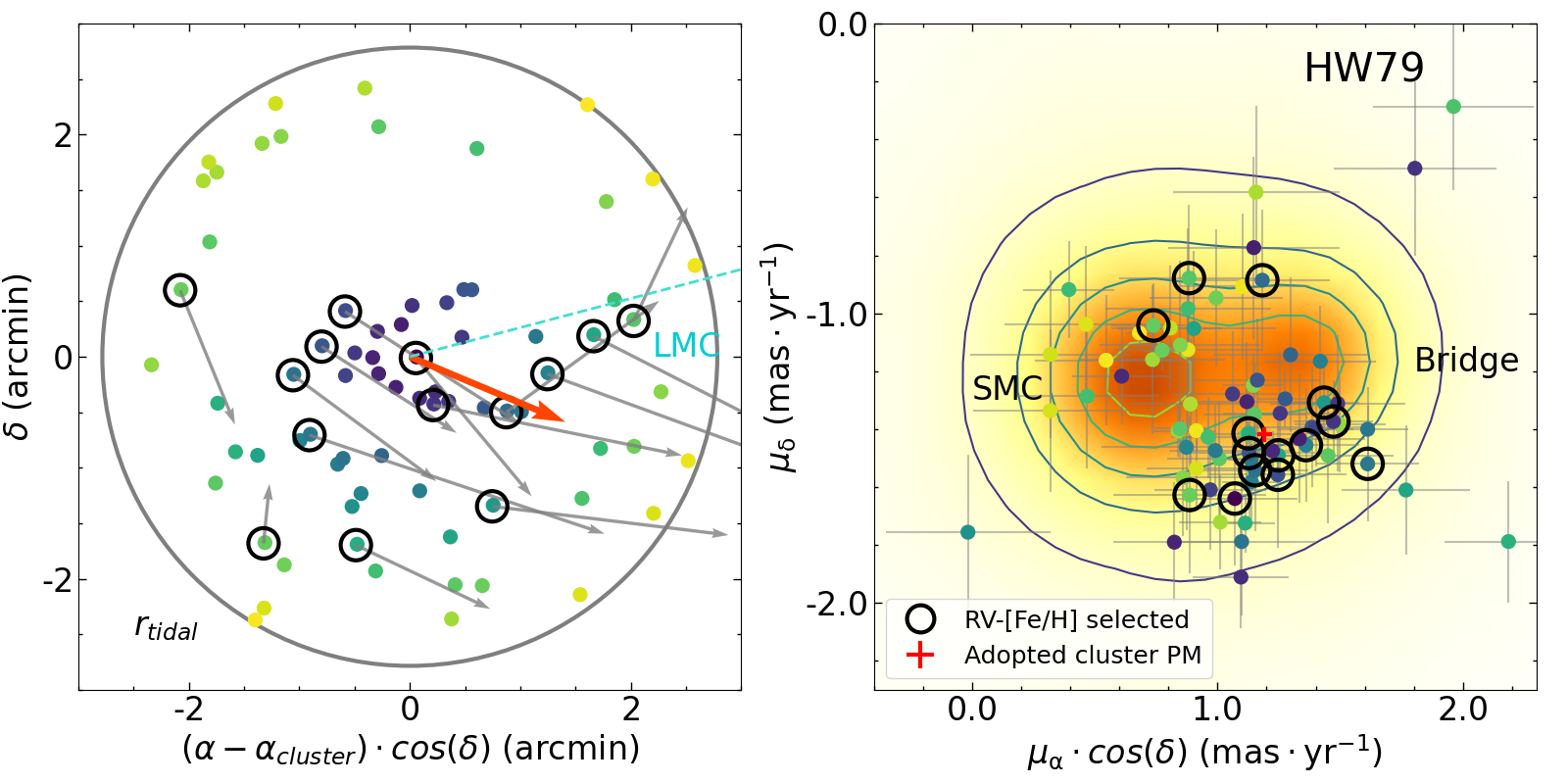}\hspace{0.5cm}
    \includegraphics[width=0.39\textwidth]{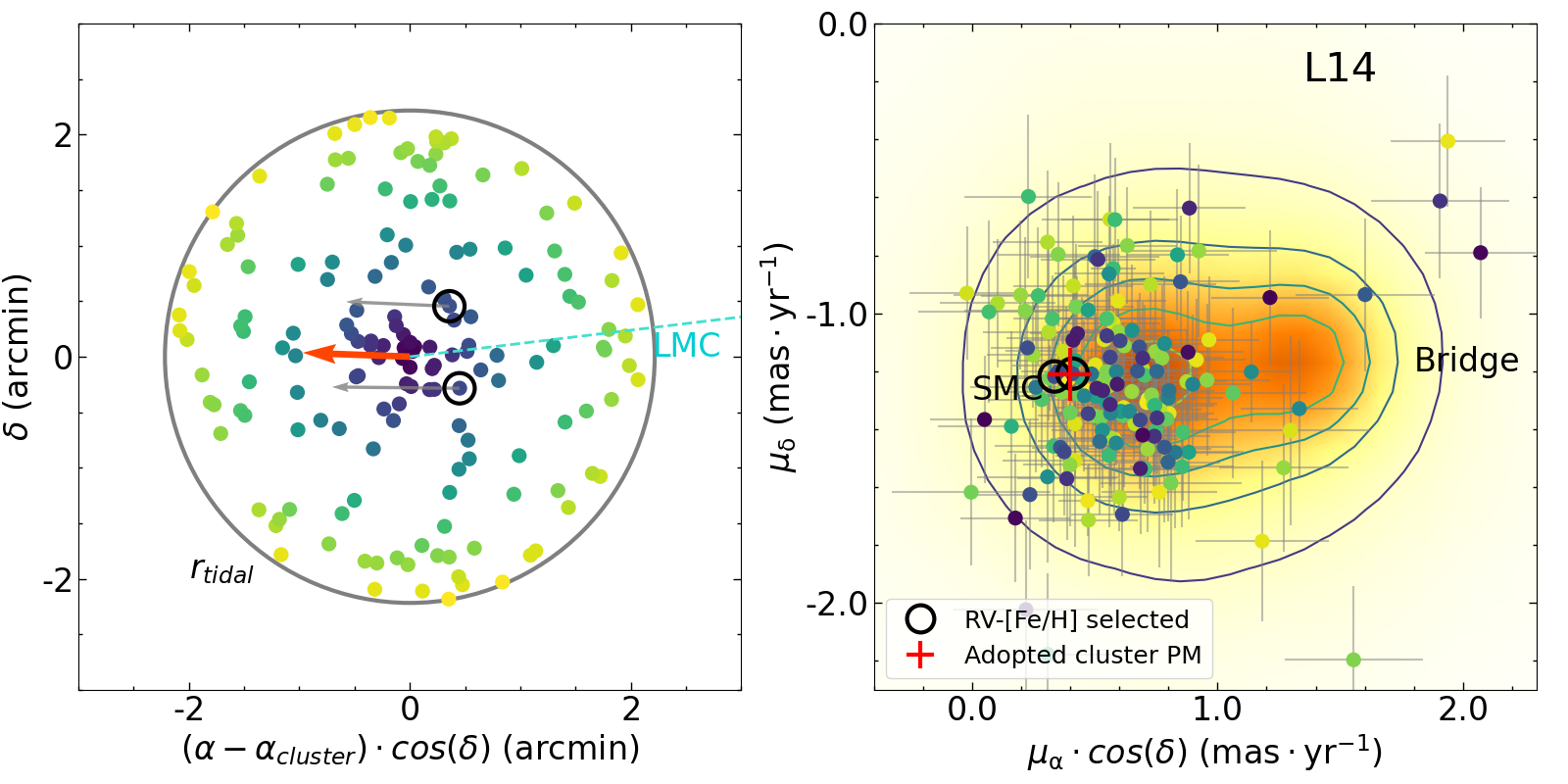}\vspace{0.2cm}
    \includegraphics[width=0.39\textwidth]{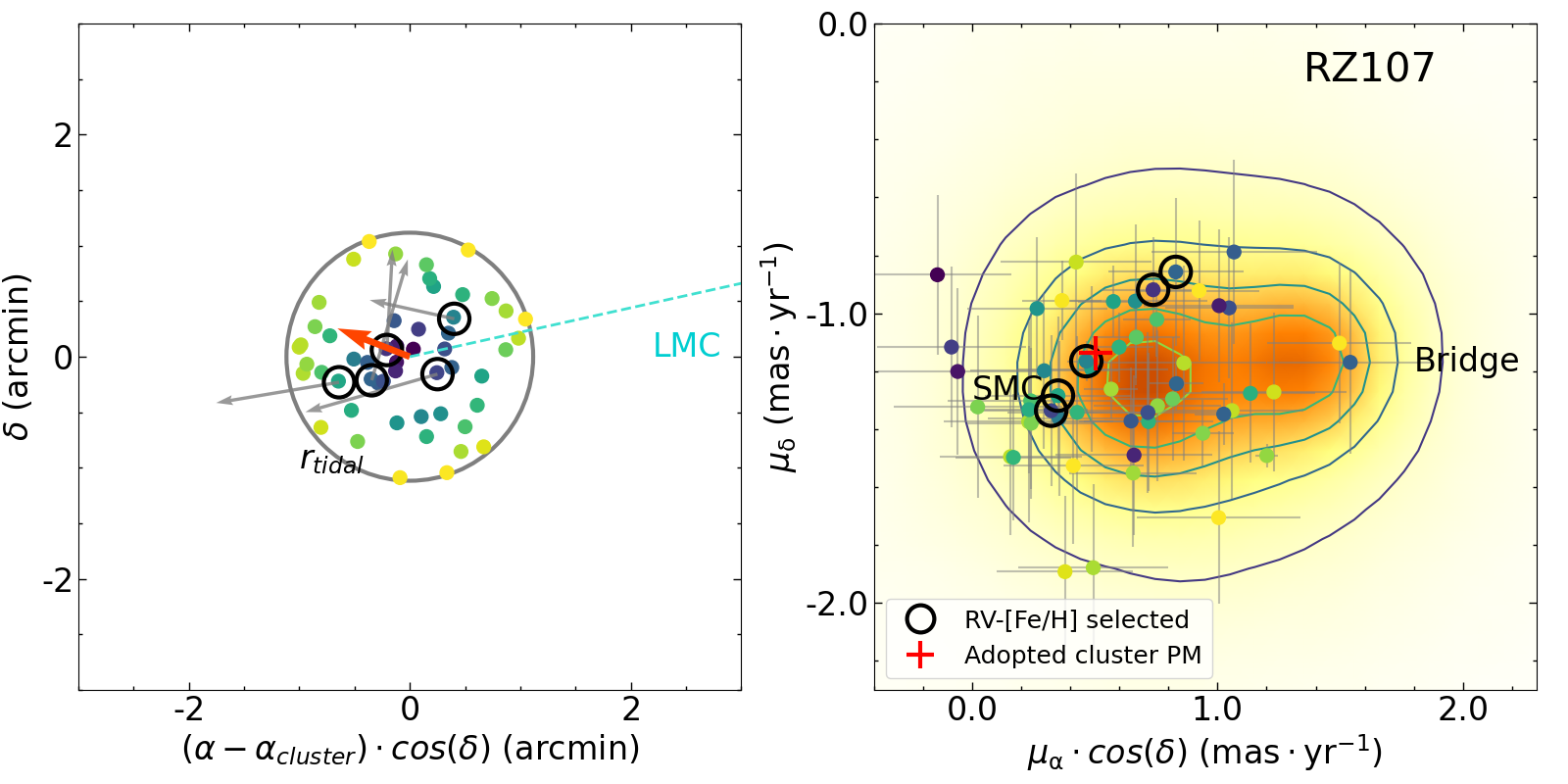}\hspace{0.5cm}
    \includegraphics[width=0.39\textwidth]{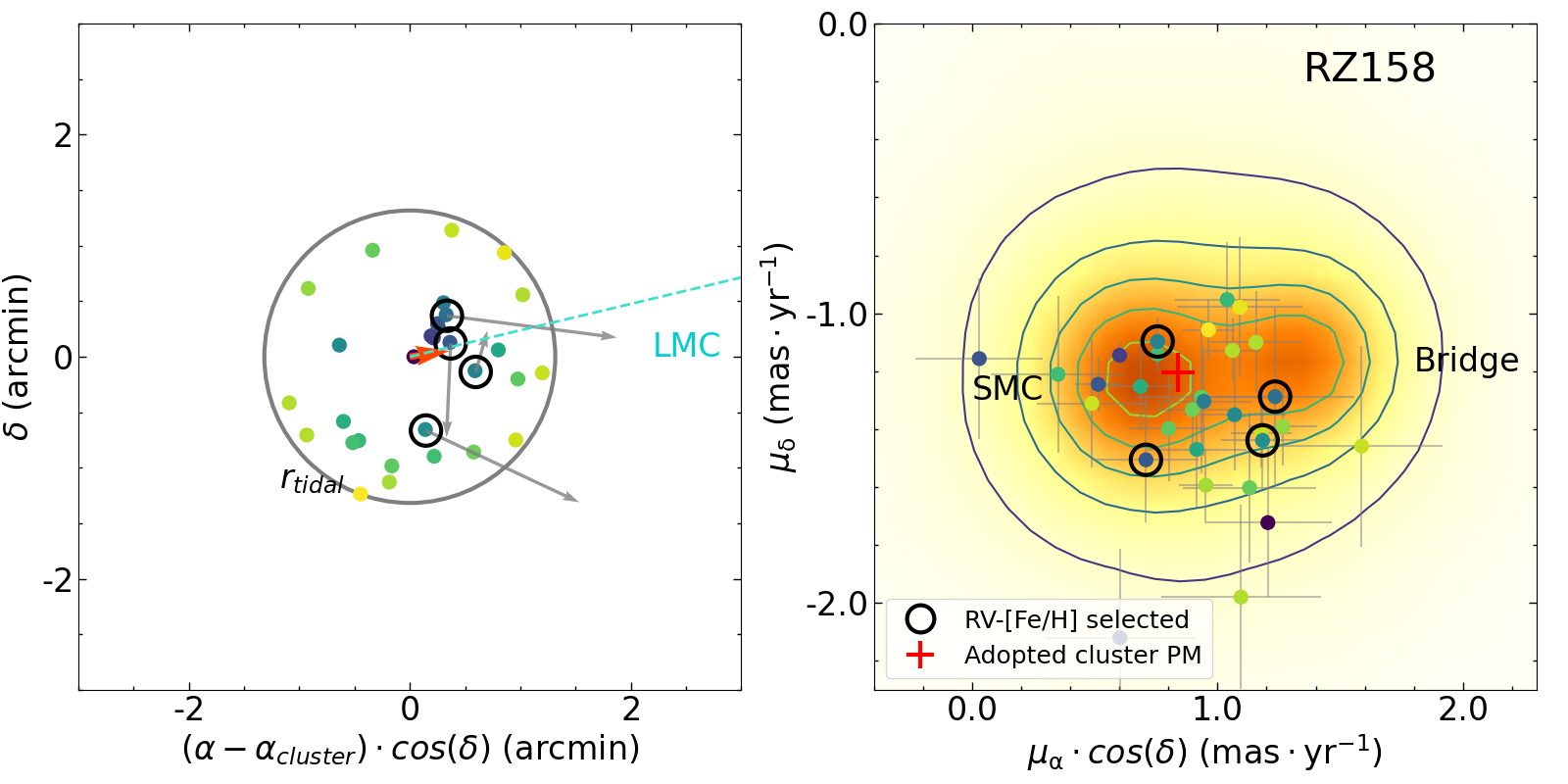} \vspace{0.2cm}   
     \caption{Vector-point diagrams from Gaia eDR3 proper motions for the 12 clusters. Left panels: Spatial distributions of Gaia stars within the tidal radii of the clusters are shown, with colours indicating the distance from the cluster centre. The spectroscopically selected members are marked with black circles. Grey arrows are Gaia eDR3 good-quality proper motions subtracted from the SMC mean proper motion for the spectroscopically selected members, and the red thick arrow is their average PM. The turquoise points towards the LMC. Right panel: The background density plot represents the locus of the SMC and Bridge based on a sample of stars. The points and their error bars are equivalents to the left panels.}
    \label{fig:vpd}
\end{figure*}

In this Appendix we include figures corresponding to the spectroscopic membership selection (Fig \ref{fig:memb}) and clusters structural parameters determinations (Fig. \ref{fig:rad_profiles}), whose respective procedures can be seen in  section \ref{subsec:met}.  Also we present the CMD isochrone fits (Fig. \ref{fig:cmds}) and the corner plots (Fig. \ref{fig:corner}) for our cluster sample, which is explained in detail in section \ref{subsec:decontam}.  The Vector Point Diagrams for our cluster sample can be seen in Fig. \ref{fig:vpd} generated as described in  Section \ref{subsec:pm}.

\bsp	
\label{lastpage}
\end{document}